\documentclass[twocolumn]{aastex63}

\def\a{\alpha}
\def\b{\beta}
\def\g{\gamma}
\def\d{\delta}
\def\D{\Delta}
\def\e{\epsilon}
\def\i{\iota}
\def\k{\kappa}
\def\l{\lambda}
\def\u{\mu}
\def\v{\nu}
\def\s{\sigma}
\def\t{\tau}

\def\x{\xi}

\def\n{\nabla}
\def\w{\omega}
\def\z{\zeta}

\def\p{\phi}

\def\L{\Lambda}

\def\O{\Omega}
\usepackage{hyperref}

\shorttitle{\textsc{Disk Model Applied to Two Recent FU Ori Outbursts}}
\shortauthors{Rodriguez \& Hillenbrand}
\graphicspath{{./}{figures/}}

\usepackage{float}
\usepackage{appendix}
\usepackage{xcolor}
\usepackage{amsmath}
\usepackage{lineno}
\usepackage{makecell}

\begin{document}

\title{Application of a Steady-State Accretion Disk Model \\ to Spectrophotometry and High-Resolution Spectra of Two Recent FU Ori Outbursts}

\correspondingauthor{Antonio C. Rodriguez}
\email{acrodrig@astro.caltech.edu}

\author[0000-0003-4189-9668]{Antonio C. Rodriguez}

\affiliation{California Institute of Technology, Department of Astronomy \\
1200 East California Blvd\\
Pasadena, CA, 91125, USA}
\affiliation{Stanford University, Department of Physics \\
382 Via Pueblo Mall\\
Stanford, CA, 94305, USA}

\author{Lynne A. Hillenbrand}
\affiliation{California Institute of Technology, Department of Astronomy \\
1200 East California Blvd\\
Pasadena, CA, 91125, USA}

\begin{abstract}


We apply a conventional accretion disk model to the FU Ori-type objects HBC 722 and Gaia 17bpi. Our base model is a steady-state, thin Keplerian disk featuring a modified Shakura-Sunyaev temperature profile, with each annulus radiating as an area-weighted spectrum given by a NextGen atmosphere at the appropriate temperature. We explore departures from the standard model by altering the temperature distribution in the innermost region of the disk to account for ``boundary region"-like effects. We consider the overall spectral energy distribution (SED) as well as medium- and high-resolution spectra in evaluating best-fit models to the data. Parameter degeneracies are studied via a Markov Chain Monte Carlo (MCMC) parameter estimation technique. Allowing all parameters to vary, we find accretion rates for HBC 722 of $\dot{M} = 10^{-4.90} M_\odot \textrm{ yr}^{-1}\; {}^{+0.99}_{-0.40} \textrm{ dex}$ and for Gaia 17bpi of $\dot{M} = 10^{-6.70} M_\odot \textrm{ yr}^{-1}\; {}^{+0.46}_{-0.36} \textrm{ dex}$; the corresponding maximum disk temperatures are $7100_{-500}^{+300}$ K and $7900_{-400}^{+900}$ K, respectively. While the accretion rate of HBC 722 is on the same order as other FU Ori-type objects, Gaia 17bpi has a lower rate than previously reported as typical, commensurate with its lower luminosity.  Alternate models that fix some disk or stellar parameters are also presented, with tighter confidence intervals on the remaining fitted parameters. 
In order to improve upon the somewhat large credible intervals for the $\dot{M}$ values,  and make progress on boundary layer characterization, flux-calibrated ultraviolet spectroscopy is needed.
\end{abstract}

\section{Introduction} \label{sec:intro}

Our understanding of star formation at the individual star level currently relies on episodic accretion as a fundamental process in 
building up the total stellar mass  \citep{hartmannreview2016}. 
Given the estimated time scales on which stars become visible as quasi-hydrostatic objects,
and the fact that most accreting young ``T Tauri"-type stars are observed in a low-accretion and low-luminosity state, 
it has been postulated that episodically occurring major accretion events are the major contributors to 
mass accumulation beyond the initial infall stage \citep{hk96}.
These accretion outbursts, during which stars undergo a temporary phase of
high-accretion and high-luminosity, as well as hotter temperature, are estimated to have decay timescales around 50-100 yr. 
Empirically, such accretion burst and large-scale outburst events
are seen as high-amplitude photometric brightenings, accompanied by specific spectroscopic changes,
and lightcurves that eventually fade slowly over decades.

FU Ori objects are a class of young stellar object (YSO) that are currently in a state of rapid accretion. The inferred accretion rates are on the order of $10^{-5} M_\odot \; \textrm{yr}^{-1}$, which is three to four orders of magnitude greater than those of quiescent-state T Tauri stars. \cite{hk96} advocate that the FU Ori phenomenon is appropriately approximated by a disk model with no central star, given that during outburst the disk should outshine the star by a factor of 100-1000.


 \cite{khh88} used a multi-temperature thin-disk model with photospheric templates to model annular regions of the inner disk. They studied the spectral energy distributions (SEDs) as well as the spectral line profiles of the three ``classic" FU Ori objects: FU Orionis, V1057 Cygni, and V1515 Cygni. More recent SED studies of these and other FU Ori sources, such as \cite{gramajo2014}, used a radiative transfer simulation code that mainly serves to capture the emission of the outer disk and circumstellar envelope. 
 
{\ In the innermost disk regions,}
\cite{popham96} argued that a vertically, and radially thick ``boundary region" could be where a significant fraction of the accretion energy is deposited. \cite{popham96} computed temperature profiles and blackbody SEDs of disks around pre-main sequence stars and found that during the high accretion rates of FU Ori events, a vertically and radially thick boundary region forms, rather than a classical thin boundary layer. They conclude that although their fits to the observed SEDs are only marginally better and perhaps not significantly so given the various uncertainties involved in the data, their work nevertheless serves as a self-consistent treatment of the boundary region that presents important improvement to the understanding of FU Ori accretion physics.  

\cite{zhu2007} add that the excess energy expected from a boundary layer could go into expanding the outer layers of the central star.
 \text{In a series of papers,} 
 Zhu et al. developed a detailed radiative transfer disk model \citep{zhu2007} 
 with a flared outer disk or infalling envelope \citep{zhu2008} studying time-dependent evolution \citep{zhu2009}, 
 and more recently magnetic fields \citep{zhu2020}. This work has served as some of the most sophisticated FU Ori disk modeling to date. Despite putting forth a more physically realistic thick disk model, however, the focus was mainly on SEDs and capturing the general trend of optical and infrared spectra -- without looking in detail at high-dispersion line profiles. \cite{khh88} showed that the basic disk model matches many aspects of high dispersion spectroscopy, while \cite{petrovherbig2008} pointed out inconsistencies and suggested the possibility of a rapidly rotating oblate star in the center of the accretion disk model. 
 
In this contribution, we recreate the general infrastructure of \cite{khh88} with the benefit of modern stellar atmosphere models. The newer models include not only updated physics with resulting improved matches to empirical spectra relative to older atmospheres,  but also broad wavelength coverage at moderate-to-high spectral resolution. We consider the FU Ori objects that we model as classical, geometrically thin and optically thick accretion disks. However, we also consider modifications to the standard temperature and velocity distributions expected from pure disks, with particular attention to the inner boundary conditions.
 
In \S2, we assemble the current knowledge of the two objects we model in this study, HBC 722 and Gaia 17bpi. In \S3, we describe the spectrophotometeric and high-resolutions spectral data sets we are working with. In \S4, we outline the steady-state disk (also referred to as the $\a$-disk for brevity) model and corresponding parameters, {\ and in \S5} the methodology we employ to extract best-fit parameters from the data via a Bayesian Markov Chain Monte Carlo (MCMC) technique. In {\ \S6}, we present the resulting best-fit spectral energy distributions (SEDs) along with {\ comparison of the model to} medium-resolution and high-resolution spectra. In {\ \S7}, we discuss the goodness-of-fit of our model to the data, and the limitations of current data in constraining the properties of the star-disk boundary. 
An Appendix provides tutorial information on the effects of various parameters in the disk model.

\section{Objects Modeled in this Study}
\label{sec:objects}


\subsection{HBC 722}

The FU Ori outburst of HBC 722 was revealed by its brightening in 2010, which was discovered independently by \cite{semkov2010} and by \cite{millerhbc722}. The source was immediately followed up by various observers around the world.  \cite{millerhbc722} noted that this object had been classified in the literature as a classical T-Tauri star, characteristically having $M_* \lesssim 1 M_\odot$.  Subsequent pre-outburst spectroscopic data has been uncovered by L. Hillenbrand \citep[appearing in][]{fang2020} and by B. Reipurth (private communication). Both spectra show that indeed, before outburst the source was a fairly normal looking M2-M3 spectral type YSO with H$\alpha$ and \ion{Ca}{2} emission lines. 
The outburst phase SED modeling by \cite{gramajo2014} resulted in an inferred accretion rate of $4\times10^{-6} M_\odot \textrm{ yr}^{-1}$, and that by \cite{kospal2016} yielded $6\times10^{-6} M_\odot \textrm{ yr}^{-1}$ in the maximum among their time series investigation.
A lightcurve can be generated using the American Association of Variable Star Observers (AAVSO) Light Curve Generator\footnote{ \url{https://www.aavso.org/LCGv2/}}. We adopt a distance of 795 pc to HBC 722 as inferred in \cite{kuhn2020}.

\subsection{Gaia 17bpi}
Gaia 17bpi underwent an outburst in 2017 in the optical, but apparently brightened earlier, in 2015, in the infrared.  
It was reported in \cite{gaia17bpi2018}, where the object was classified as an FU Ori-type object based on its light-curve and wavelength-dependent spectral type. 
This source has not been analyzed in the context of a disk model, as we present in this study,
and thus there is no previous estimate of the accretion rate in outburst.
The only other study of Gaia 17bpi is by \cite{kuhn2019} who constrained the X-ray luminosity. A \textit{Gaia} lightcurve can be generated using the Cambridge Photometry Calibration Server\footnote{ \url{https://gsaweb.ast.cam.ac.uk/alerts/alert/Gaia17bpi}}. We adopt a distance of 1.27 kpc to Gaia 17bpi, as inferred in \citep{gaia17bpi2018}.



\section{Spectral Data}
For our modeling purposes, we constructed spectrophotometric SEDs from 0.39--2.2 $\mu$m. Both optical and near-infrared medium-resolution spectra were collected for the two objects in this study, as reported in Table \ref{tab:datalog}. We acquired new optical spectra, and we combined each with previously existing near-infrared spectra.  We also make use of high resolution optical spectroscopy that is used to check the details of our modeling work and its predictions for the line profiles. 

\begin{table*}
    \caption{Observing Log of New and Previously Published Data } 
    \centering 
    \begin{tabular}{|c|c|c|c|c|c|} 
    \hline\hline 
    Object & Observatory/Instrument & Date (UTC) & $\lambda$ Range & R ($\lambda/\Delta\lambda$) & Reference\\ [0.5ex] 
    \hline 
    Gaia 17bpi & Palomar 5m/DBSP & 27 Jul. 2019 & 3850 - 9000 \AA & 4,000 \& 6,000 & this paper\\
    Gaia 17bpi & IRTF/SpeX &  25 Jun. 2019 & 2.0 - 2.4 $\mu$m & 2,000 & Connelley 2022, in preparation\\
    Gaia 17bpi & Keck 10m/HIRES & 3 Nov. 2018 & 4800 - 9200\AA & 36,000 & Hillenbrand 2022, in preparation\\
    
    HBC 722 & Palomar 5m/DBSP & 27 Jul. 2019 & 3950 - 9000 \AA& 4,000 \& 6,000 & this paper\\
    HBC 722 & Lick 3m/Kast & 16 Sept. 2010 & 3550 - 10000 \AA & 6,000 & \cite{millerhbc722} \\
    HBC 722 & Palomar 5m/TripleSpec & 23 Sept. 2010 &  1.2 - 2.4 $\mu$m & 2,700 & \cite{millerhbc722}\\
    HBC 722 & WHT 4.2m/LIRIS & 11 Jul. 2011 &  0.9 - 2.4 $\mu$m & 2,000 & \cite{kospal2016}\\
    HBC 722 & Keck 10m/HIRES & 20 May 2016 & 4800 - 9200 \AA & 36,000  & Hillenbrand 2022, in preparation \\
    \hline
    \end{tabular}\\
    \label{tab:datalog}
\end{table*}
\subsection{New Optical Spectrophotometric Observations}

Our new data were acquired on July 27, 2019 (UTC), using the Double Spectrograph (DBSP) on the 200-inch Hale telescope at Palomar Observatory.
Seeing was reported to be around 1.5", and a 1" slitwidth was used.
There was no Moon in the sky during the entire observation sequence.
The blue side of the spectrograph used a 600 lines/mm grating blazed at 3780 \AA, centered at 5350 \AA. The red side used a 1200 lines/mm grating blazed at 7100 \AA, centered at 8200 \AA. 
The D68 dichroic splitting the blue and red beams effectively 
cuts off the blue spectrum redwards of $\sim 6800$ \AA, and creates a break in our optical data.
We began the observations with a spectrum of HD 338808 (BD +25 3941), a B 1.5V standard star from the list of IRAF \citep{tody1986} 
standards. 
In addition to observing both HBC 722 and Gaia 17bpi, we also obtained spectrophotometry of V1057 Cyg and V1515 Cyg as FU Ori type comparison objects.

To reduce the data, we used the \texttt{pyraf-dbsp} pipeline\footnote{\url{https://pyraf-dbsp.readthedocs.io/en/latest/}}
developed by E. Bellm and B. Sesar. 
To calibrate the wavelength solution, iron-argon lamps were used for the blue side of DBSP, and helium-neon-argon lamps for the red side. The pipeline is capable of bias- and flat-subtracting the images, extracting the one-dimensional spectra, applying the pixel to wavelength calibration, and flux-calibrating the spectra, all using the standard IRAF/PyRAF tools.

The lack of flux-calibrated points in the bluest and reddest ends of the spectrum of our spectrophotometric standard star required an additional calibration of our data. Three stars from the LAMOST catalog were used for this step. While the observation of the spectroscopic standard and additional stars lets us achieve the relative flux calibration across the wavelength range, the absolute level of the fluxing can still be off, given that the seeing exceeded the slit width. Our additional calibration step puts the data onto a reliable flux scale, but because seeing varies with wavelength, we cannot account for wavelength-dependent slit losses.

We also use an optical spectrum for HBC 722 taken using the Shane 3m telescope at Lick Observatory as reported in \cite{millerhbc722}. The flux calibration for this data was performed using a standard star at high airmass while HBC 722 was observed at low airmass; thus, the absolute flux calibration for this spectrum is not adequate for the SED analysis of our work. We limit the use of this data to examination of spectral lines, especially towards the blue.

\subsection{Near-Infrared Spectrophotometry }

Near-infrared spectra for HBC 722 were reported in \cite{millerhbc722} and \cite{kospal2016}. The former was taken using the TripleSpec spectrograph on the Palomar 200-inch Hale telescope, and reduced with an IDL-based data reduction pipeline developed by P. Muirhead. It was sky-subtracted and divided by a flat field using standard techniques, and flux calibrated using an AO star.
The latter was obtained using LIRIS and the WHT and reduced and flux calibrated using standard techniques.

A near-infrared spectrum for Gaia 17bpi was reported in \cite{gaia17bpi2018}.  A more recent spectrum was 
acquired by M. Connelley on 27 June 2019 using the NASA Infrared Telescope Facility (IRTF)  and SpeX \citep{rayner2003} in its SXD mode. 
The individual frames were sky-subtracted, divided by a flat field, extracted, wavelength calibrated, and flux calibrated using standard techniques.  These data will be more fully reported elsewhere.

\subsection{{\ Creating Coherent SEDs}}
\label{sec:corrections}

The lightcurves of FU Ori objects after their initial outburst are not static. Since the acquisition of our optical and near-infrared spectrophotometry is not simultaneous, we inspected the lightcurves of both HBC 722 and Gaia 17bpi to determine if there were significant differences in brightness between the time of each observation. If so, we applied a multiplicative correction (a wavelength-independent shifting of flux for a data set sometimes referred to as a  ``grayshift") to the dimmer observation to better represent the overall SED for our modeling purposes.

The photometric monitoring of HBC 722 reveals a change in the I-band brightness between the time of the near-infrared and optical data sets that we use
of around 0.25 mag (for the Sept. 2010 spectrum) and 1.0 mag (for the Jul. 2011 spectrum). In the lightcurve, we find the point nearest in time to each of our observations and rescale our non-contemporaneous datasets to match based on the photometric flux. 

We implement this grayshift to the TripleSpec (Sept. 2010) near-infrared spectrum as a factor of 1.2 correction and to the LIRIS (Jul. 2011) near-infrared spectrum as a factor of 2.5 correction.  
Any slight differences from the ideal correction should be on the order of the error due to flux calibration. 
We also call attention to the fact that the LIRIS spectrum, that was published in \cite{kospal2016},
has more pronounced water absorption and is also brighter in H-band and K-band 
than the TripleSpec spectrum we initially model, which is that of \cite{millerhbc722}. 

The near-infrared spectra for Gaia 17bpi were acquired in June 2019, when the object was at its maximum brightness. 
Our optical spectra, however, were acquired in the noticeable dip in the lightcurve of Gaia 17bpi, occurring 
between July and August 2019. 
As a consequence, we use a grayshift to correct the optical spectra by a factor of 1.25 in order to match the flux level of the near-infrared spectrum. 

\subsection{Optical High Dispersion Spectra}


The high dispersion spectra used in this analysis were acquired at the W.M. Keck Observatory
Keck I telescope with HIRES \citep{vogt94}, the facility high dispersion spectrograph.
The grating angle setup and C5 decker resulted in wavelength coverage from 4800--9200 \AA, 
with some gaps between the redder orders, at resolving power R=36,000.

The two-dimensional spectral images were processed with the MAKEE 
pipeline reduction package, written by Tom Barlow.   
The signal-to-noise ratio of these data is
$\sim$10 at 5000 \AA\ and $\sim$25 at 8000 \AA\  for Gaia 17bpi, and 
$\sim$30 and $\sim$75 at the same wavelengths for HBC 722.

\section{Disk Model Components}

The model that we adopt in this study is a {\ standard} steady-state \text{(no radial variation of accretion rate)}, geometrically thin \text{(no vertical structure)}, optically thick \text{(no radiative transfer)}, accretion disk.  The luminosity is purely thermal and due to viscous shear in the disk. The following sub-sections describe the various components that we consider in our calculations, and the model parameters that we wish to constrain through comparisons to data.

{\
Two key elements of the disk model are the adopted temperature profile and the adopted velocity profile, both as a function of stellocentric radius.

The model inputs include 
the stellar parameters $M*$, mass, and $R_*$, radius. 
The main disk parameter is $\dot{M}$, the rate of mass accretion through the disk and onto the star.
Additional disk parameters that we consider are $i$, the inclination with respect to our line of sight, 
and $A_V$, the line-of-sight extinction.
Finally there is $\gamma$, the power law exponent governing the temperature distribution 
in the boundary region of the disk.
Appendix \S \ref{sec:appendix_tutorial} demonstrates the effect of modifying individual parameters on the model SED, and the various parameter degeneracies. 
}

\subsection{Temperature Profile}

We consider both a standard disk temperature profile, and a modified profile that accounts for possible variations in the innermost disk temperature profile in a region we refer to as the ``boundary region".

\subsubsection{Shakura-Sunyaev Profile}

The fundamental physical equation that governs the flux output of the disk is the temperature profile as a function of radial distance from the central star. \cite{ss1973} derive such a profile for a geometrically thin accretion disk around a black hole, known as an $\alpha$-disk for the single parameter needed to describe the efficiency of the angular momentum transport mechanism. In their study,  \cite{ss1973} break up the disk into concentric annuli, whose viscous shear with one another lead to the thermal luminosity. A calculation balancing momentum and energy, in which they assume that $\alpha \ll 1$ leads to a temperature profile described in terms of
stellar mass $M_*$, \text{disk accretion rate $\dot{M}$ (constant throughout the disk)}, and $r$ the radial distance in the disk, as
\begin{equation}
     T^4_{\text{eff}}(r) = \frac{3GM_*\dot{M}}{8\pi \sigma r^3}\Big[1 - \Big(\frac{r_i}{r}\Big)^{1/2}\Big]
     \label{eq:temp_proifle}
\end{equation}
Here, $r_i$ is intended by \cite{ss1973} as the innermost stable circular orbit of the black hole: the point where viscous shear terminates due to the lack of stably rotating material inside that radius; $r_i$ is several times the Schwarzschild radius. 
\subsubsection{Considerations for FU Ori Objects}

As it is believed that FU Ori accretion is so strong that it breaks through the conventional magnetospheric accretion of T Tauri stars and dumps material directly onto the surface of the central star, we set $r_i = R_*$ \citep[following e.g.][]{hk96}. We place caveats on this choice, as has been done since \cite{khh88}. Given that the material from the disk flows directly onto the star, it may be unclear how to even define $R_*$ during an FU Ori outburst.

Importantly, the  \cite{ss1973} temperature profile features a characteristic $T_\text{max}$ which is obtained at $r = 1.361R_* $, with the temperature decreasing in both radial directions away from $r(T_\text{max})$. Thus, \cite{khh88} chose to set the temperature at radial distances less than $1.361R_*$ to remain at $T_\text{max}$, so as to avoid the unphysical nature of the temperature going to zero at the stellar surface. However, as mentioned in \cite{khh88}, this constitutes an inconsistent treatment of the boundary layer region. To address this, we investigate if we can place reasonable constraints on a new parameter we introduce to govern the behavior of the temperature profile of the inner disk, given our current data. 

\subsubsection{Our \text{Adopted} Temperature Profile}\label{sec:gamma_parameter}
\begin{figure}
    \centering
    \includegraphics[scale=0.38]{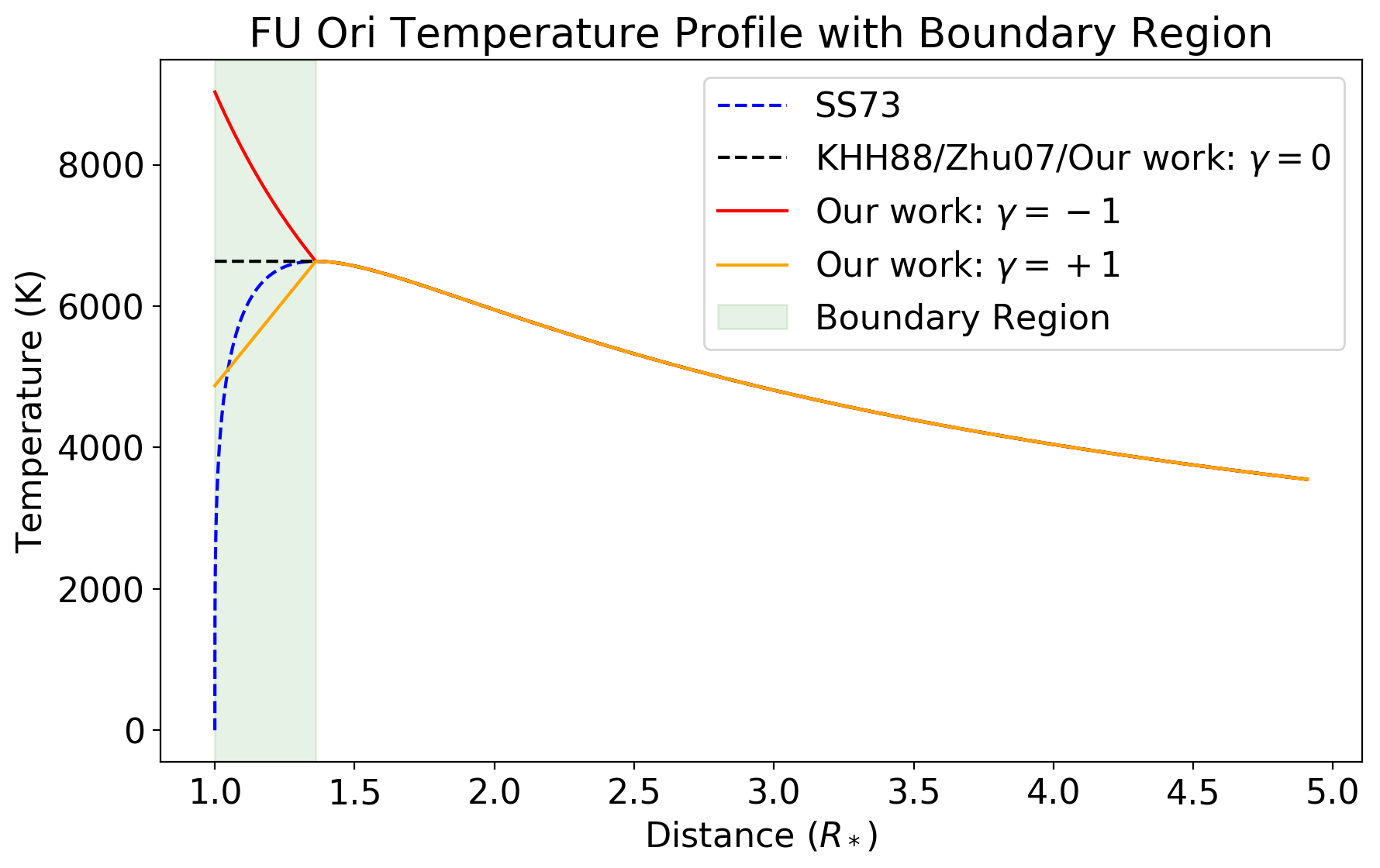}\\
    \caption{The temperature profiles shown are: the one originally developed by \cite{ss1973}, the one similar to SS73 but with modification made for FU Ori disks by \cite{khh88}, and the ones we develop and explore in this work. Fiducial values of $M_*, R_*, \textrm{ and } \dot{M}$ are used as inputs to the temperature profile. The single parameter $\gamma$, as defined in Equation \ref{eq:new_temp_profiles}, 
    is the power law factor governing the behavior of the temperature profile in the innermost disk or ``boundary region".}
    \label{fig:new_temp_profiles}
\end{figure}

In the spirit of \cite{popham96}, we introduce a parameter, $\gamma$, to capture the behavior of the temperature profile in the ``boundary region" between $1R_*$ and $1.361R_*$. Our adopted temperature profile is defined by Equation \ref{eq:new_temp_profiles}, 
\begin{gather}
    T^4_{\text{eff}}(r) = \frac{3GM_*\dot{M}}{8\pi \sigma r^3}\Big[1 - \Big(\frac{R_*}{r}\Big)^{1/2}\Big] \text{ for } r > 1.361R_{*}\notag
    \\
    T_{\text{eff}} (r) = T_{\text{max}}\left(\frac{r}{1.361 R_*}\right)^\gamma \text{ for } R_{*} \leq r \leq 1.361R_{*}
    \label{eq:new_temp_profiles}.
\end{gather}
As above, we have the parameters $\dot{M}$, the accretion rate and $M_*$, the mass of the star, but now also $R_*$, the radius of the star, and $\gamma$, the boundary region power law slope.  Consistent with previous work, 
we use $T_\textrm{max}$ to indicate the temperature at $r = 1.36 R_*$, but allow the temperature to increase or decrease within that radius, depending on the choice of $\gamma$. Figure \ref{fig:new_temp_profiles} provides a visual illustration. 

While our work emulates the intuition behind \cite{popham96} in adding an additional parameter to characterize the boundary region, such a simple modification to the temperature profile is by no means rigorously derived. However, we believe that this approach can add to our qualitative understanding of the boundary region of FU Ori objects, while maintaining the previously established $\a$-disk disk behavior at $r > 1.361R_{*}$.

\subsection{Rotation Profile and Disk Broadening}
Assuming simple Keplerian rotation, the specific luminosity, $L_\l$, must be broadened by a kernel corresponding to a rotating disk. As adopted in \cite{khh88}, we convolve the following kernel 
\begin{equation}
    \phi(\l) = \Big[1 - \Big(\frac{\l - \l_0}{\l_{\text{max}}}\Big)^2\Big]^{-1/2}
    \label{eq:disk_broaden}
\end{equation}
where $\l_0$ is the central wavelength of the range of $L_\l$ in question, and
\begin{equation}
    \l_{\text{max}} = \l_0\frac{v_{\text{Kep}}(r)\sin i}{c} = \l_0\frac{\sin i}{c}\sqrt{\frac{GM_*}{r}} 
    \label{eq:disk_broaden_2}
\end{equation}
where $r$ corresponds to the mean radial distance of the annulus from the central star, and $i$ is the inclination of the disk, 
with the specific flux (set to a stellar atmosphere) of each annulus.
We perform rotational broadening in our modeling only when comparing the accretion disk models to spectra, either moderate or high resolution, and not when considering the broad SEDs. 



\subsection{Extinction}

The final element in our model is extinction, arising from the absorption and scattering of light by dust grains in the interstellar medium, as well as from any circumstellar structure that obscures optical and infrared continuum photons.
FU Ori stars, being located in star-forming regions replete with dust grains,
as well as being surrounded by circumstellar disks and sometimes envelopes, 
typically have high extinction or $A_V$, as detailed and derived in \cite{connelley2018}. 

We make use of the \texttt{PyAstronomy} package \footnote{\url{https://pyastronomy.readthedocs.io/en/latest/index.html}}
developed by S. Czesla and S. Schröter, to implement the \cite{fitzpatrick99} extinction correction. 
We adopt the interstellar medium average of $R_V = 3.1$ throughout our study. While star-forming regions can have different,
typically larger values of this parameter, 
we choose to fix it at 3.1 so as to avoid introducing another parameter, especially one with redundancies with our other parameters in terms of effect on the computed SEDs. 

\section{Disk Model Methodology}

\begin{figure*}[htp]
    \centering
    \includegraphics[scale=0.48]{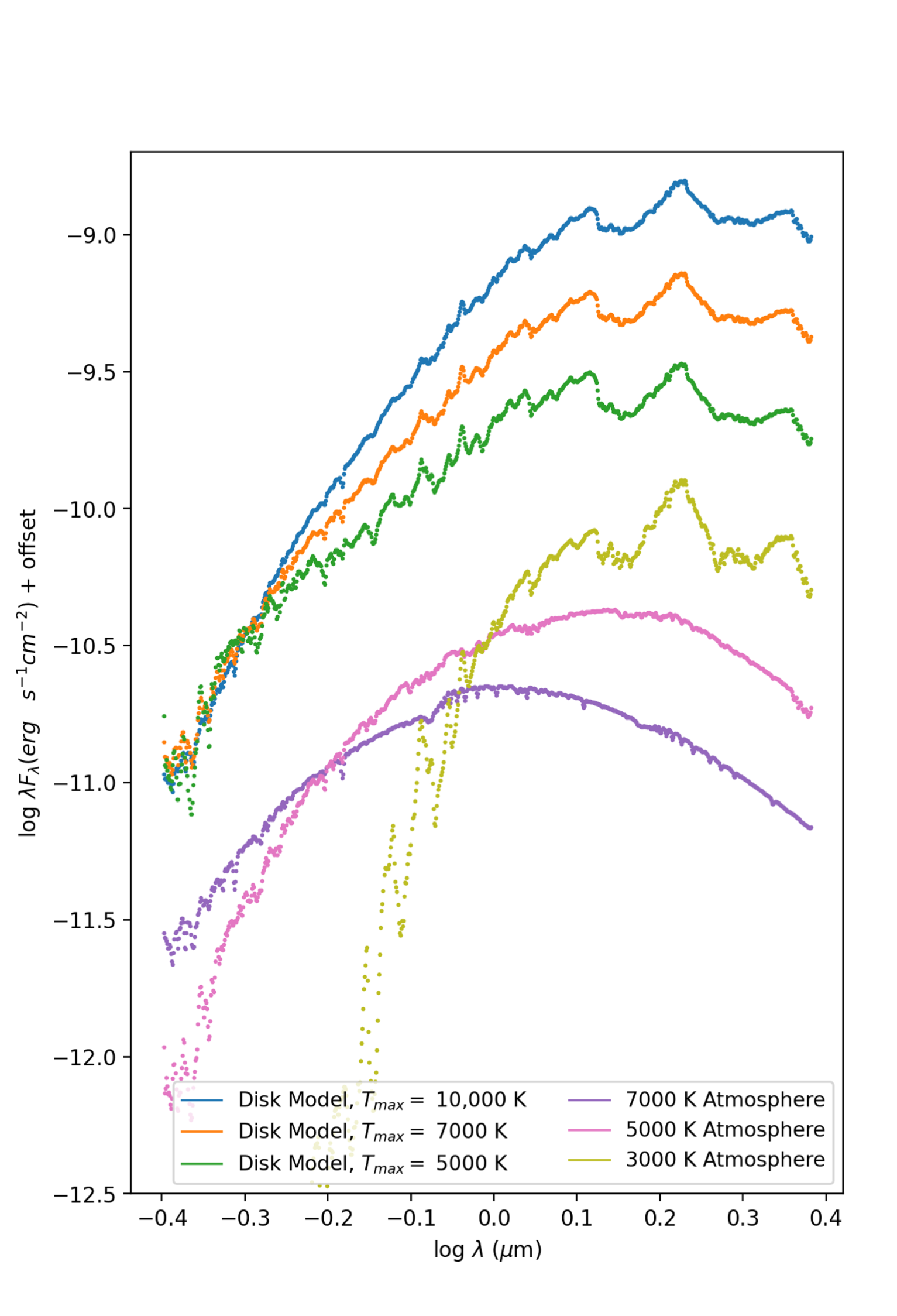}\includegraphics[scale=0.45]{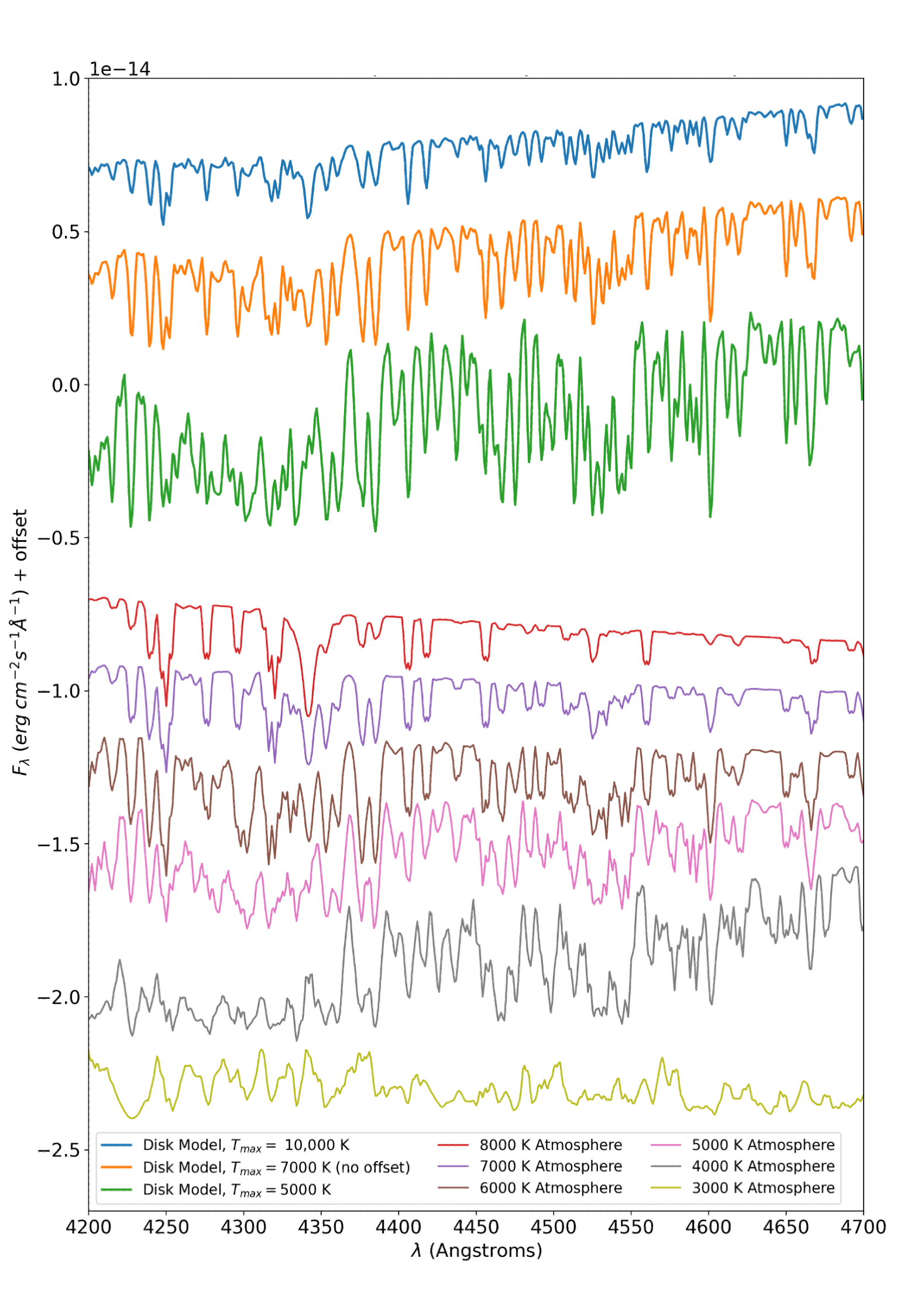}
    \caption{Three disk models {\ with both temperature and velocity gradients}, and several single-temperature models {\ broadened by a single-velocity disk kernel},
    presented as SEDs (left) and medium-resolution spectra (right). The three disk models have the same fiducial parameters, namely Model 2a in Table \ref{tab:hbc722_table}, with the accretion rate increased or decreased to lead to different values of $T_\textrm{max}$ in the disk. $A_V$ is adjusted accordingly to match the continuum level in the bluest optical wavelengths. The comparison to the suite of single temperature atmospheres illustrates the composite nature of the multi-temperature disk models.}
    \label{fig:model_demo}
\end{figure*}

\subsection{Annuli Creation with Stellar Atmospheres}

We discretize the disk as a series of concentric annuli, each radiating at a temperature according to Equation \ref{eq:new_temp_profiles}. We set the temperature of an annulus as the mean of the temperatures at its outer and inner radii.   The luminosity contribution of each annulus is then summed to give the total luminosity of the disk. 

Some care must be taken in discretizing the disk, since mathematically, what is being calculated is 
\begin{equation}
    L^{\text{TOT}} = \sum_{i \text{ annuli}} F[T(\overline{r}_i)]\cdot A(r_{a,i}, r_{b,i})
    \label{eq:disk_luminosity}
\end{equation}
where $F(T(\overline{r}_i))$ is a flux density that depends on temperature (which depends on the average radial distance from the center of the disk), and $A(r_{a,i}, r_{b,i})$ is the area of an annulus as a function of its inner and outer radius. 
Once the values of $\overline{r}_i$ corresponding to each temperature in the grid are found within the disk, the collection of all $\overline{r}_i$s is used to break the disk up into annuli, each with a characteristic $r_{a,i}$ and $r_{b,i}$ with $\overline{r}_i$ as the average of the two.

Previous studies such as \cite{khh88} used empirical photospheric templates to build a disk model.  We choose to use modern, theoretically computed stellar atmospheres. 
Advantages of our approach include that we can sample from a finer temperature grid, and have access to updated physics.
We employ a pre-computed grid of NextGen\footnote{BT-NextGen (AGSS 2009) available at \url{http://svo2.cab.inta-csic.es/theory/newov2/}} 
stellar atmospheres \citep{hauschildt99} at fixed temperatures between 2000 K and 10,000 K with 100 K intervals  (200 K for high temperatures).
For surface gravity, we adopt the $\log(g) =$ 1.5 models, as previous studies such as \cite{khh88}, \cite{hk96} and \cite{zhu2007} have empirically found that such low surface gravities, those characteristic of giant and supergiant photospheres, are a more adequate fit to FU Ori-type atmospheres than a solar-type surface gravity. 

The wavelength coverage of the model grid is from $\sim$1500 \AA$\;$ in the ultraviolet to $\sim$5 $\mu$m in the mid-infrared, which is suitable for our purposes. 
The native wavelength resolution is 0.02 \AA.

The details of the disk annuli creation are as follows. First, we start with model parameters $M_*, R_*, \dot{M}$, and compute the temperature profile as in Equation \ref{eq:new_temp_profiles}, assuming a value of $\gamma$ for the boundary region. Next, we identify where the disk temperature reaches 2000 K, the lowest temperature at which NextGen atmospheres are available. Between the star and this radial extent, we use stellar atmospheres to model the luminosity, but between this radial extent and some parameter $R_{\text{outer}}$, we use blackbodies.  In practice we fix $R_{\text{outer}}= 100 R_\odot$.

In the boundary region and inner disk region out to 2000 K, we divide the disk into annuli according to their mean radiating temperature and set $F(T)$ to the corresponding NextGen stellar atmosphere. We sample evenly in $T$, constrained by the temperature resolution of the NextGen tables, which amounts to 100 K at 2000 K $<$ T $<$ 7000K and 200 K at hotter temperatures. We find the radial distance $\overline{r}_i$ corresponding to each temperature in the grid, and the collection of all $\overline{r}_i$s is used to break the disk up into annuli. Each has a characteristic $r_{a,i}$ and $r_{b,i}$ with $\overline{r}_i$ as the average of the two, and the area of the annulus can now be calculated and used  in Equation \ref{eq:disk_luminosity}. In the disk regions beyond 2000 K and out to $R_{\text{outer}}$, we divide the disk into annuli according to their mean radial distance from the center of the disk and set $F(T)$ to the corresponding blackbody flux density. We sample the outer annuli in this manner to avoid oversampling at cooler temperatures. Just as in the previous step, $\overline{r}_i$, $r_{a,i}$, and $r_{b,i}$ for all annuli are calculated.

\subsection{Preparing Models for Comparison to Data}\label{sec:model_to_data}
Our stellar atmosphere models and blackbody calculations output specific luminosities, so we use the familiar formula applicable to geometrically thin disks as stated in \cite{hartmannbook}, to compute specific fluxes for spectra and SEDs:
\begin{equation}
    F_\lambda = \frac{L_\lambda}{2\pi d^2}\cos i.
    \label{eq:disk_flux}
\end{equation}

In order to adequately compare our model to data, we must ensure that the effective resolutions are similar. We take advantage of the full 0.02 \AA\ resolution of the model atmospheres when considering our high dispersion spectroscopy, but otherwise we down-sample the atmospheres when considering our medium dispersion spectroscopy, and the 0.4-2.4 $\mu$m spectrophotometry.

For the SEDs that we fit, it is unnecessary to keep \text{the model atmosphere spectra} at their full-resolution, especially considering the wavelength range of more than a factor of six that we model. 
We make use of \texttt{SpectRes}, a Python module\footnote{\url{https://spectres.readthedocs.io/en/latest}} developed by A. Carnall, 
in combination with a standard low-pass filter, to resample spectra by eliminating small features but keeping the continuum at its true level. 

When qualitatively analyzing the medium-resolution spectra and spectral features, we downsample the model atmospheres to $\Delta \lambda_\textrm{med-res} \sim 2$\AA, since this resolution can approximate well both our optical ($\sim$ 1.08\AA) and near-infrared ($\sim$ 3.5\AA) data. 

Finally, when qualitatively looking at high-resolution spectra, we perform boxcar smoothing on the data and compare to the high-resolution models directly, without
any additional adjustments for instrumental resolution since, in the high-resolution case, the data have comparable resolution to the native resolution ($\Delta \lambda_\textrm{high-res} \sim 0.02$\AA) of the model atmospheres. For both the medium and the high spectral resolution cases, the rotational broadening from the disk Keplerian assumption dominates any potential instrumental broadening effects. 


Figure \ref{fig:model_demo} illustrates both the SEDs and the corresponding 
medium-resolution spectra {\ that can be produced by our models.  Shown are} a set of accretion disk models that incorporate
all of the parameters described above, plus for comparison, a set of simple stellar atmospheres 
of single temperature, but broadened by the rotational kernel appropriate to disks in Equation \ref{eq:disk_broaden}, {\ albeit using a single velocity}. 
The point to note is the distinction and spectral complexity of the multi-temperature disk models
relative to the single-temperature stellar models.
The disk model parameters 
represent a disk with a maximum disk temperature of 7000 K and an extinction value of $A_V = 3.16$, 
as well as somewhat hotter and cooler disks:
a model at $T_\textrm{max} = 10,000$ K  with $A_V = 5$, 
and a model at $T_\textrm{max} = 5,000$ K  with $A_V = 1$. $\gamma$ is set to 0 for all of these models.
The other model parameters correspond to those of Model 2a in Table \ref{tab:hbc722_table},  
which is introduced more completely in a later section. 
The varying extinction values are needed in order to adjust the continuum flux levels to comparable values over the plotted wavelength range. 

We direct  the reader to the Appendix  for examples of the effects of modifying various parameters 
in the disk model, where we treat them one at a time.  In the following section, we describe  how we obtain sets 
of disk parameters that appropriately fit the observed SED formed by a complete optical and near-infrared flux-calibrated spectral data set.

\subsection{MCMC Parameter Exploration from SEDs}

Our fitting process is geared towards determining the fundamental parameters in the disk model,
namely $M_*$, $R_*$, and $\dot{M}$.  In order to fit these parameters accurately, we must also constrain
the disk inclination, $i$, and the total line-of-sight extinction, $A_V$.
In this section, we employ MCMC techniques to see how well they perform at this task.

Given the accretion disk model described above, we estimate the credible intervals of the parameters relevant to the model via the following two-fold approach:
\begin{enumerate}
    \item We synthesize flux-calibrated model SEDs to compare against the optical-infrared spectrophotometry, resampling both models and the flux-calibrated spectra onto the same low resolution  grid, as described in \S \ref{sec:model_to_data}. This ensures we match the flux level across the entire range of our data. We then use a series of MCMC parameter exploration techniques to obtain "best-fit" sets of parameters.
    \item We then create model spectra at the full resolution of the atmospheres grids, to qualitatively compare to \text{the observed} spectra at moderate and high spectral resolution ($R\sim 5,000$ and $R\sim 36,000$), based on the model parameter sets that result from the \text{best fitting} MCMC runs on the SEDs.
\end{enumerate}

\subsubsection{\text{MCMC Outline}}

\begin{figure*}[htp]
    \centering
    \includegraphics[scale=0.7]{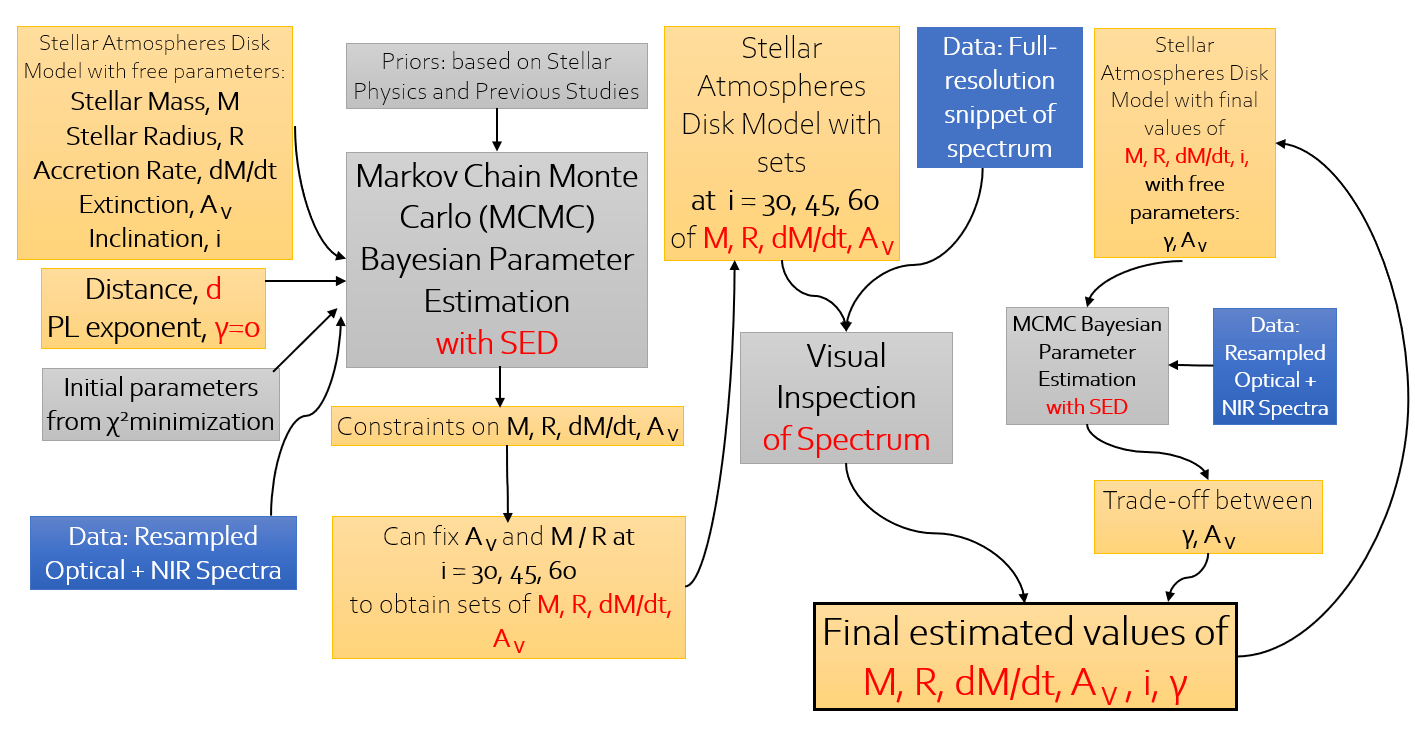}\\
    \caption{Flow chart showing the method used to obtain sets of parameters with credible intervals from disk models based on comparison to SEDs and spectra. While a simple $\chi^2$ minimization would give us \textit{one} set of best-fit parameters, we exploit the Bayesian nature of MCMC to obtain a more complete set of values based on physical priors and better understanding of degeneracies among multiple parameters.}
    \label{fig:final_flow_chart}
\end{figure*}

We carry out the Markov Chain Monte Carlo (MCMC) analysis as follows.  First we explore the parameter space of our model SED to fit the data, in order to obtain quantitative estimates on our parameters. We then perform a qualitative fit (by eye) to the spectra to tune the inclination of the objects, though we do not explicitly perform quantitative spectral fits in this study. We thus ask and answer the question: is fitting broadband SEDs sufficient to also describe the spectral properties of FU Ori objects in the framework of the $\a$-disk model?

In order to sample the parameter space and place constraints on our free parameters based on the data, we start by performing a simple $\chi^2$ minimization to obtain best-fit parameters given a reasonable initial guess based on typical young star parameters. While this serves as a way to get \textit{a single set} of best-fit parameters, we choose to explore how pairs of parameters trade off with one another through corner plots, and get a sense of which \textit{various sets} of parameters can serve as good fits to the data. Thus, we perform a MCMC parameter estimation to place constraints on the free parameters. We deem the MCMC approach appropriate because we do not use it to search the entire parameter space or optimize the likelihood or posterior probability distribution functions (pdfs). Crucially, we use MCMC to simply sample the complex pdf created once we develop the full accretion disk model and obtain credible intervals for our parameters individually. These principles are all in accordance with rules-of-thumb laid out by astrostatisticians regarding the (over-)usage of MCMC \citep{hogg2018}.

\begin{table*}
    \centering
    \begin{tabular}{c|c|c|c|c|c|c|c|c}
         Model & $M_*$ & $R_*$ & $\dot{M}$ & $A_V$ & $i$ & $\gamma$ & $d$ & $R_\textrm{outer}$\\\hline
         1 & 0.1-1~$M_\odot$ & 0.2-8~$R_\odot$ & $10^{-7.5} - 10^{-3.5}~M_\odot$/yr & (0.5-2)$\times$lit.value & $0^\circ - 90^\circ$ & \textit{0} & \textit{lit. value} & \it{ 100 $\mathit{R_\odot}$}\\
         2a & 0.1-1~$M_\odot$ & 0.2-8 $R_\odot$ & $10^{-7.5} - 10^{-3.5}~M_\odot$/yr & \textit{Mod. 1 fit} & $\mathit{30^\circ}$ & \textit{0} & \textit{lit. value} & \it{ 100 $\mathit{R_\odot}$}\\
         2b & 0.1-1~$M_\odot$ & 0.2-8 $R_\odot$ & $10^{-7.5} - 10^{-3.5}~M_\odot$/yr & \textit{Mod. 1 fit} & $\mathit{45^\circ}$ & \textit{0} & \textit{lit. value} & \it{ 100 $\mathit{R_\odot}$}\\
         2c & 0.1-1~$M_\odot$ & 0.2-8 $R_\odot$ & $10^{-7.5} - 10^{-3.5}~M_\odot$/yr & \textit{Mod. 1 fit} & $\mathit{60^\circ}$ & \textit{0} & \textit{lit. value} & \it{ 100 $\mathit{R_\odot}$}\\
         3 & $\dfrac{M_* / M_\odot}{R_* / R_\odot}= 0.5$ & \textit{Mod. 2a fit} & \textit{Mod. 2a fit} & \textit{Mod. 1 fit} & $\mathit{30^\circ}$ & \textit{0} & \textit{lit. value} & \it{ 100 $\mathit{R_\odot}$}\\
         4 & \textit{Mod. 2a fit} & \textit{Mod. 2a fit} & \textit{Mod. 2a fit} &  (0.5-2)$\times$lit.value & $\mathit{30^\circ}$ & $-1 - +1$ & \textit{lit. value} & \it{ 100 $\mathit{R_\odot}$}\\
         5 & \textit{Mod. 2 fit} & \textit{Mod. 2 fit} & \textit{Mod. 2 fit} & \textit{Mod. 4 fit} & $\mathit{45^\circ}$ & \textit{-1, 0, +1} & \textit{lit. value} & \it{ 100 $\mathit{R_\odot}$}\\
    \end{tabular}
    \caption{{\ Description of model parameter free ranges and set values for all models. Values in italic font are fixed. The MCMC-derived parameters resulting from each model appear in Tables \ref{tab:hbc722_table} and \ref{tab:gaia17bpi_table}.}}
    \label{tab:model_list}
\end{table*}

\subsubsection{\text{Step-by-Step MCMC Process}}
A summary of our steps in the full $\alpha$-disk parameter determination are summarized below, with each described in more detail as we implement it in the next section.
A tutorial flow chart is provided in Figure \ref{fig:final_flow_chart}.

We set the distance, $d$ to our objects, the outer radius, $R_{\text{outer}}$, and the boundary region power law exponent, $\gamma$, to be fixed parameters. The distance to each source is fixed at the value from the literature as reported in \S \ref{sec:objects}.
For  $R_\textrm{outer}$ we adopt a value of $100 R_\odot$, which is beyond the distance in the disk where the temperatures become cool enough so that they do not contribute to the observations in the optical-NIR range. For $\gamma$ we assume a value of 0 (flat profile in the boundary region) as a default based on previous studies such as \cite{khh88} and \cite{zhu2007}, but explore variations on this parameter in \S \ref{sec:results}. 
    
Next, we create model SEDs using stellar atmospheres, downsampled onto the same grid as the combined optical and infrared spectra as described earlier. We run a simple $\chi^2$ minimization using a Nelder-Mead method through the $\texttt{scipy}$ module in Python to obtain starting parameters for the Bayesian parameter exploration. All spectrophotometry is weighted with uniform errors across all wavelengths. At this point, we explore the space of the remaining free parameters through MCMC: the mass of the central star $M_*$, radius of the central star $R_*$, accretion rate $\dot{M}$, extinction $A_V$, and inclination $i$, to be free parameters.  Using the initial set of parameters from the $\chi^2$ fit, we then run our MCMC algorithm using the $\texttt{emcee}$ package to perform affine-invariant MCMC to sample the parameter space.

For the stellar parameters, we set a uniform prior on $M_*$ between $0.1M_\odot$ and $1 M_\odot$, and a uniform prior on $R_*$ defined between $0.2 R_\odot$ and $8.0 R_\odot$. For the disk parameters, we choose a log-uniform prior on $\dot{M}$  defined between $10^{-7.5} M_\odot \textrm{ yr}^{-1}$ and  $10^{-3.5} M_\odot \textrm{ yr}^{-1}$ and
we set a uniform prior on the cosine of the inclination, $\cos i$ between 0 and 1.  
Regarding extinction, we initially use a uniform prior on $A_V$, centered around the value reported in the literature for each of our sources, with a range of roughly $\sim 50\%$ of that value on either side. 

We proceed with the multi-parameter fitting as follows. After an initial MCMC run, we fix $A_V$ at the value we obtain and run three MCMC routines, with $i$ fixed at 30, 45, and 60 degrees. $R_*, M_*$, and $\dot{M}$ are left as the free parameters. From the corner plots we generate in this step, we can see the trade-off between $M_*$ and $\dot{M}$ very clearly, leading to neither parameter being able to be constrained individually as expected from Equation \ref{eq:new_temp_profiles}. $R_*$, however, is well-constrained in each of the three MCMC routines due to the fixed inclination. At this point, we fix $\dfrac{M_* / M_\odot}{R_* / R_\odot}= 0.5$ to obtain a final set values that we can use as model parameters. {\ Our method also allows us to start with $M_* \dot{M}$ from the beginning, and fit for the product of those two parameters, which we also pursue in \S \ref{sec:combined_param_explore}.} 

The above process results in several sets of model parameters that best-fit the 
$0.39-2.4 \mu$m spectrophotometry (SEDs). We then produce and
visually analyze the fit of model spectra from each of the sets of model parameters in comparison to the moderate and high-resolution spectral data. We try our various parameter sets generated from the past steps to visually determine which of three models (inclination at $30^\circ, 45^\circ, 60^\circ$) provides the optimal fit of spectral features within various ranges of 500 \AA$\;$and fix the value of inclination accordingly. 

As a final step in our MCMC routine, we explore different values of $\gamma$. We are limited to values $\gamma > -1$ so as to not exceed the maximum temperature of the model atmospheres. Given our set of parameters for a given inclination, say $i = 45^\circ$, we fix all parameters other than $A_V$ and $\gamma$ to explore that parameter space while fitting the SED. As above, we then generate model spectra using values values of $\gamma$ and $A_V$ to inspect the effect of varying those parameters in the full-resolution models.

In each parameter estimation routine, we run the MCMC parameter explorations with four chains for each parameter in our affine-invariant sampling. We run the chains for $4 \times 10^5$ steps and take 50 percent of the steps as the burn-in period to avoid autocorrelation in our MCMC techniques. After we run our sampler, we test for convergence using the Gelman-Rubin statistic \citep{vats2018},  $\hat R$, obtain values of $\hat R$ for the free parameters. The higher values of the statistic are to be expected for parameters that are strongly correlated in the corner plot, and do not converge to sharply peaked values individually. The $\hat R$ values for well-constrained parameters, however, we expect to be below 1.1. If our MCMC iteration is to converge after infinite iterations, then $\hat R$ should converge to 1, so we set a reasonable standard for $\hat R$, say, not so high as 1.5, but not so low as 1.001 \citep{vats2018}. Choosing to adopt a standard for convergence of $\hat R = 1.1$, then, allows us to say that the MCMC chains have converged adequately for parameters whose $\hat{R}$ values are approximately equal to or below that value.

\section{Results: Models vs Data}
\label{sec:results}

\begin{figure*}[h]
    \centering
    \includegraphics[scale=0.54]{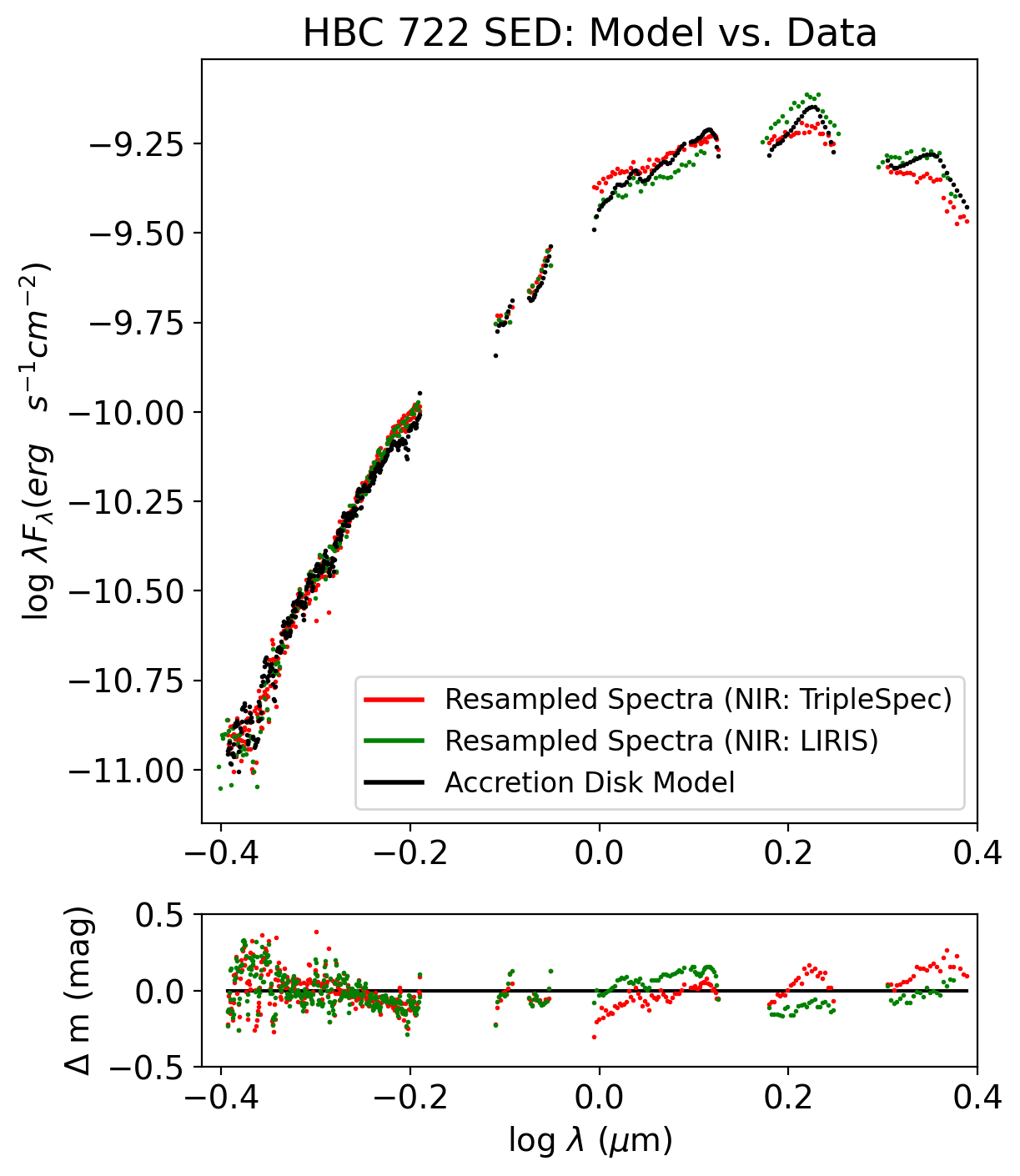}\includegraphics[scale=0.32]{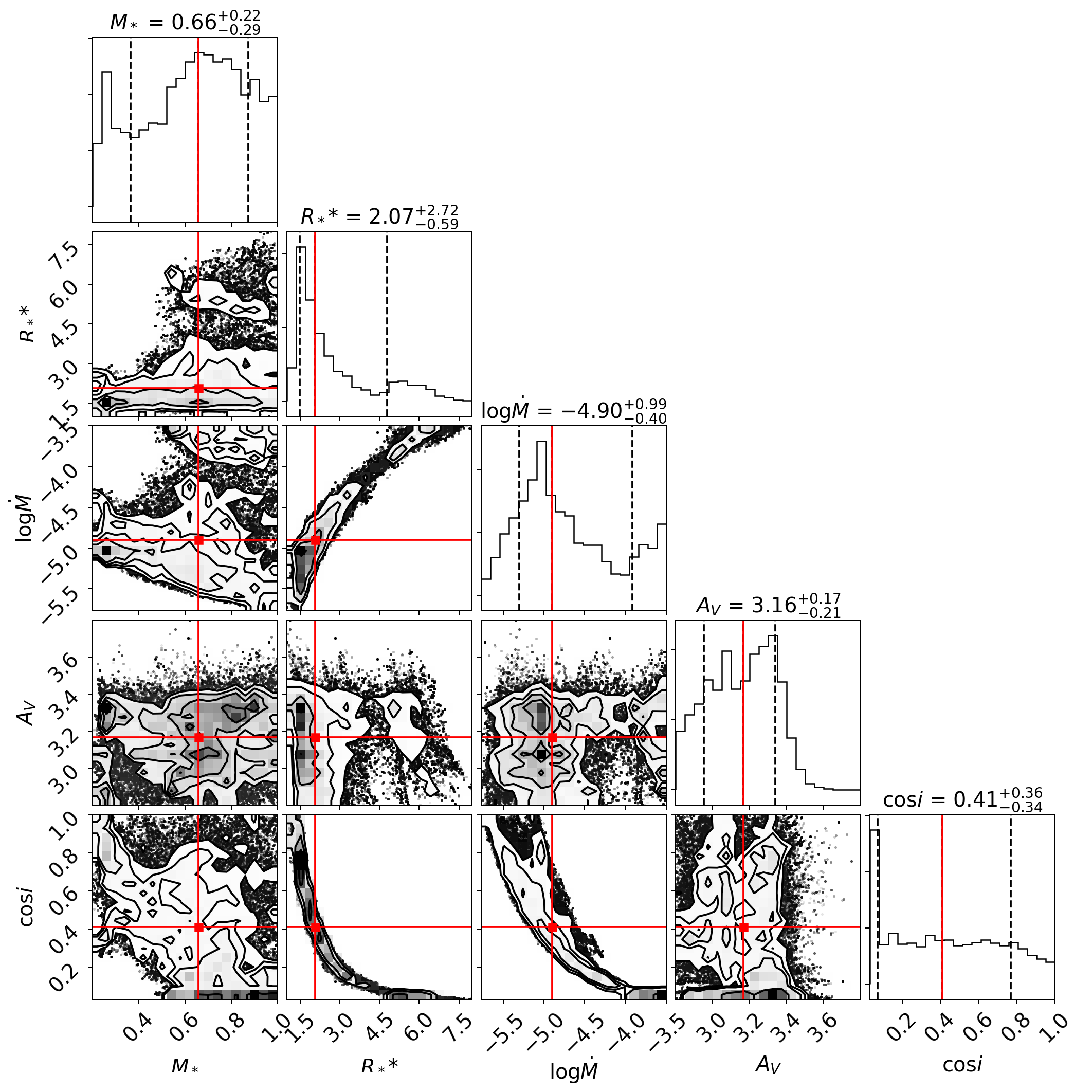}\\
    \includegraphics[scale=0.54]{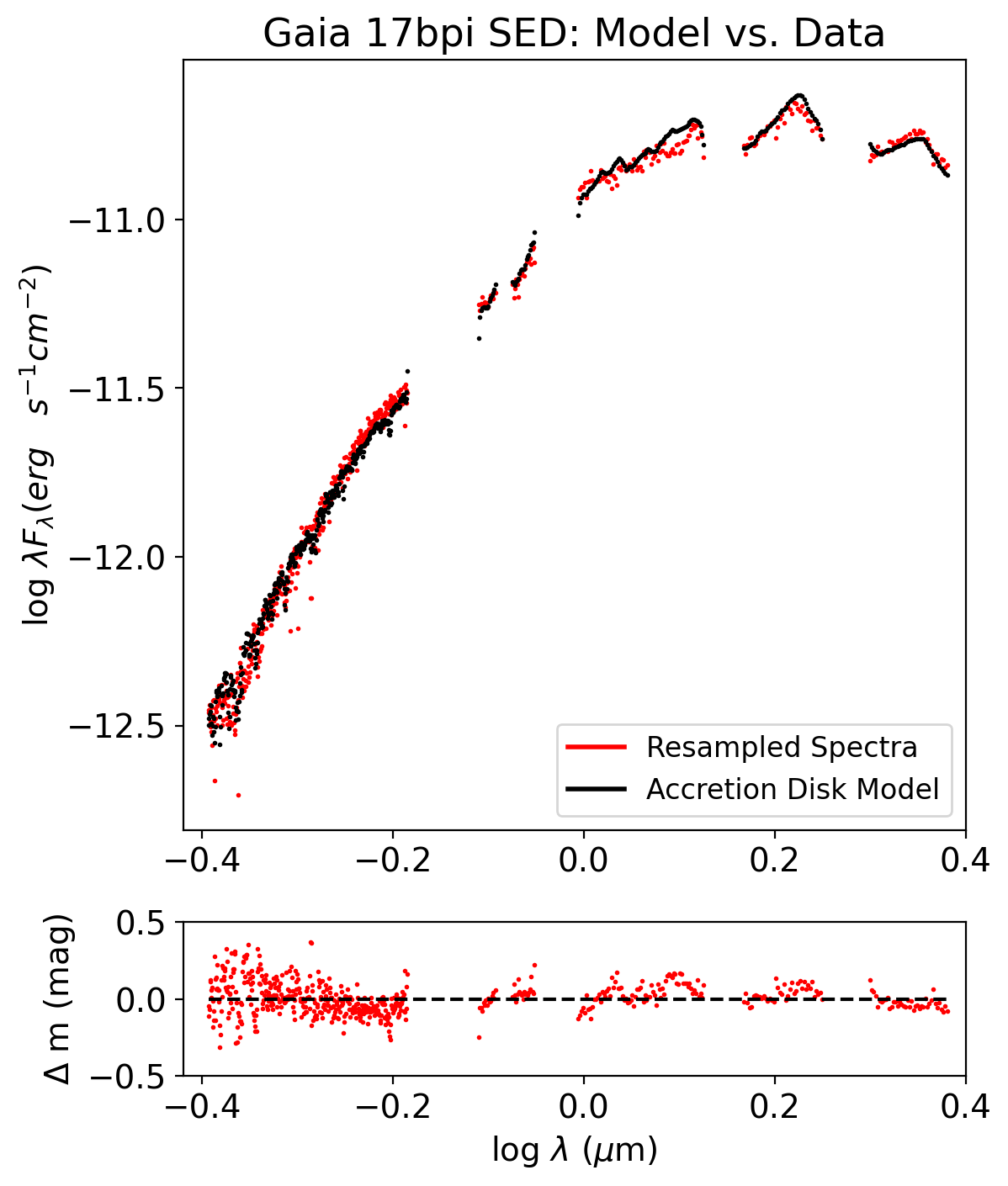}\includegraphics[scale=0.32]{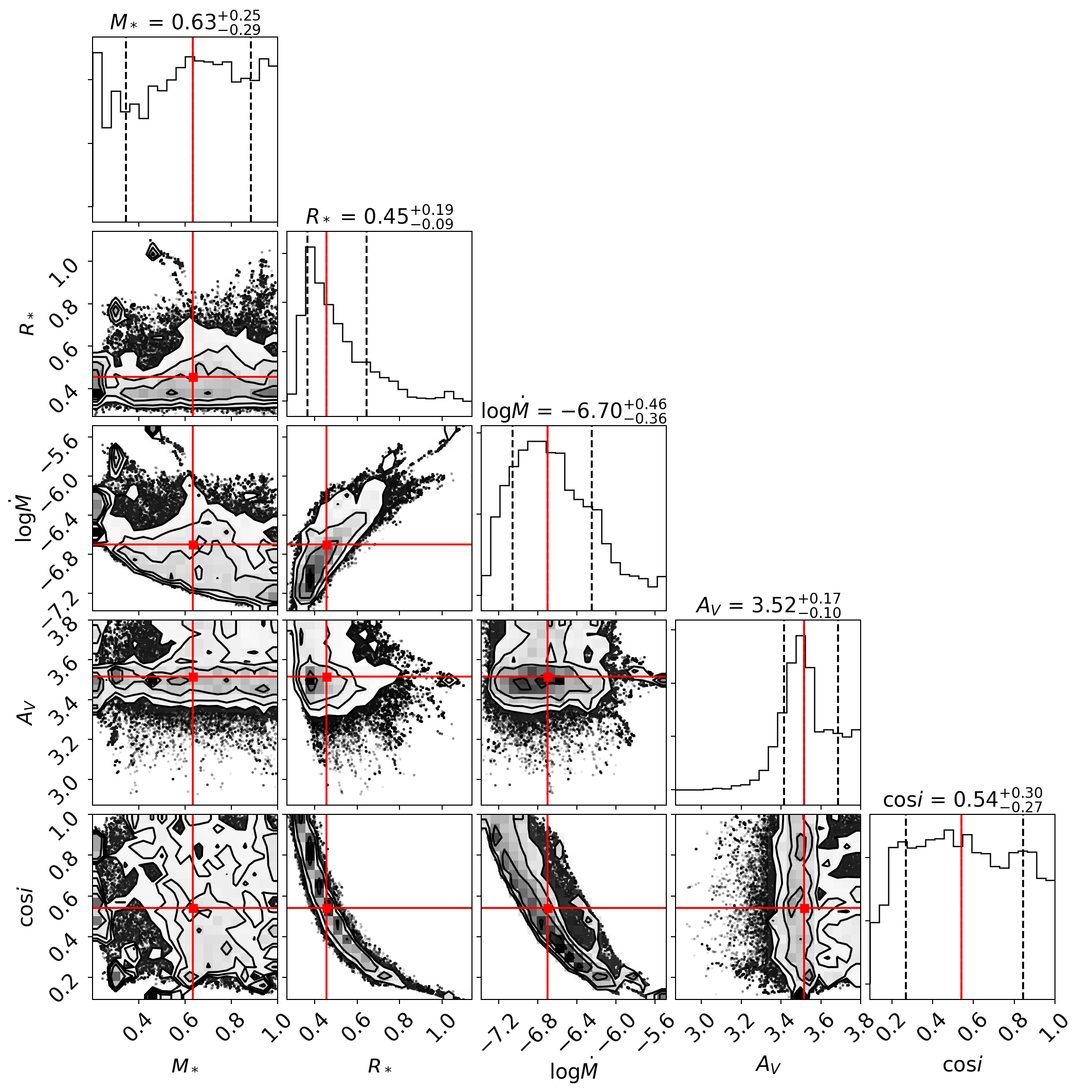}\\
    \caption{Best-fit SEDs and corresponding MCMC corner plots with maximal parameter exploration in the accretion disk model.
   Upper panels: HBC 722,  lower panels: Gaia 17bpi. For each source, in the left panels the source SED is plotted with the best-fit model whose parameters minimize the $\chi^2$-value. Two infrared spectra are shown for HBC 722, one from 2010 and one from 2011, demonstrating the spectral evolution of this object.
   The residual panels show that the model accurately predicts the continuum level to within a few tenths of a magnitude from 4000 \AA\ to 2.4 $\mu$m. For each source, the right panels show the corner plot with posterior distributions and covariance plots of the free parameters. This analysis, despite its inability to constrain some parameters such as $i$ and $M_*$, nevertheless serves to provide the most conservative credible intervals for each individual model parameter, crucially $\dot{M}$.}
    \label{fig:results_all}
\end{figure*}

In this section, we present model SEDs and spectra of the two FU Ori objects HBC 722 and Gaia 17bpi, using the procedures described above to generate model parameters with MCMC techniques. \text{Table \ref{tab:model_list} describes each model.}

\subsection{Full Parameter Exploration}
\text{The results of Model 1 are shown in}
Figure \ref{fig:results_all}, illustrating the corner plots exploring $M_*$, $R_*$, $\dot{M}$, $A_V$, and $i$. Visually, it is clear that $A_V$, $\dot{M}$, and $R_*$ are the only parameters with sharply peaked marginalized distributions.  The panel exploring
$M_*$ and $\cos i$ clearly did not converge, for example.

This model provides the largest credible intervals when marginalizing over all over parameters.
This MCMC run is crucial, however, in showing the entire range of likely parameter values given our priors. Even allowing the inclination to vary across its entire range of possible values, and $M_*$ to vary across a factor of ten range of physically reasonable values, we still produce credible intervals for the other free parameters. We report these intervals in Tables \ref{tab:hbc722_table} and \ref{tab:gaia17bpi_table}. {\ The fitting details for individual parameters are as follows}:
\begin{itemize}
    \item $M_*$ does not converge, with the entire range set by the prior explored. In the next set of MCMC runs, we find (consistent with expectations) that this parameter directly trades off with $\dot{M}$.
    \item $R_*$ converges, albeit with a large credible interval, particularly skewed towards large values. The corner plot reveals its strong anticorrelation with $\cos i$, as well as correlation with $\dot{M}$. The latter is resolved in the next MCMC run; 
    once we fix $i$, $R_*$ becomes well-constrained. 
    \item $\dot{M}$ converges, and is strongly correlated with $R_*$ and $\cos i$. As above, we note that the correlation with $R_*$ disappears once we fix $i$ and $R_*$ becomes constrained.  $\dot{M}$ is perhaps the most interesting parameter from a physical standpoint for us to be able to estimate. 
    With the generous parameter exploration, we see that the accretion rate of HBC 722 is constrained with 16 and 84 percent confidence intervals to 
    $\dot{M} = 10^{-4.90} M_\odot \textrm{ yr}^{-1}\; {}^{+0.99}_{-0.40} \textrm{ dex}$. These are high accretion rates, typical of other FU Ori objects \citep{hk96,hartmannreview2016}. The corresponding value for Gaia 17bpi is $\dot{M} = 10^{-6.70} M_\odot \textrm{ yr}^{-1}\; {}^{+0.46}_{-0.36} \textrm{ dex}$, which is on the lower end of FU Ori accretion rates.
    \item $A_V$ converges, due to its distinct effect on the SED which no other parameter can replicate. Constraining this parameter early in our process is useful, and we fix it for the remaining MCMC trials, save a final run where we explore its correlation with $\gamma$.
    \item $\cos i$ does not converge, as expected given the uniform prior on this parameter and the degeneracy with other parameters. There is no way to constrain inclination without spatially-resolved observations, or estimates of $v\sin i$ from high-resolution spectra. Thus, allowing it to be fully explored gives us the most conservative credible intervals for all other parameters.
\end{itemize}
The resulting model SEDs, as seen in Figure \ref{fig:results_all}, accurately predict the spectrophotometric data to within a tenth of a magnitude over the majority of the wavelength range, though reaching a maximum deviation of nearly 0.4 mag (40\%) in the bluest optical ($\lambda \lesssim 4300$\AA) region.  This modeling success continues support for the $\a$-disk model for FU Ori objects, as it is able to produce broad-band flux-calibrated spectroscopy that matches observations in the optical to near-infrared wavelength range of these objects.

\subsection{Fixing Inclination}
The initial (Model 1) MCMC run motivates the need to constrain the inclination, as this parameter strongly trades off with the stellar radius and accretion rate. In the next set of MCMC runs (Model 2), we fix the inclination at values of $i = 30^\circ$, $i = 45^\circ$, and $i = 60^\circ$ and we fix $A_V$ at the median value from the initial Model 1 fit. We thus study here only $M_*$, $R_*$, and $\dot{M}$ as free parameters. 

\begin{figure}
    \centering
    \includegraphics[scale=0.45]{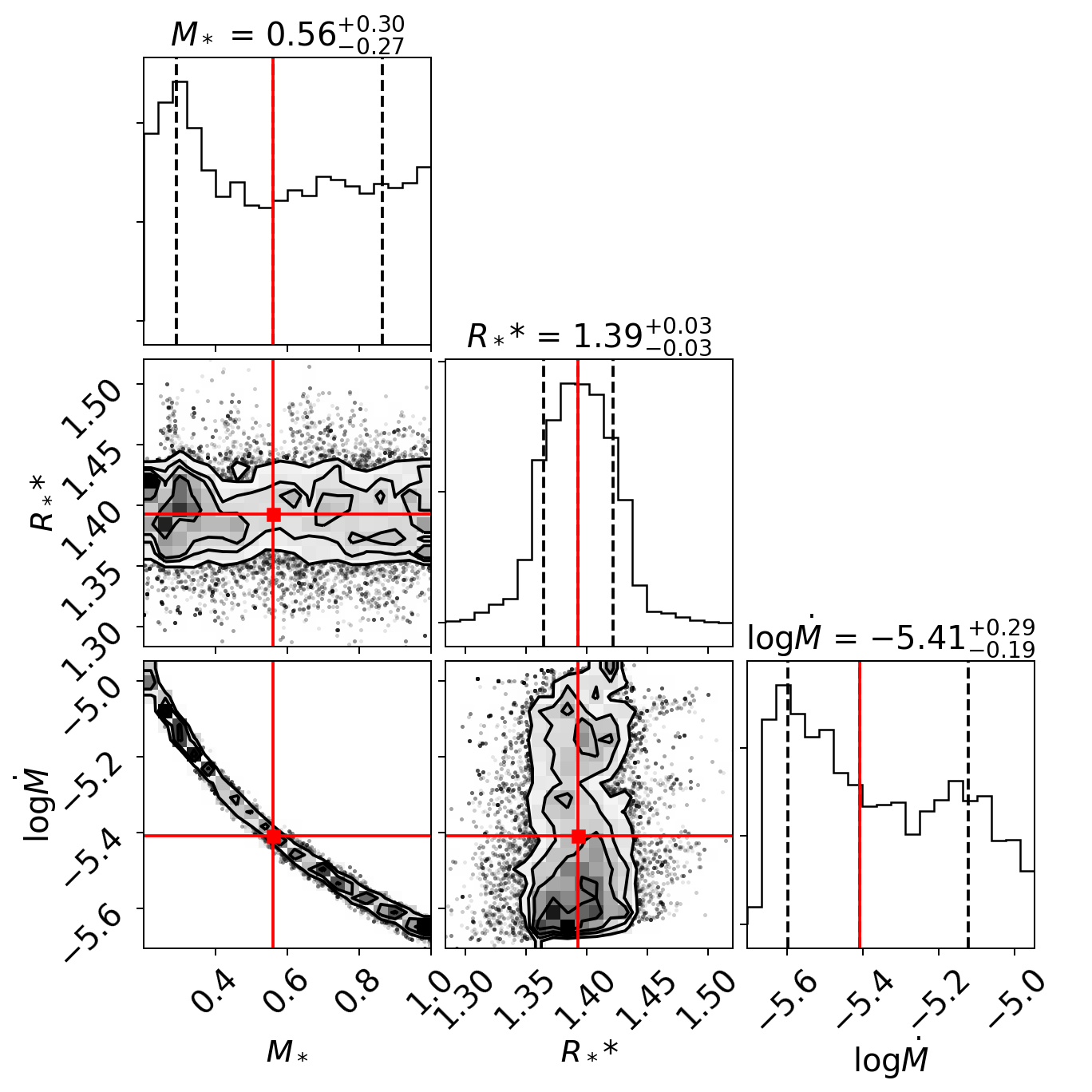}
    \includegraphics[scale=0.45]{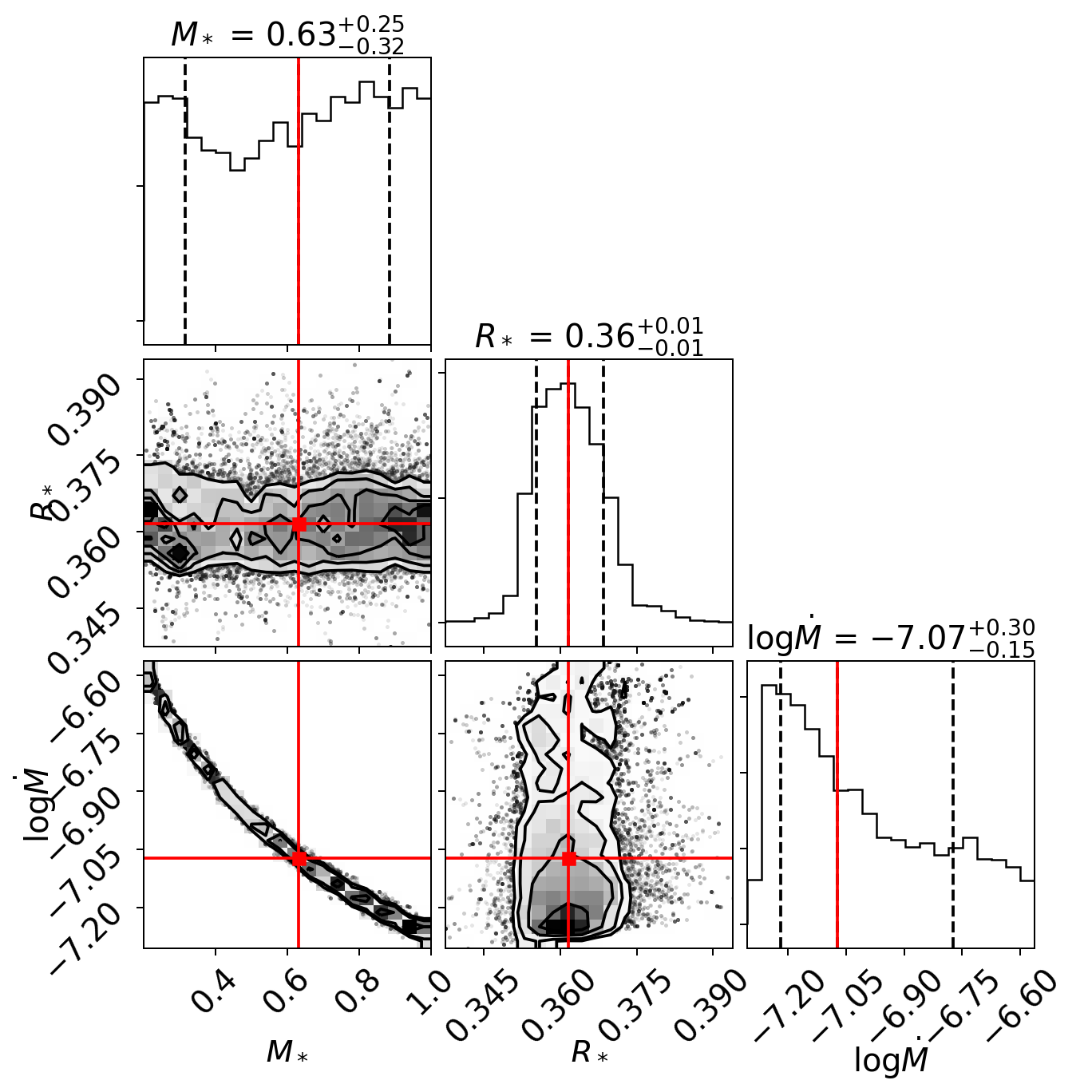}
    \caption{Corner plots leading to the Model 2a, having inclination fixed at $i = 30^\circ$. Upper panel: HBC 722, Lower panel: Gaia 17bpi. With this parameter fixed, $R_*$ is sharply constrained, while the direct trade-off between $M_*$ and $\dot{M}$ is elucidated.}
    \label{fig:results_inclinations}
\end{figure}

Figure \ref{fig:results_inclinations} shows the resulting corner plots from these MCMC runs. They demonstrate that $R_*$ is strongly dependent on the inclination, as expected, but its distribution converges very sharply once inclination is fixed. Moreover, the corner plots show that, with the strong influence of the inclination out of the way, $M_*$ and $\dot{M}$ are inversely correlated, as expected by the temperature profile in Equation \ref{eq:new_temp_profiles}.  It is thus straightforward to use the relationship between $M_*$ and $\dot{M}$ to jointly constrain those parameters,
when $A_V$ and $i$ are fixed, so $R_*$ becomes well constrained.  There is no way to constrain $\dot{M}$ or $M_*$ beyond this step, unless we fix one of those parameters, or fix one of those parameters in combination with a parameter that is already constrained {\ (see \S6.3) or fit only for their product (see \S6.5)}.

The remainder of this section is devoted to taking the parameters from these Model 2 fits, and assessing how they perform with respect to medium- and high-resolution optical spectra. As we show below, relative to SEDs, spectra are more sensitive to certain parameters, such as $T_{max}$ towards the blue optical, and $M_*/R_*$ as well as inclination, towards higher spectral resolution.

\subsection{Insights from Comparison to Spectra}
Using the 
sets of parameters obtained from the previous MCMC runs derived from SED fitting, we take a closer look at the 
predicted spectra to see how well the disk model captures spectral lines, as opposed to just the overall continuum level measured by the SED. 
We consider 500 \AA\ snippets in medium resolution data, and $\sim$100 \AA\ snippets in high resolution data. 

\subsubsection{Medium Resolution}

\begin{figure}
    \centering
    \includegraphics[scale=0.30]{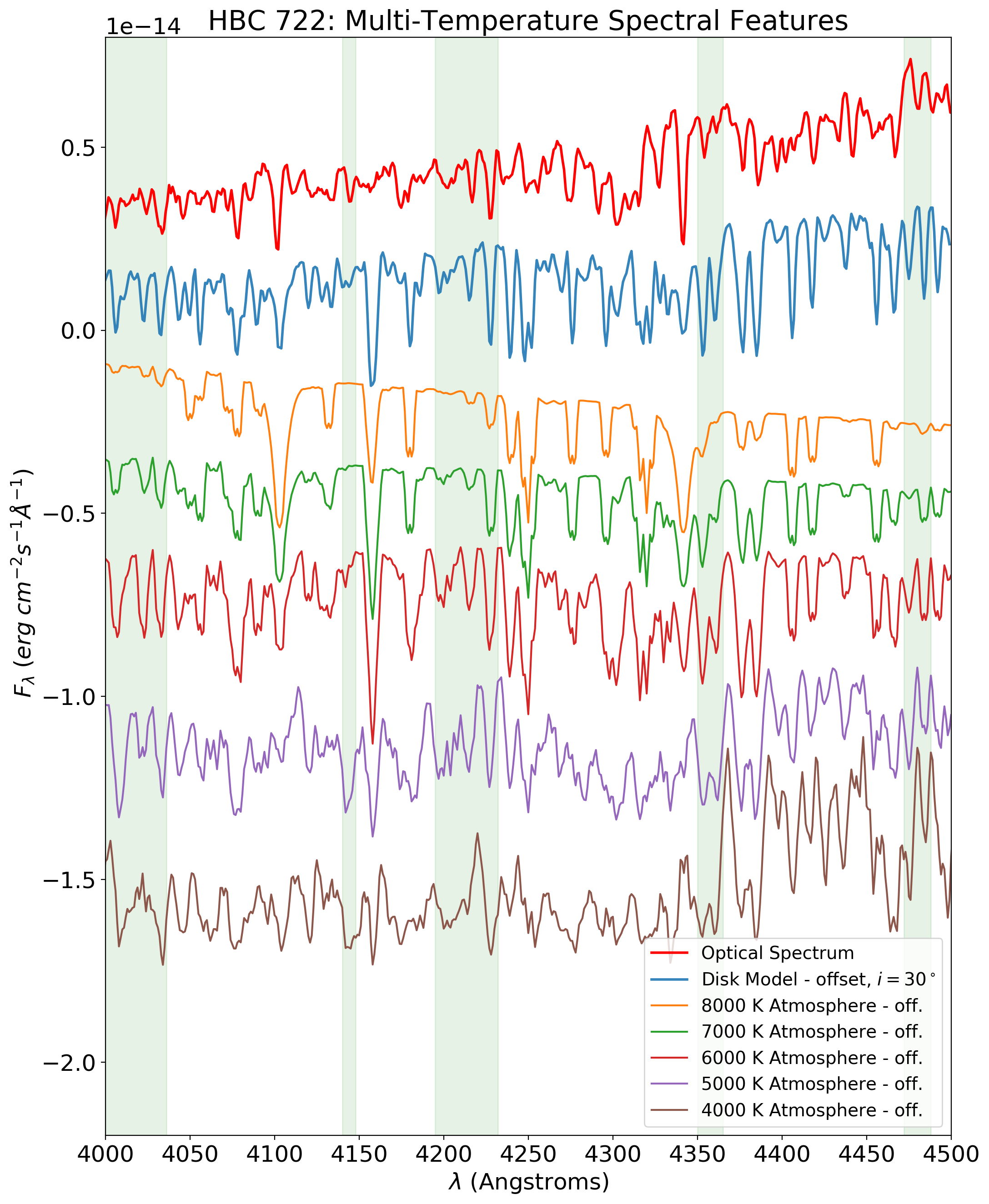}
    \includegraphics[scale=0.30]{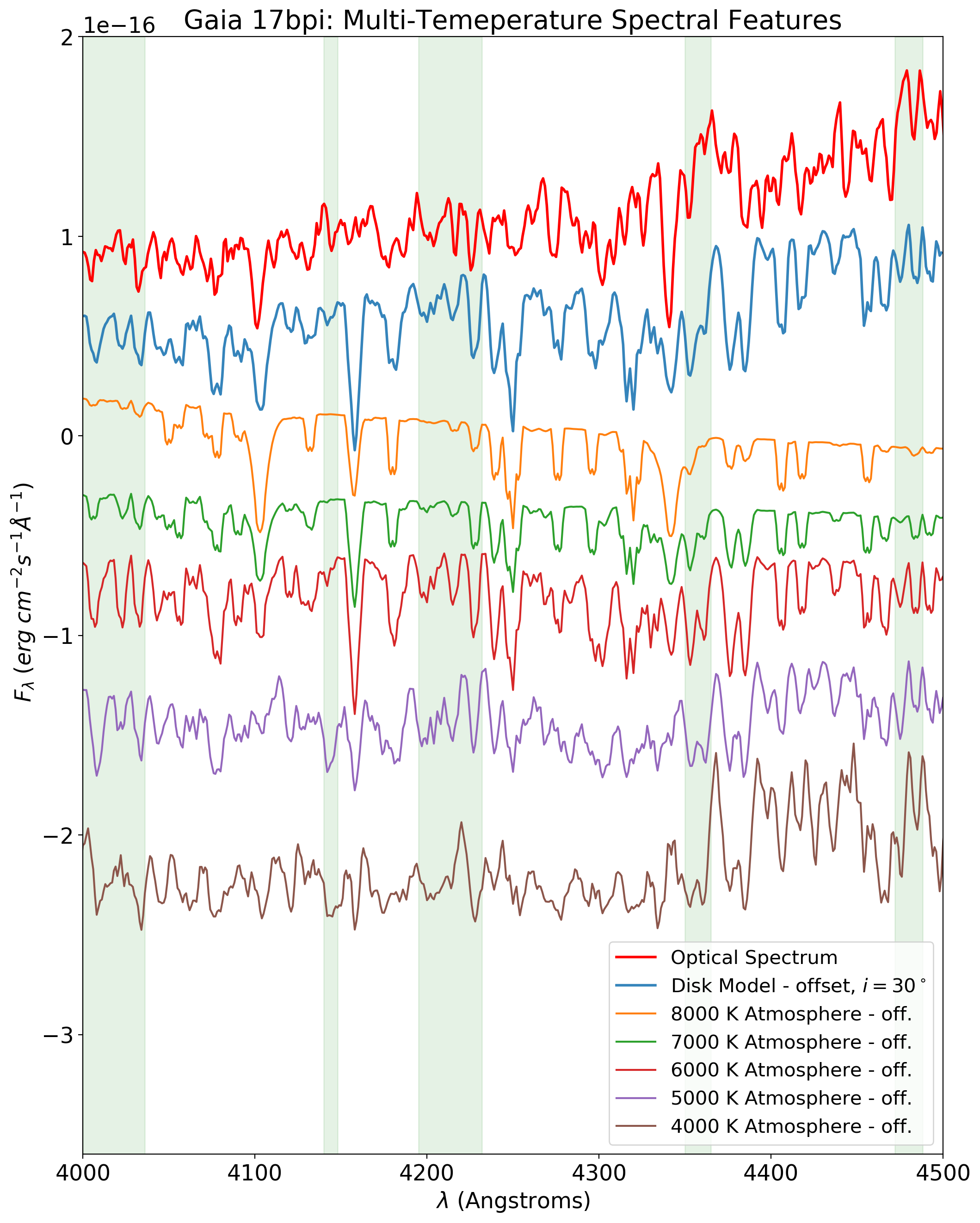}\\
    \caption{The optical spectrum and a disk model are plotted alongside various single-temperature stellar atmospheres broadened with a disk rotational profile as in Equation \ref{eq:disk_broaden}, rotating roughly at the same velocity as the innermost annulus of the disk, $\sim 180 \textrm{ km s}^{-1}$. The green regions highlight spectral features that are temperature-sensitive, and confirm that both HBC 722 and Gaia 17bpi are indeed multi-temperature extended objects.}
    \label{fig:hbc722_disk_atmospheres}
\end{figure}

\text{
Figure \ref{fig:hbc722_disk_atmospheres} }
shows a portion of the observed optical spectrum for each of HBC 722 and Gaia 17bpi compared to disk models, to demonstrate the spectral evidence for their multi-temperature nature.
We show the observed spectrum between 4000-4500 \AA\ and disk model 2a, alongside various single-temperature stellar atmospheres that are broadened with a disk rotational profile. In the highlighted region between 4000-4040 \AA, the relative line depths most closely resemble those of a 7000 K atmosphere, while at 4140-4150 \AA, the highlighted absorption line is present only in $<$ 6000 K atmospheres. At around 4470-4490 \AA, the features seen are more closely reminiscent of a 5000 K atmosphere. 

\begin{figure}
    \centering
    \includegraphics[scale=0.32]{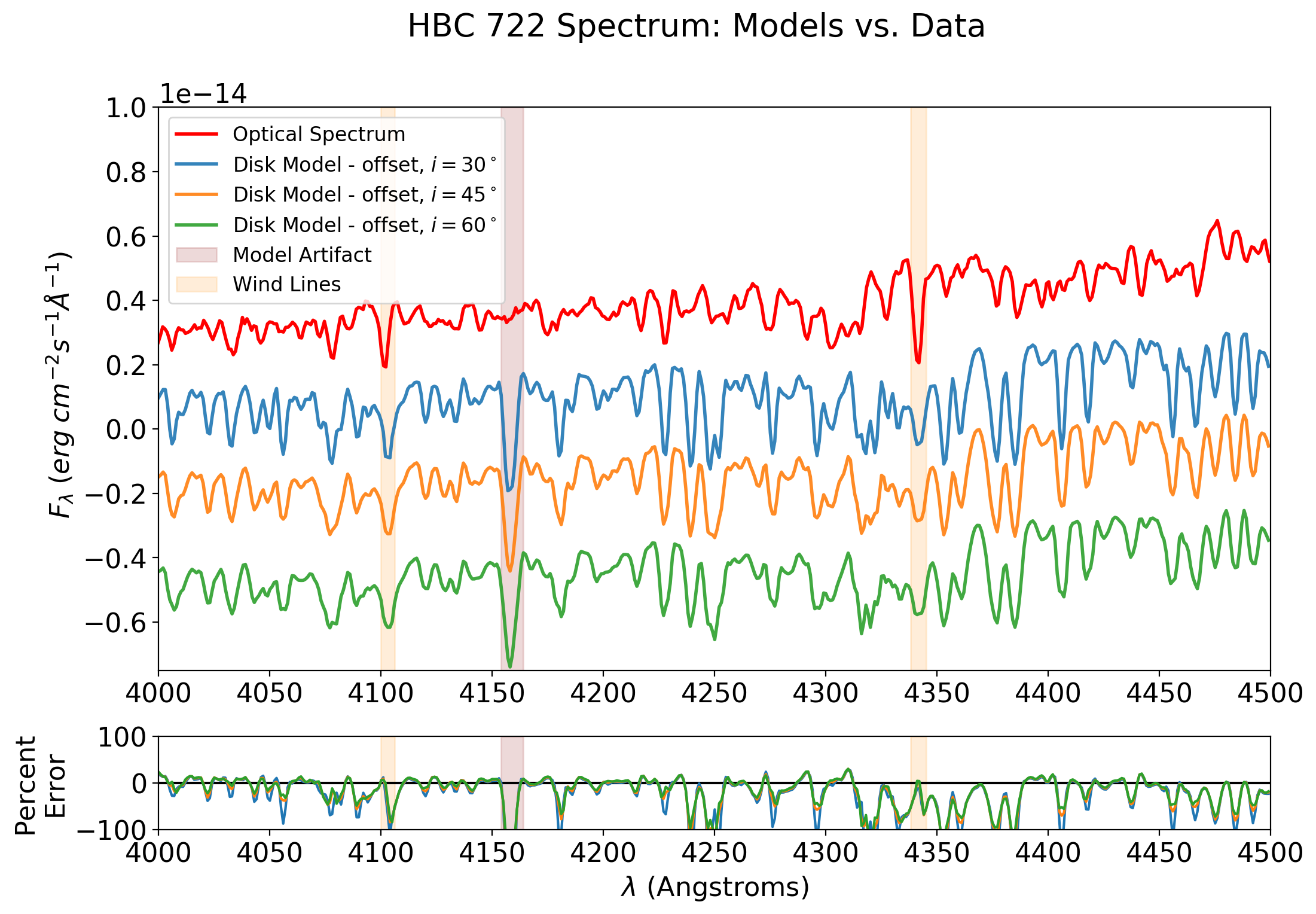}
    \includegraphics[scale=0.32]{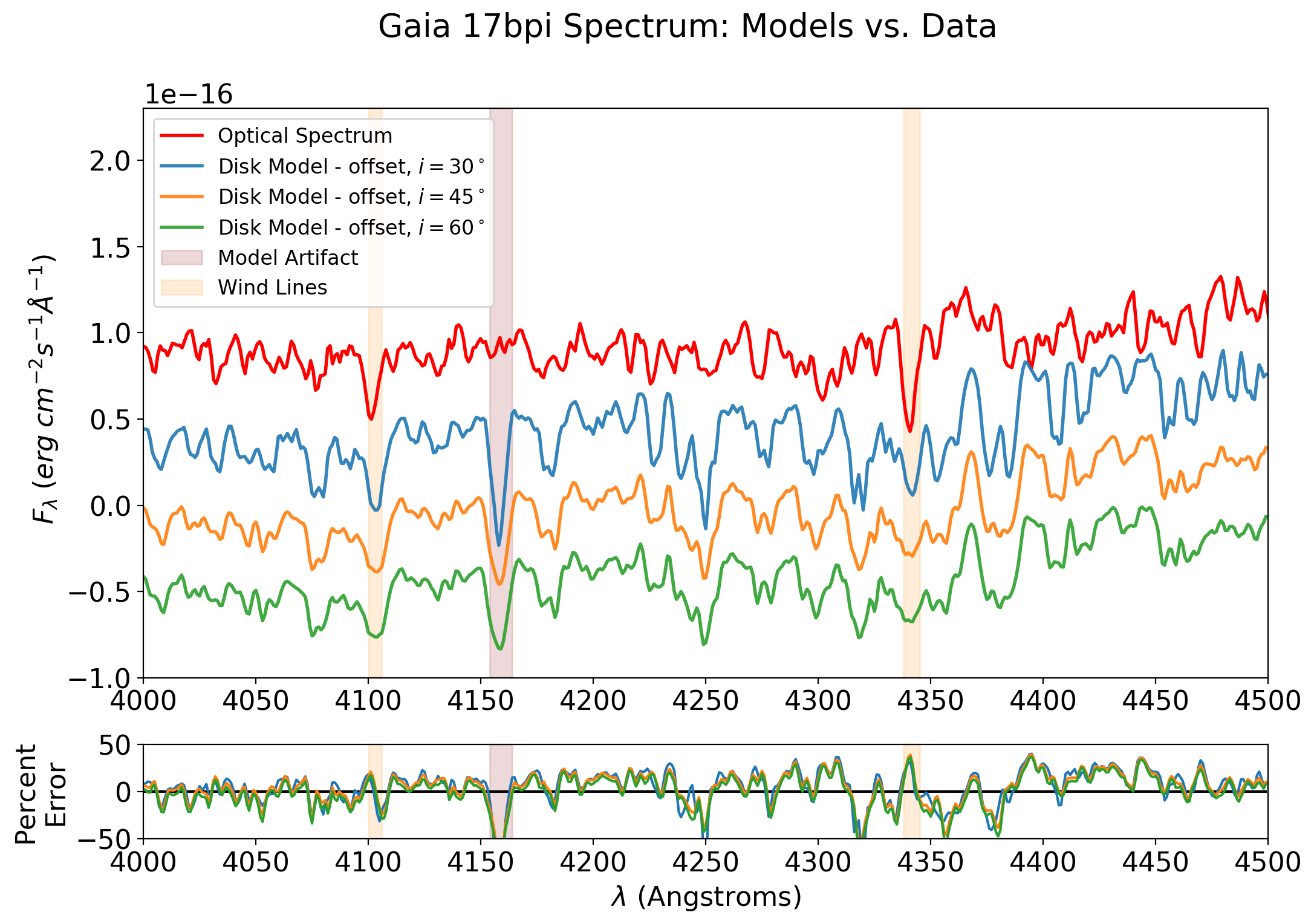}\\
    \caption{The three model spectra corresponding Models 2a, 2b, 2c, with inclinations of $i = 30^\circ, 45^\circ, 60^\circ$ respectively, are shown compared to the data. Upper panel: HBC 722, lower panel: Gaia 17bpi. The data is shown at the true flux level, and the models are vertically shifted  for easier visual comparison. The residual plot shows that the overall flux level is matched to within 10 percent, yet some line depths are overpredicted.}
    \label{fig:results_spectrum}
\end{figure}

In Figure \ref{fig:results_spectrum} we show the influence of inclination on the spectra.
The disk models for each of HBC 722 and Gaia 17bpi 
use the MCMC best-fit parameters obtained from the SED after fixing the inclination at $i = 30^\circ$, $i = 45^\circ$, and $i = 60^\circ$ (Models 2a, 2b, 2c). 

Visual inspection suggests good overall agreement in the line positions and profiles. The residual plot shows that the continuum level is matched well, within 10 percent, as expected from the SED analysis. However, it is clear that the majority of line depths are over-predicted by our models. At the same time, the $i = 60^\circ$ model appears to be too broadened to fit the data well. While not quantitatively apparent at medium resolution, analysis of the high-dispersion spectra (see below) reveals that HBC 722 likely has an inclination close to, or less than $30^\circ$. 

In this work, we are modeling only the spectrophotometry of the sources SEDs, and then demonstrating post facto that the spectral line predictions of the same models are a reasonable match to observed spectra.  We are not undertaking spectral line fitting here.  However, in this context, we highlight two further aspects of Figure \ref{fig:results_spectrum} that will need to be considered when performing future model fitting to spectra.

First, The hydrogen Balmer lines (indicated in orange shading) 
are likely produced in a strong wind surrounding these FU Ori sources \citep{milliner2019}. Our models capture only the lines created in the self-luminous disk, and thus are unable to reproduce the depths and velocity structure of the \ion{H}{1} lines or other lines affected by wind (more of an issue at high dispersion than low).  Second,  we call attention to a particularly strong feature (indicated by brown shading) appearing in the models just redward of 4150 \AA. A line of such strength is not found in standard spectral atlases or libraries, and highlights the presence of possible issues with model linelists or other deficiencies and artifacts that our disk model inherits from the set of NextGen stellar atmosphere models.

\subsubsection{High Resolution}
\begin{figure*}
    \centering
    \includegraphics[scale=0.33]{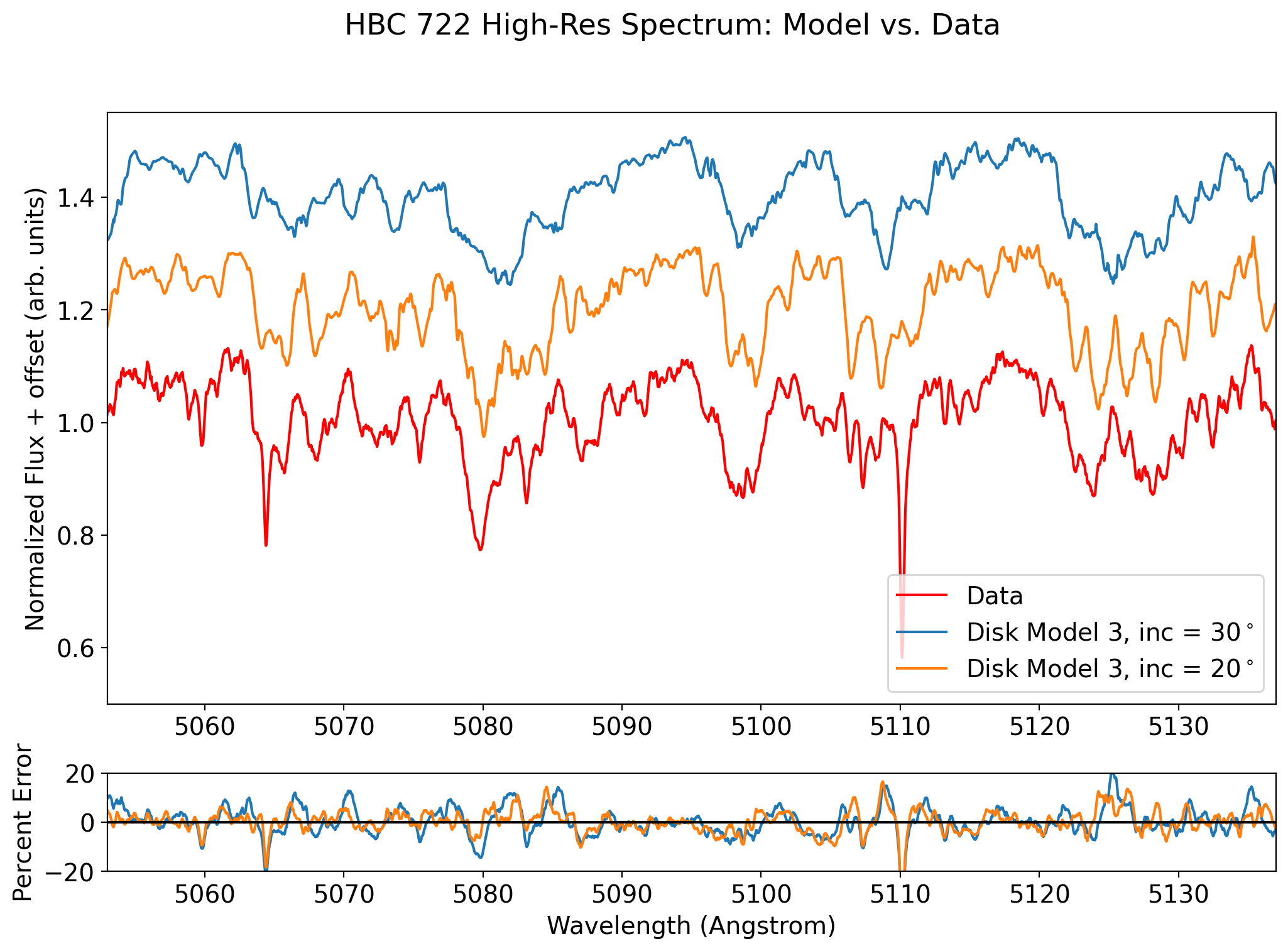}\includegraphics[scale=0.33]{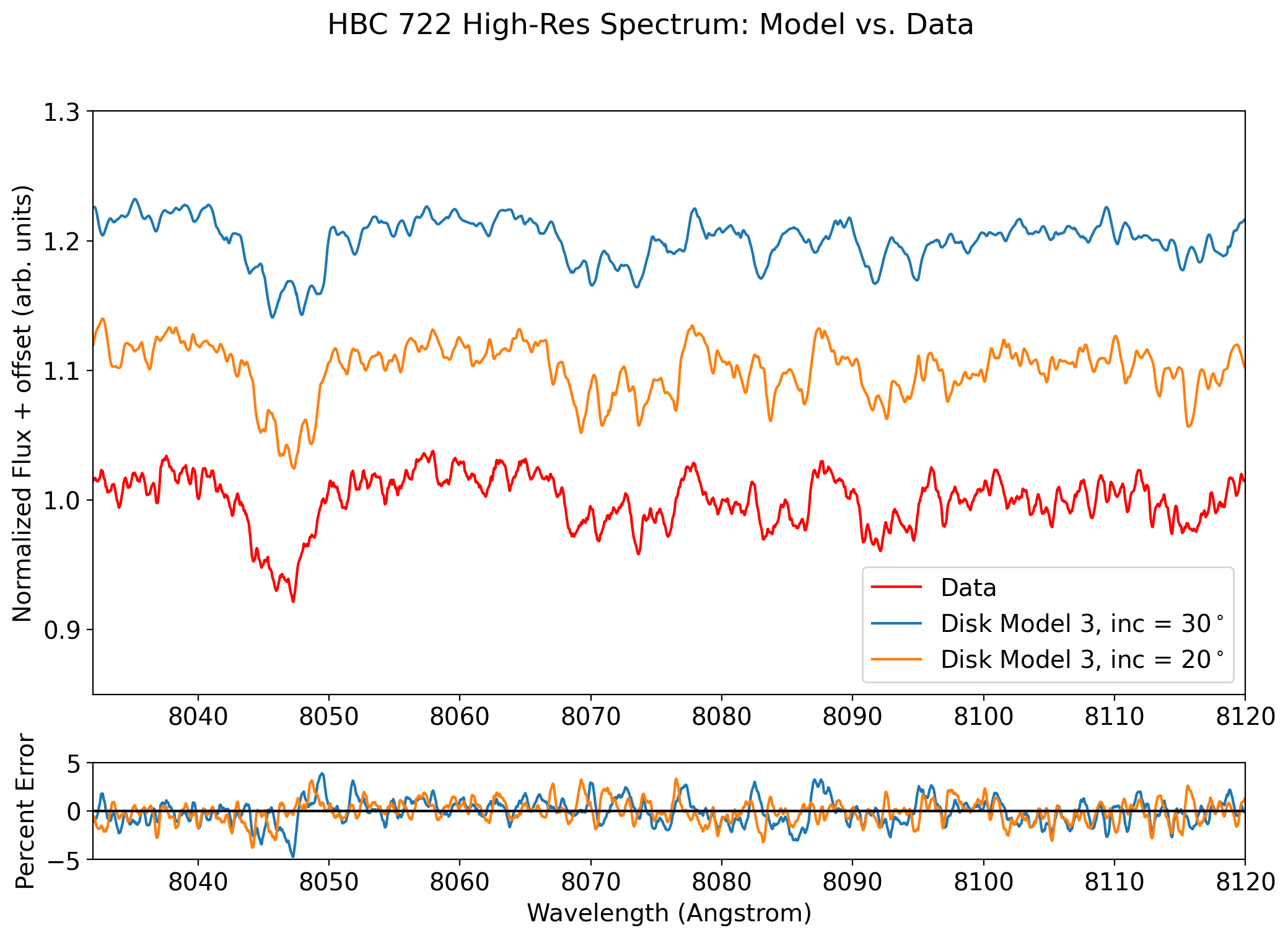}\\
    \includegraphics[scale=0.33]{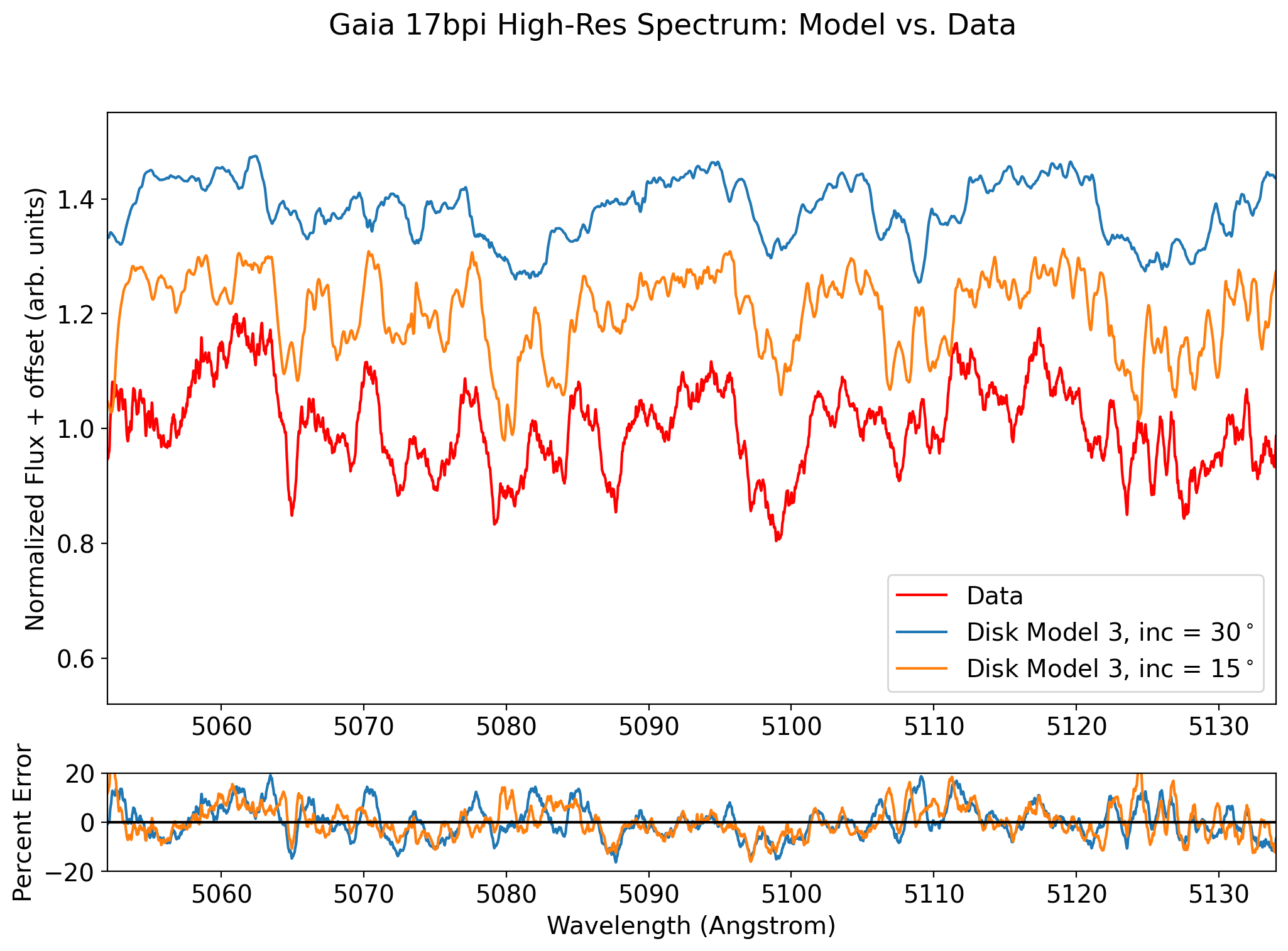}\includegraphics[scale=0.33]{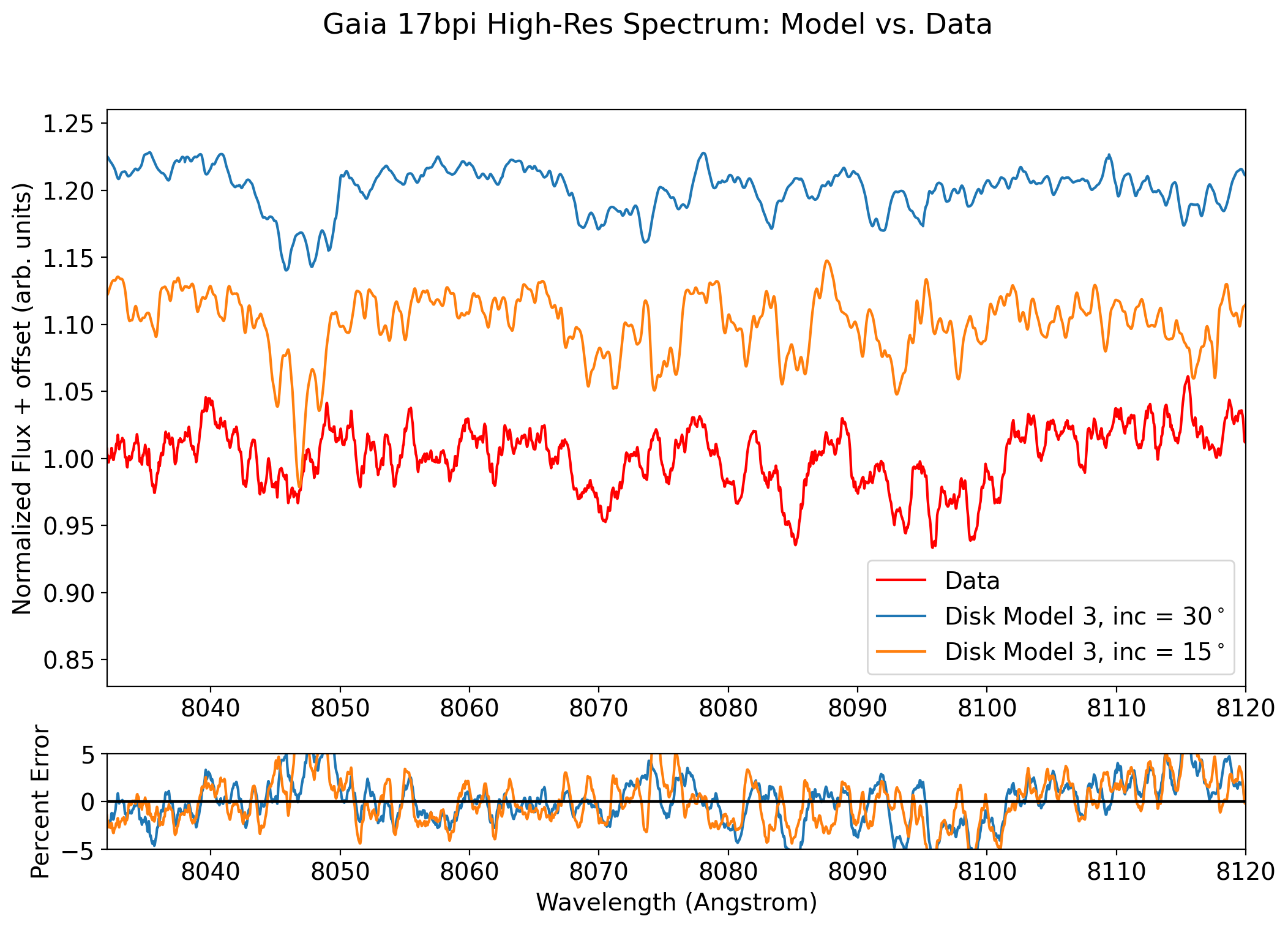}
    \caption{High-resolution ($R\sim 36,000$) 
    spectra of HBC 722 (upper panels) and Gaia 17bpi (lower panels) versus Model 3 of each object, with parameters given in Tables \ref{tab:hbc722_table} and \ref{tab:gaia17bpi_table}. We also include a better-fitting model with reduced rotational broadening, which is achieved via lower inclination, lower central mass, lower velocity in the inner disk, or combination of all three. The agreement between models and data is quite good on the whole, comfortably within 20 percent in the 5000 \AA $\;$regime, and within 5 percent in the 8000 \AA$\;$regime.}
    \label{fig:results_hires_spectrum}
\end{figure*}

Next,  with the additional challenges of matching spectral models to data in mind, we consider comparisons of our disk models to high-resolution spectra.  
In doing so, we consider that the ratio $M_*/R_*$ can be obtained from evolutionary models of young pre-main sequence stars \citep[e.g.][]{dartmouthmodels}, 
and we adopt this strategy to fix $\dfrac{M_* / M_\odot}{R_* / R_\odot}= 0.5$ in creating Model 3.   
We acknowledge that fixing this ratio may still carry uncertainties, as the value of $M_*/R_*$ is both slowly evolving and poorly constrained in young stars, and may even differ for the central objects of FU Ori disks.
We present the results of this Model 3 in Tables \ref{tab:hbc722_table} and \ref{tab:gaia17bpi_table}, as well as visually in Figure \ref{fig:results_hires_spectrum}. In Figure \ref{fig:results_hires_spectrum}, the same model is also shown with a reduced line broadening that leads to a better by-eye fit to the individual line profiles. While we represent the reduced broadening with a lower inclination value, keeping $M_*/R_*$ fixed, given Equation \ref{eq:disk_broaden_2} we could have chosen a lower value of $\frac{M_*}{R_*}$ to represent this effect, or point to non-Keplerian rotation in the inner disk as a possible source of the reduced broadening needed to fit the data. 

We emphasize in Figure \ref{fig:results_hires_spectrum} that the parameters derived from the fit of an $\a$-disk model 
to the broadband SED -- with additional guidance on inclination and $M_*/R_*$ from detailed examination of the spectra --
translate to a good qualitative fit to high-resolution spectra as well. There are obvious defects in the fit, however.  These could be attributed to the limitations of our simple model, which is not a full radiative transfer treatment as performed in modern theoretical studies (e.g. \citet{zhu2007}, \cite{zhu2020}). Nevertheless, certain aspects of the simple $\a$-disk model, such as the doubling of absorption line profiles
(e.g. around 5060, 5110, 8035, and 8070 \AA) are suggested in the data as well.

Overall, we find agreement at the 5-10\% level between high-resolution spectra of our two FU Ori stars and 
the model parameters derived above from the SED fit.

\subsection{Varying the Inner Disk Temperature Profile: Evidence for an Inner Disk Boundary Region?}

In considering the spectra, we can also explore the temperature profile in the innermost disk regions.  Appendix Figure \ref{fig:gamma_and_inc} illustrates the influence of the temperature profile exponent on the overall SED.

\begin{figure}
    \centering
    \includegraphics[scale=0.32]{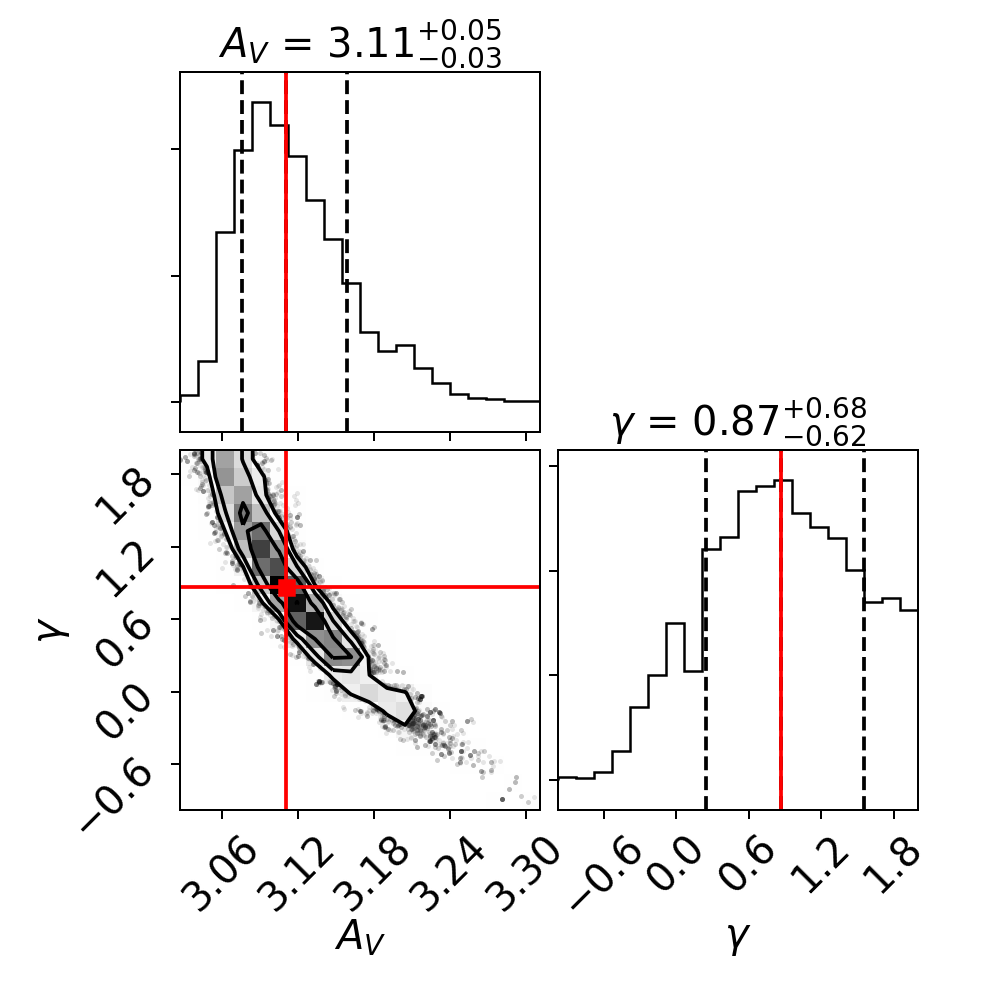}\includegraphics[scale=0.32]{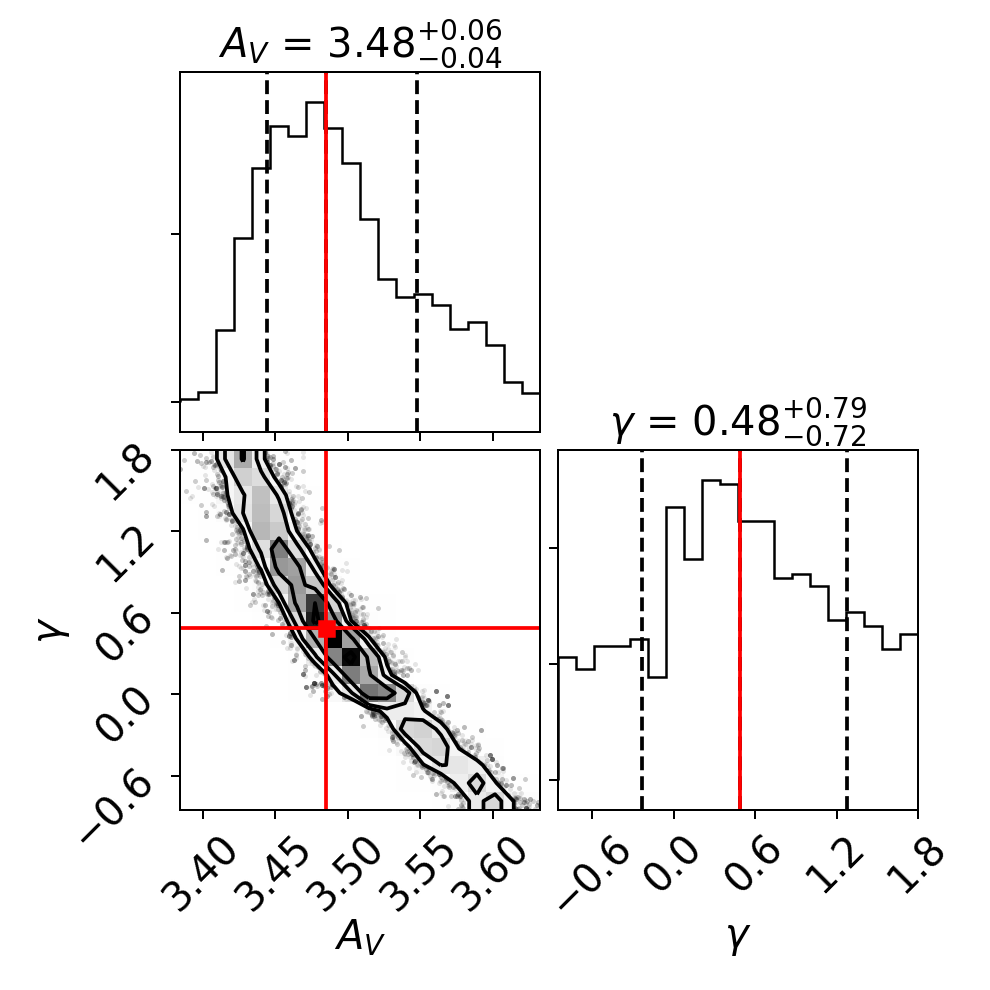} \\
    \caption{ The trade-off between extinction, $A_V$, and the power-law index of the boundary region, $\gamma$, is shown. Left panel: HBC 722, right panel: Gaia 17bpi. The higher the extinction, the lower the power-law index, and therefore the higher the boundary region temperature needed to fit the continuum level of the data.}
    \label{fig:gamma_trade}
\end{figure}

To explore this parameter, in our (final) MCMC run we keep all parameters fixed except for $A_V$ and $\gamma$, and study the trade off between these two parameters alone. A power law index $\gamma < 0$ leads to temperatures in the boundary region being higher than $T(r = 1.361R_*)$, the maximum temperature of a Shakura-Sunyaev disk. As expected, higher values of $A_V$ imply lower values of $\gamma$, and vice versa. In the corner plots shown in Figure \ref{fig:gamma_trade}, we show that even with $A_V$ constrained to a very small range, $\gamma$ still has a large credible interval. We conclude that at this point, while we can explore the trade-off between other parameters such as $A_V$ and $\gamma$ based on an SED analysis, the effects on the SED resulting from varying $\gamma$ are small enough to lie well within our errors for flux calibration, distance estimation to the object, inclination of the object, and $A_V$. Our corner plot result suggests the trade-off we can expect if one of those parameters becomes better constrained.

\begin{figure}
    \centering
    \includegraphics[scale=0.23]{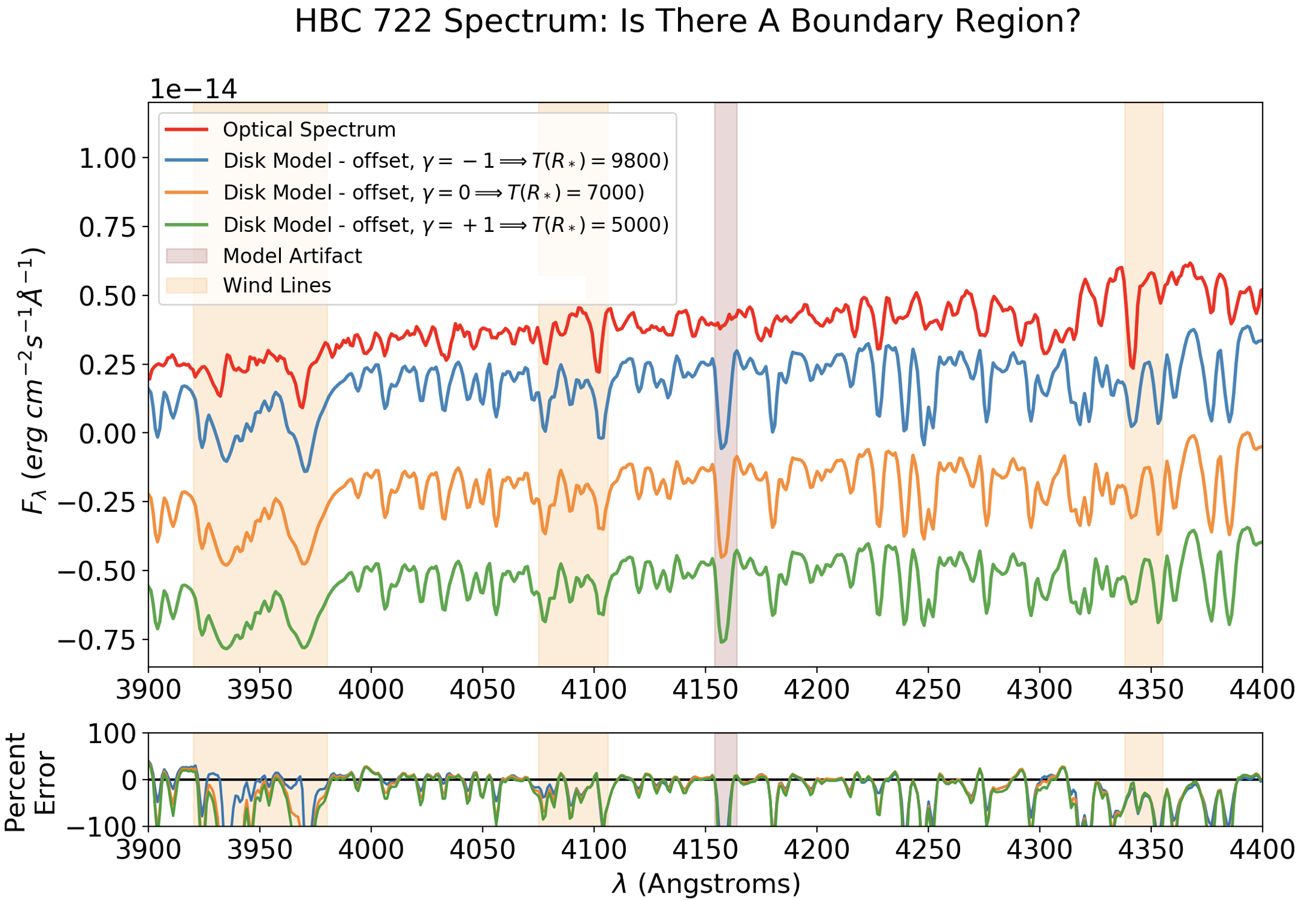}
    \includegraphics[scale=0.23]{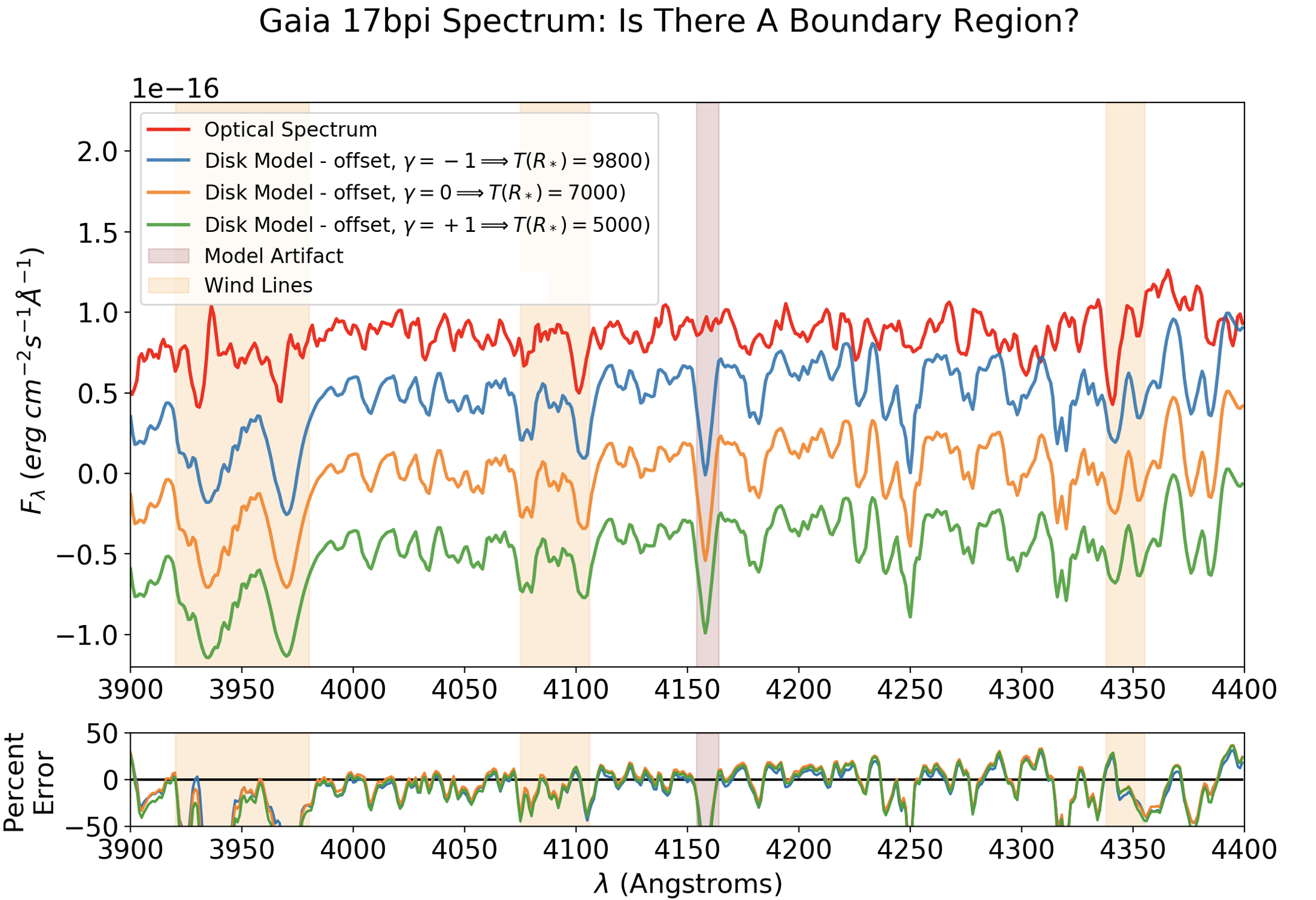}\\
    \caption{The three model spectra created using $\gamma = -1, 0, +1$ are compared to the data. Upper panel: HBC 722, lower panel: Gaia 17bpi. The data is shown at the true flux level, and the models are vertically shifted for easier visual comparison. The most diagnostic lines of the temperature structure in the boundary region are unfortunately the same as the "wind lines" discussed earlier. Bluer spectra are needed in order to constrain $\gamma$.
    }
    \label{fig:results_boundary_spectrum}
\end{figure}

To assess how deviations from the standard $\a$-disk temperature distribution might manifest, we present in Figure \ref{fig:results_boundary_spectrum} the effects of changing $\gamma$ on the bluest parts of our optical spectrum, between 3900-4400 \AA\  where the presence of hotter emission is most apparent. The three model spectra with $\gamma = -1, 0, +1$ indicate rising, flat, or decreasing temperature profiles in a boundary region. Highlighted are the spectral features that are most distinctly different between the various models. However, since the temperature range that is probed is above $\sim 7000$ K, with the $\gamma = -1$ model reaching 9600 K, it is primarily the continuum which begins to dominate over all lines other than the hydrogen Balmer series and the calcium "H" and "K" lines. These lines, however, are ``wind lines" that arise mostly from a hot outer wind that we do not model in our framework.

In summary, we find that we cannot use optical spectra as a suitable diagnostic for probing the innermost disk, and must appeal to (currently non-existent) ultraviolet spectra, where hotter emission will both affect the continuum level and produce spectral features that are more diagnostic of the boundary region temperature.

{
\
\subsection{Combined Parameter Exploration}
\label{sec:combined_param_explore}

In our modeling routine described above, we chose to fit for the five physical parameters $M_*, R_*, \dot{M}, A_V, i$.  
A more commonly adopted approach 
is to fit  an accretion disk SED using only two parameters, $T_\textrm{max}$ and $L_\textrm{acc}$, 
with $i$ and $A_V$ usually adopted rather than fitted. 

In Figure \ref{fig:lacc_corner_all}, we show the corner plot trading off the four parameters $L_\textrm{acc}, T_\textrm{max}, A_V, i$. 
We use 16 walkers (4 for each parameter), and run for 15,000 steps, taking half as the burn-in period. 
This exercise allows us to place constraints on $L_\textrm{acc}, T_\textrm{max}$, as reported in Tables \ref{tab:hbc722_table} and \ref{tab:gaia17bpi_table}. 

\begin{figure*}
    \centering
    \includegraphics[scale=0.35]{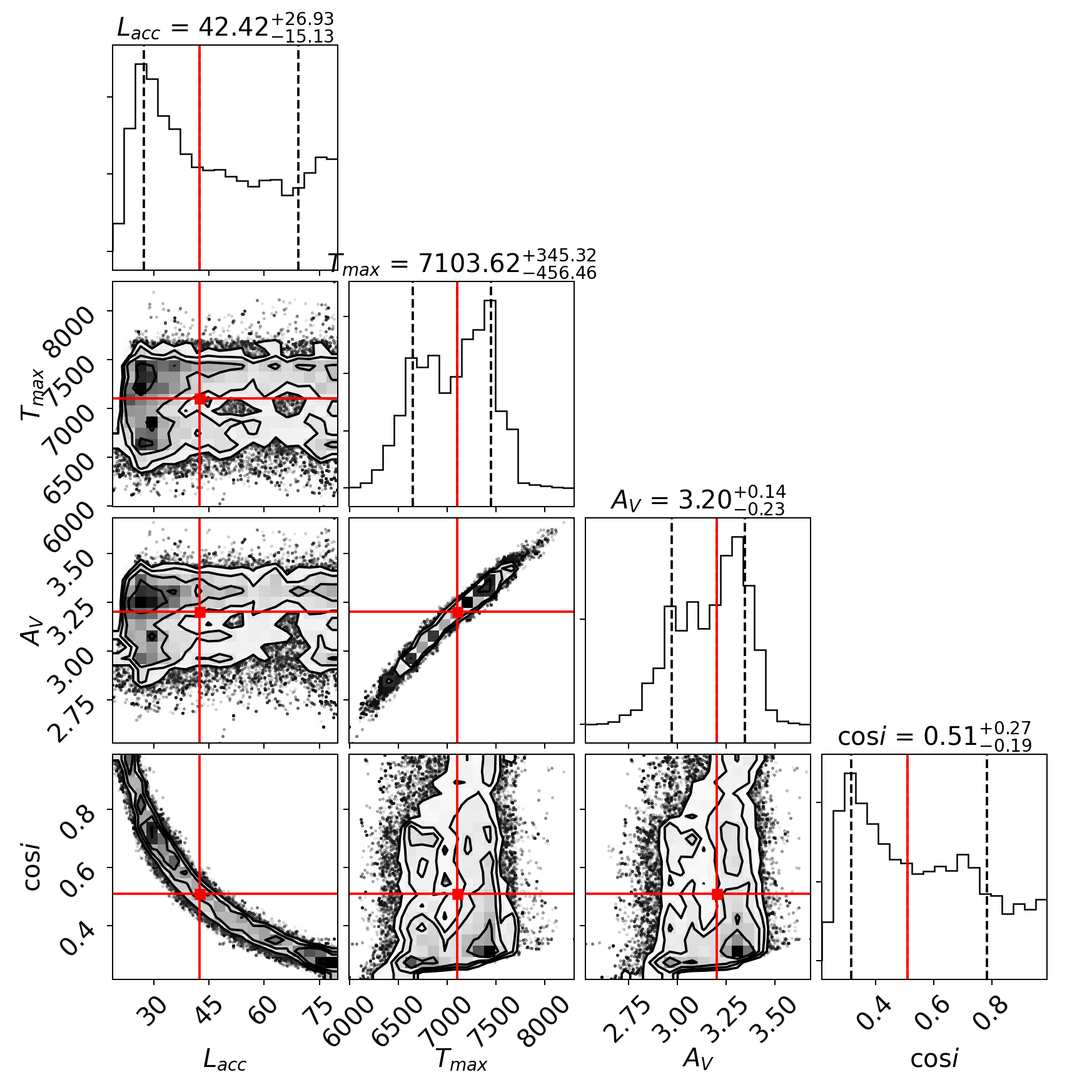}
    \includegraphics[scale=0.35]{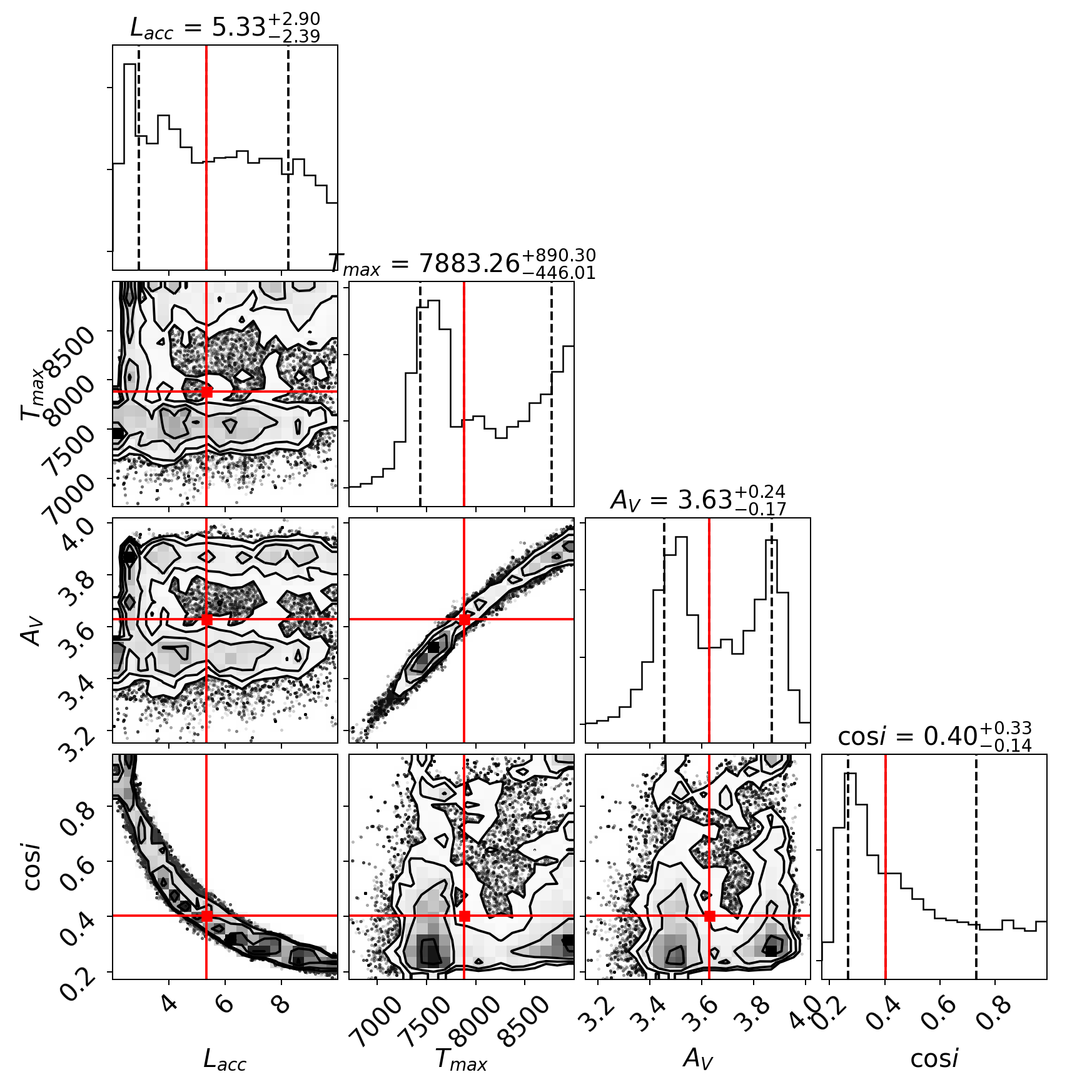}
    \caption{MCMC corner plots trading off parameters $L_\textrm{acc}, T_\textrm{max}, A_V, i$. Left: HBC 722, Right: Gaia 17bpi. $L_\textrm{acc}$ and $T_\textrm{max}$ are well-constrained, and the credible intervals (16th and 84th quartiles) obtained here are reported in Tables \ref{tab:hbc722_table} and \ref{tab:gaia17bpi_table}.}
    \label{fig:lacc_corner_all}
\end{figure*}

In Figure \ref{fig:lacc_corner_select}, we show the corner plot after fixing the extinction, $A_V$, and inclination, $i$. We use 8 walkers (4 for each parameter) and run for 8,000 steps, taking half as the burn in period.
The resulting fits for $T_\textrm{max}$ and $L_\textrm{acc}$ are now much better constrained. $L_\textrm{acc}$, however, still strongly depends on the distance to the source. In the case of Gaia 17bpi, which is more distant and thus has suboptimal distance constraints, if its distance is underestimated, then $L_\textrm{acc}$ will be higher. Since $L_\textrm{acc}/T_\textrm{max}^4 \propto R_*^2$, an underestimated distance would lead to a larger best-fit stellar radius.

\begin{figure}
    \centering
    \includegraphics[scale=0.3]{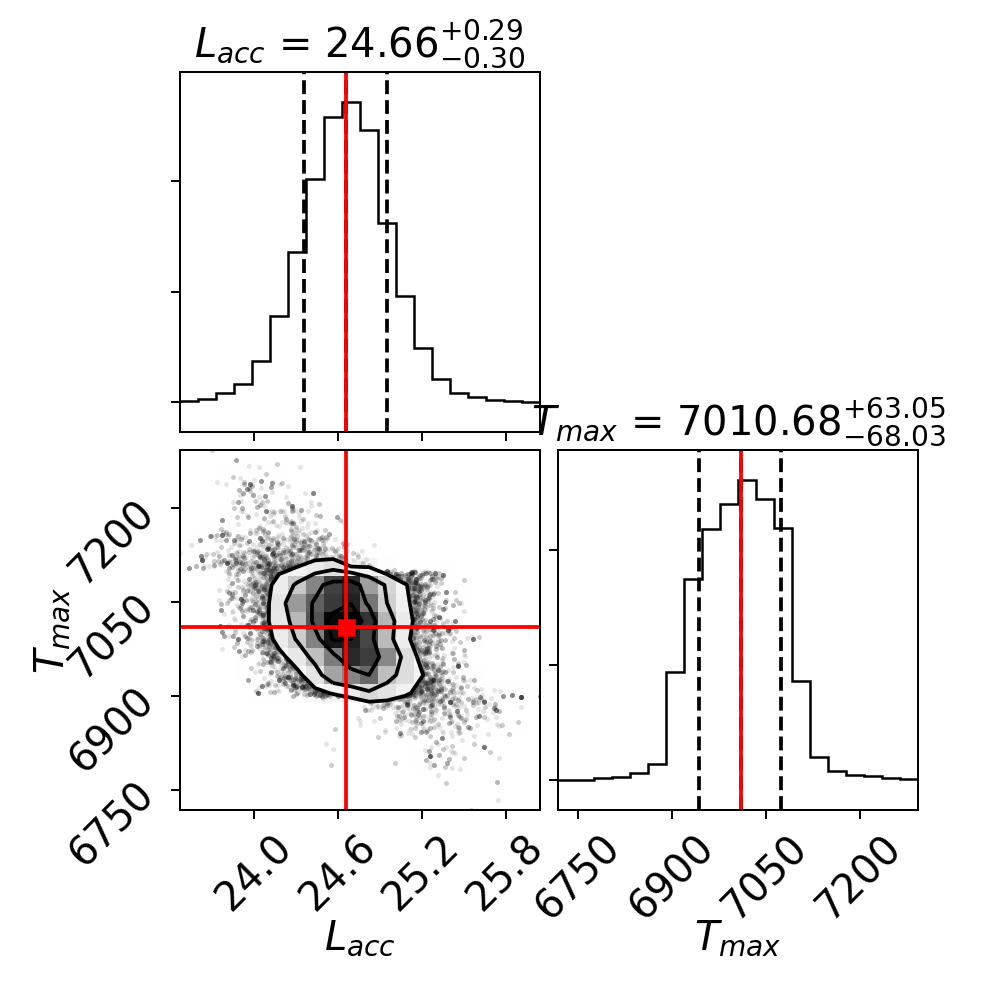}
    \includegraphics[scale=0.3]{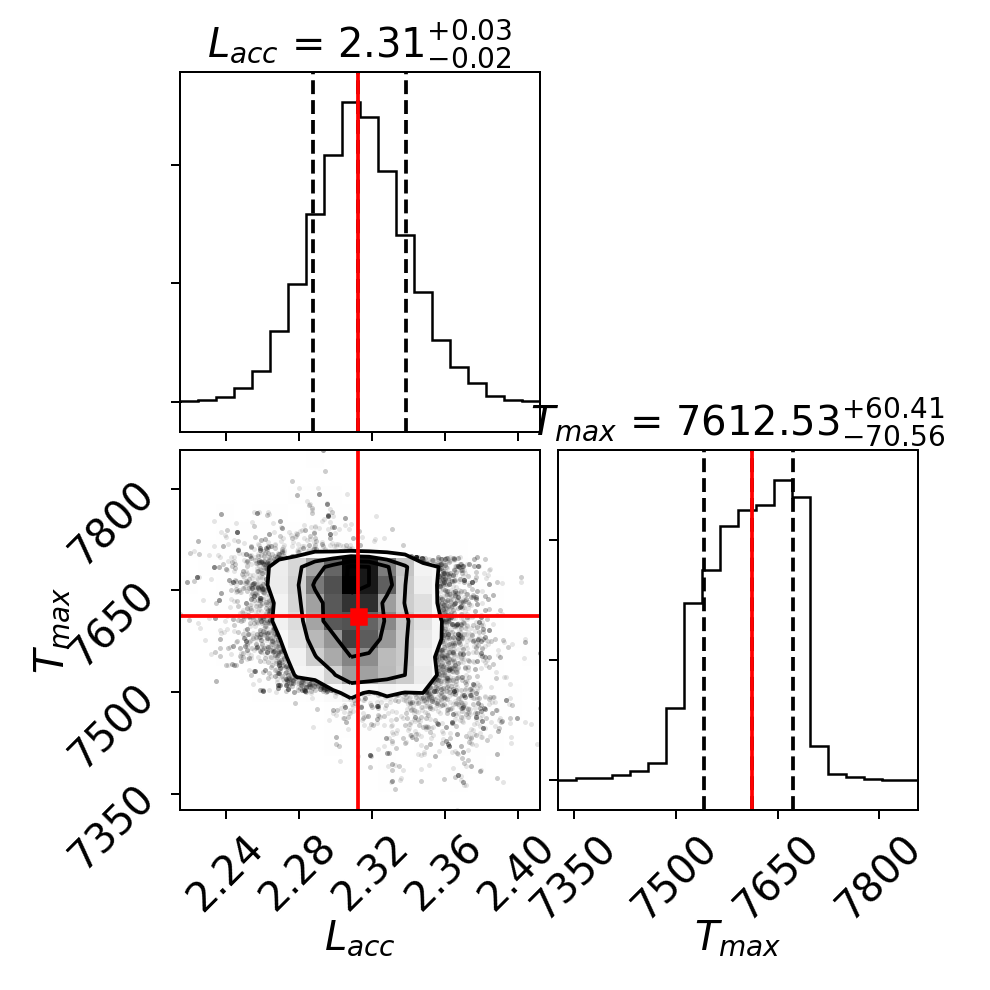}
    \caption{MCMC corner plots trading off parameters $L_\textrm{acc}, T_\textrm{max}$. Left: HBC 722, Right: Gaia 17bpi. $L_\textrm{acc}$ and $T_\textrm{max}$ are better constrained than in \ref{fig:lacc_corner_all}, and the credible intervals (16th and 84th quartiles) obtained here are reported in Tables \ref{tab:hbc722_table} and \ref{tab:gaia17bpi_table}.}
    \label{fig:lacc_corner_select}
\end{figure}

}

\subsection{Summary of Disk Model Parameters}
Tables \ref{tab:hbc722_table} and \ref{tab:gaia17bpi_table} summarize the lists of model parameters based on all of the above corner plots, for both HBC 722 and Gaia 17bpi. 
Model 1, our most conservative model with the highest number of free parameters is chosen for reporting in the abstract and conclusion. This model is the one with the largest credible intervals, and therefore allows for the greatest uncertainty in flux calibration, distance estimates, etc.

\begin{table*}
    \centering
    \caption{Model Parameters for HBC 722 with Credible Intervals}
    \begin{tabular}{|c||c|c|c|c|c|c||c|c|}
    \hline
        Model No. & $M_*$ $(M_{\odot})$ & $R_*$ $(R_{\odot})$ & $\log\dot{M}$ $(M_{\odot} \textrm{ yr}^{-1})$ & $A_V$ & $\cos i$ & $\gamma$&  $T_{\textrm{max}}$ (K) & 
        $L$ $(L_{\odot})$\\
        \hline
        1 & $0.66^{+0.22}_{-0.29}$ & $2.07^{+2.72}_{-0.56}$& $-4.90^{+0.99}_{-0.40}$ & $3.16^{+0.17}_{-0.21}$& $0.41^{+0.36}_{-0.34}$& 0 & $7100_{-500}^{+300}$ & $42^{+27}_{-16}$\\
        2a & $0.56^{+0.30}_{-0.27}$ & $1.39^{+0.03}_{-0.03}$& $-5.41^{+0.29}_{-0.19}$ & $3.16$ & $\sqrt{3}/2$& 0 & $7010_{-70}^{+60}$ & $24.7_{-0.3}^{+0.3}$\\
        2b & $0.60^{+0.29}_{-0.28}$ & $1.54^{+0.03}_{-0.03}$& $-5.30^{+0.27}_{-0.17}$ & $3.16$ & $\sqrt{2}/2$& 0 & $7010_{-70}^{+60}$  &$29.6_{-0.3}^{+0.3}$\\
        2c & $0.58^{+0.27}_{-0.27}$ & $1.82^{+0.04}_{-0.03}$& $-5.06^{+0.27}_{-0.17}$ & $3.16$ & $1/2$ & 0 & $7010_{-70}^{+60}$ & $39.5_{-0.3}^{+0.3}$\\
        3 & $0.70$ & $1.39$& $-5.50$ & $3.16$& $\sqrt{3}/2$& 0  & $7000$ & $24$\\
        4 & $0.60$ & $1.54$& $-5.30$ & $3.16^{+0.05}_{-0.03}$ & $\sqrt{2}/2$ & $0.87^{+0.68}_{-0.62}$ & $7000$ & $30$\\
        \hline
    \end{tabular}\\
    \textrm{The sets of parameters, along with 16 and 84 percent credible intervals (if parameter is free), that fit the SED of HBC 722. 
    }
    \label{tab:hbc722_table}
\end{table*}

\begin{table*}
    \centering
    \caption{Model Parameters for Gaia 17bpi with Credible Intervals}
    \begin{tabular}{|c||c|c|c|c|c|c||c|c|}
    \hline
        Model No. & $M_*$ $(M_{\odot})$ & $R_*$ $(R_{\odot})$ & $\log\dot{M}$ $(M_{\odot} \textrm{ yr}^{-1})$ & $A_V$ & $\cos i$ & $\gamma$&  $T_{\textrm{max}}$ (K) & 
        $L$ $(L_{\odot})$\\
        \hline
        1 & $0.63^{+0.25}_{-0.29}$ & $0.45^{+0.19}_{-0.09}$& $-6.70^{+0.46}_{-0.36}$ & $3.52^{+0.17}_{-0.10}$& $0.54^{+0.30}_{-0.27}$& 0 & $7900_{-400}^{+900}$ & $5.3^{+2.9}_{-2.4}$\\
        2a & $0.63^{+0.25}_{-0.32}$ & $0.36^{+0.01}_{-0.01}$& $-7.07^{+0.30}_{-0.15}$ & $3.52$& $\sqrt{3}/2$& 0 & $7610_{-70}^{+60}$ & $2.31^{+0.03}_{-0.02}$\\
        2b & $0.66^{+0.25}_{-0.27}$ & $0.40^{+0.01}_{-0.01}$& $-6.96^{+0.23}_{-0.14}$ & $3.52$& $\sqrt{2}/2$& 0 & $7610_{-70}^{+60}$ & $2.81^{+0.03}_{-0.02}$\\
        2c & $0.56^{+0.29}_{-0.27}$ & $0.48^{+0.01}_{-0.01}$& $-6.67^{+0.28}_{-0.18}$ & $3.52$& $1/2$& 0 & $7610_{-70}^{+60}$ & $3.92^{+0.03}_{-0.02}$\\
        3 & $0.18$ & $0.36$& $-6.53$ & $3.52$& $\sqrt{3}/2$& 0 & $7600$ & $2.3$\\
        4 & $0.66$ & $0.40$& $-6.96$ & $3.52^{+0.06}_{-0.04}$& $\sqrt{2}/2$& $0.48^{+0.79}_{-0.72}$ & $7600$ & $2.8$\\
        \hline
    \end{tabular}\\
    \textrm{The sets of parameters, along with 16 and 84 percent credible intervals (if parameter is free), that fit the SED of Gaia 17bpi.} 
    \label{tab:gaia17bpi_table}
\end{table*}



\begin{figure*}
    \centering
    \includegraphics[scale=0.33]{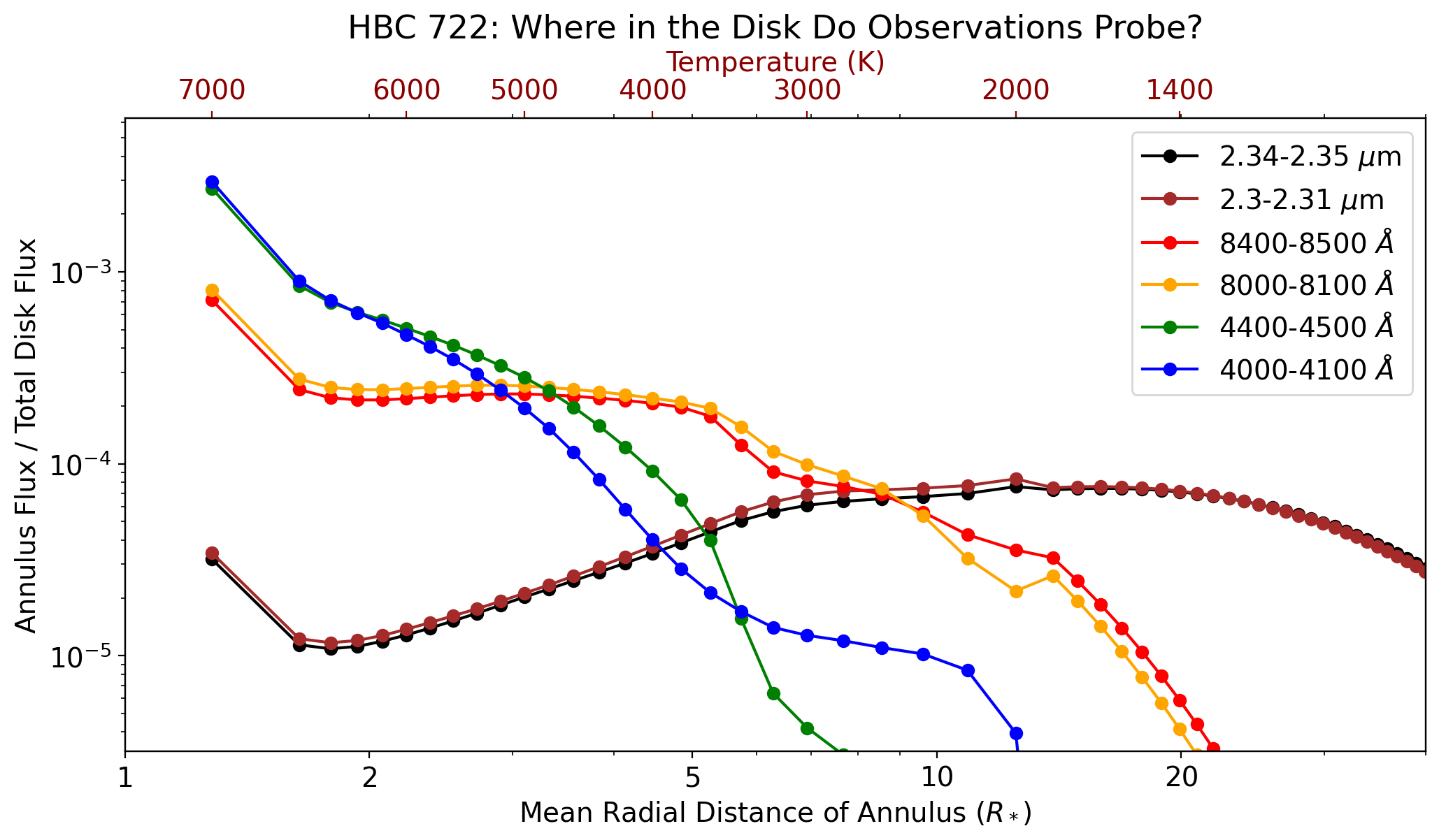} \includegraphics[scale=0.33]{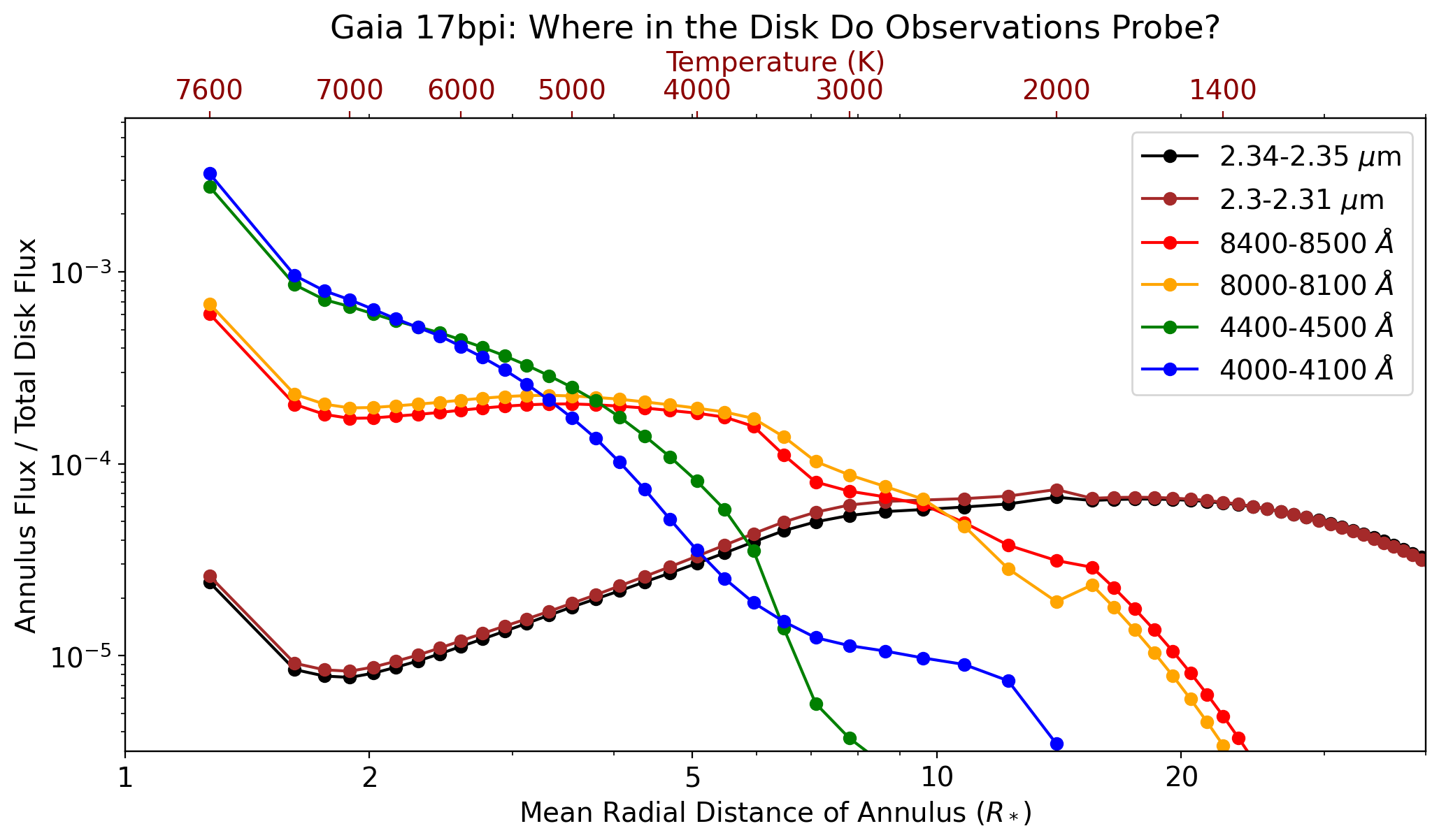}\\
    \caption{Visualizations of the structure of the multi-temperature disk model shown for HBC 722 (left) and Gaia 17bpi (right). The fractional contribution of each annulus to the total flux of the disk integrated over various wavelength bands is illustrated, detailing how the hotter regions of the disk closer to the star are probed by optical observations, and cooler regions farther out are probed by near-infrared observations.
    }
    \label{fig:fracflux}
\end{figure*}

Another way of visualizing the disk model is presented in Figure \ref{fig:fracflux}, showing
of the fractional contribution of each annulus to the total flux of the disk, 
as integrated over various wavelength bands. Here we have adopted Model 3. To create these figures for any given wavelength range, we integrate $F_\lambda$ over that range, and divide by the total luminosity of the disk to obtain the fractional contribution of that annulus. Recall that in our modeling routine, for annuli cooler than 2000 K, we assign blackbodies instead of stellar atmospheres binned by radius, not by temperature. Thus, in order for the fractional flux contribution of these outer annuli to be plotted according to equal temperature sampling, we weigh their flux contribution by $\Delta T_\textrm{inner, fixed}/\Delta T_\textrm{outer, variable}$. Note that the fractional flux plots have a small jump at this point due to the transition from stellar atmospheres to blackbodies.

\begin{figure*}
    \centering
    \includegraphics[scale=0.34 ]{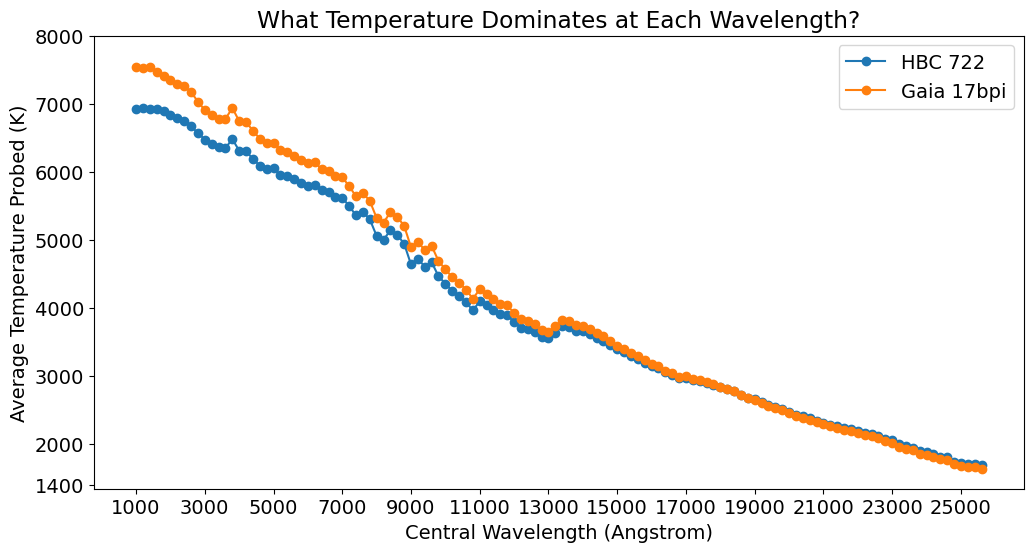}\includegraphics[scale=0.34]{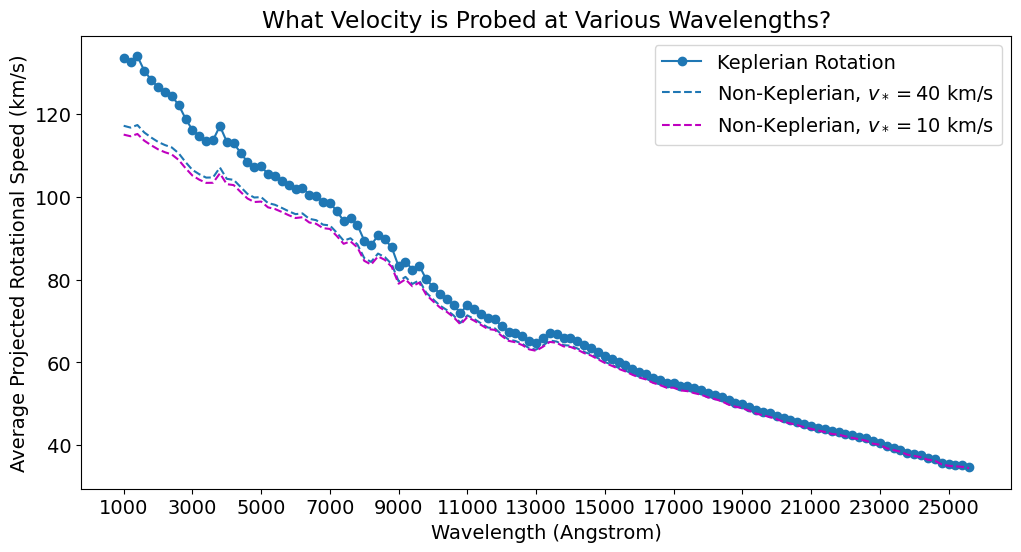}\\
    \caption{
    Left: Recasting of the information in Figure \ref{fig:fracflux}, now showing the weighted average temperature of all annuli in the disk model at wavelengths from the UV to the NIR. Right: Same as left, but showing the weighted average rotational speed (both Keplerian and a toy non-Keplerian profile with a stellar rotational speed of either 10 km/s or 40 km/s) of all annuli in the disk. These figures highlight the importance of the UV in probing the hottest temperatures and fastest rotating regions of the disk. Note that both plots are made for Model 3 and that the right plot is independent of object since $M_*/R_*$ is kept fixed.
    }
    \label{fig:lambda_t_v}    
\end{figure*}

The plots in Figure \ref{fig:fracflux} show quantitatively how the bluest optical wavelengths are dominated by the hottest, innermost regions of the disk, whereas near-infrared wavelengths are dominated by the coolest, outermost regions of the disk. The plots in Figure \ref{fig:lambda_t_v} show this property in a different way, with the contribution of each annulus within 200 \AA\ wavelength bands weighted by the relative contribution of the annulus from Figure \ref{fig:fracflux}, summed across the entire disk. Both the temperature and the velocity profiles with wavelength are given in Figure \ref{fig:lambda_t_v}. The former shows the dominant temperature contribution in a given wavelength band, and the latter shows the dominant rotational broadening. For illustrative purposes, in Figure \ref{fig:lambda_t_v}, we also consider a departure from Keplerian rotation for the disk to match a hypothetical 10 km/s or 40 km/s stellar rotational speed at the star-disk boundary region. The non-Keplerian velocity profile we adopt is a simple toy model to match the disk rotation to the slower rotation of the star at the star-disk interface, following derivations found in a standard text such as \cite{trittonfluidbook}:
\begin{gather}
    u_\phi(r) = \Omega_Kr + (\Omega_* - \Omega_K)R_*^2/r
\end{gather}
where $\Omega_K$ is the Keplerian orbital frequency and $\Omega_*$ the rotational frequency of the star. At large radial distances, it converges to the standard Keplerian model.


As expected, going to shorter wavelengths increases the contribution from hotter material closer to the innermost disk regions.
Howevever, what matters for the derivation of physical parameters such as $T_\textrm{max}$ is the spectral content at these blue wavelengths, not just the fractional flux.

\section{Discussion}

\subsection{HBC 722 and Gaia 17bpi in Context}

For both HBC 722 and Gaia 17bpi, we have derived relatively narrow constraints on the accretion rate, despite the possible leeway in other parameters. For  HBC 722 the accretion rate is constrained to within 1.5 orders of magnitude (full error range), and for Gaia 17bpi it is to within 0.8 dex (full error range). 
Gaia 17bpi would have one of the lowest FU Ori accretion rates,
while HBC 722 has an accretion rate comparable to other FU Ori objects \citep[e.g.][]{hartmannbook,gramajo2014}.

For Gaia 17bpi, our work is the first attempt to model the disk and derive an accretion rate, as well as to constrain the other source parameters.
The low luminosity implies both a relatively low accretion rate, and also a small radius with $R_* = 0.45^{+0.19}_{-0.09}R_\odot$, about 4-5 times lower than that of HBC 722. This is significantly lower than expected of typical pre-main sequence stars assumed to be capable of undergoing an FU Ori-type outburst, though could imply
a young substellar mass object or brown dwarf.  The modeling constraints on source mass are poor, and while such low masses are outside the reported credible intervals in our modeling, they can not be excluded given the corner plots.  With a very low mass, the very low luminosity becomes more acceptable in the context of an HR diagram, as being mainly accretion driven.
In order to find a model with a much larger source radius that is more consistent with young stars, the disk inclination would have to be much higher (see Figure \ref{fig:results_all} illustrating the parameter space trade), above $\sim$75 deg.  The high dispersion spectrum (Figure~\ref{fig:results_hires_spectrum}), however, seems more consistent with very low inclinations, more like $\sim$15 deg.

For HBC 722, our median accretion rate $10^{-4.9} M_\odot \textrm{ yr}^{-1}$ (equivalently written as $1.3 \times 10^{-5} M_\odot \textrm{ yr}^{-1}$ ), is only slightly lower (factor of 3) than the updated accretion rate of FU Ori of $3.8 \times 10^{-5} M_\odot \textrm{ yr}^{-1}$ as reported by \cite{zhu2020}.

We have shown two infrared spectra for HBC 722 in Figure~\ref{fig:results_all}, 
one from 2010 \citep{millerhbc722} and one from 2011 \citep{kospal2016}. 
Although our model fitting was conducted using the former spectrum, the latter is in some respects 
a better match to the resulting accretion disk model. 
As mentioned earlier and also in \cite{kospal2016}, spectral evolution is apparent over this 9 month period following the lightcurve peak
and temporary flux decline by about 1 mag.   The later spectrum shows development of a redder infrared SED and more pronounced water absorption in the H-band and K-band.  These features match our model well.

Our spectrophotometric fitting results for HBC 722 compare well with the accretion rate inferred by \cite{kospal2016},
who performed blackbody modeling of time series BVRI$+$JHK$+$IRAC1,2 photometry.
Our accretion rate is similar, though higher by a factor of 2-3, as is the estimated inclination, within 10 deg.
However, we have also demonstrated based on examination of the spectral lines, that the inclination is
likely much lower than is implied in the spectrophotometric or SED fitting.
Another difference is that \cite{kospal2016} find that they need to truncate the outer disk whereas we do not.
The argument in that paper was that the outburst development proceeded from the inside-out, 
with both the temperature increasing and the emitting area growing (from 0.066 to 0.146 AU) over time. 
We note that in the Zhu et al. models, however, the outburst starts in the 0.15-1 AU range and proceeds inward. 
Our disk model fitting also provides more robust derivations of the other parameters 
relative to the assumptions in \cite{kospal2016}, e.g. regarding stellar mass and radius, and line-of-sight extinction,
though the results are generally consistent.   

Comparing the two sources we have modeled here, we find that the median temperature of Gaia 17bpi is 600-800 K higher than that of HBC 722 despite being much less luminous and having a lower accretion rate. We note that since optical and near-infrared spectrophotometry were acquired at different times for this object, we were forced to use a lightcurve to guide our relative calibration between the two data sets. Therefore, we acknowledge the possibility that the median temperature of Gaia 17bpi could be somewhat lower, while still being within the credible interval in Model 1. 

\subsection{The Star-Disk Boundary Region}

In our study, we use the term ``boundary region" to refer to the region of the accretion disk between the surface of the star and the radial distance at which the Shakura-Sunyaev temperature profile reaches a maximum, $R = 1.361R_{\odot}$. While \cite{kh89} report no evidence of a classical, thin boundary layer as seen in cataclysmic variables from IUE ultraviolet spectra, the more recent work by \cite{kravtsova07}, using Hubble Space Telescope spectra, finds signatures of $T_{\textrm{eff}} \sim 9000 \textrm{ K}$ gas at wavelengths shorter than 2600 \AA. Thus, in our modeling, we introduce the parameter $\gamma$ as outlined in \S \ref{sec:gamma_parameter} to allow the entire boundary region to be governed by a temperature profile with a power-law exponent of $\gamma$ and thus allowing for higher or lower temperatures in the innermost region of the disk
than in the classical $\alpha$-disk model. 

As illustrated in the previous sub-section, optical and near-infrared spectra are not sufficient to see traces of a hotter (or cooler) boundary region than predicted by the simple empirical disk model of \cite{khh88}, so we can neither demonstrate nor discard evidence of such a region. The continuum level changes by very small amounts, and constraining $\gamma$ from SEDs alone requires very precise constraints on, $A_V$, $d$, and $i$ that we cannot obtain given our current data sets of FU Ori objects. 

Furthermore, the spectral lines that seem to be most diagnostic of the variation of $\gamma$ in our models, the calcium H and K lines, as well as the hydrogen Balmer series, are dominated by the kinematics of outflowing wind, rather than the self-luminous rotating disk we are modeling. Thus, we must appeal to future quantitative analyses of high-resolution spectra, or ultraviolet spectra, to more precisely determine temperature characteristics. High resolution spectra allow for temperature sensitivity to be probed at the level of individual spectral lines. Ultraviolet spectra directly measure hotter disk emission, as demonstrated in Figure \ref{fig:lambda_t_v}.

\subsection{Does the Gaia 17bpi Disk Really Outshine Its Central Star?}
\label{sec:gaia17bpi_star_and_disk}
The conventional model of FU Ori objects leaves out the central star simply because the luminosity of the entire accretion disk system seems to be on the order of 100 to 1000 times the luminosity of typical pre-main sequence stars. In the case of FU Ori itself, the total luminosity is estimated at $340-466 L_{\odot}$, and in the case of V1057 Cyg, another ``classic" FU Ori, $170-370 L_{\odot}$ \citep{gramajo2014}. HBC 722, which is studied here, was reported in the same work using the previously calculated distance of 520 pc, with luminosity $8-12 L_{\odot}$; scaled using the new distance of 795 pc obtained from \cite{kuhn2020}, this becomes $18-28 L_{\odot}$. Note that uncertainties in extinction still allow for reasonably higher or lower values in all of these cases.

As characterized from a pre-outburst SED, the central star of Gaia 17bpi is estimated to have $L_* = 0.3 L_{\odot}$ \citep{gaia17bpi2018}. In the models presented above, the luminosity of the outbursting inner disk is estimated to be around $3-8L_{\odot}$. This puts the outburst luminosity only $\sim 10-30$ times the luminosity of the central star, calling into question one of the initial assumptions of this work, that the luminosity of the central star is negligible and the SED is disk-dominated.  
Could the correct interpretation of Gaia 17bpi be that we are observing \textit{both} the self-luminous accretion disk as well as a minor contribution from the central star? In this scenario, Gaia 17bpi would be unique among FU Oris in that the luminosity of the central star would not be able to be ignored.  The object would thus present the opportunity to study the properties of the central star and the star-disk interaction during FU Ori-type accretion events.

\subsection{Needs for Future Modeling Progress}

Any physical model of an astrophysical source relies on parameters that are estimated from observations, 
and thus known to some level of accuracy.  Beyond the typical distance and extinction uncertainties, 
the main uncertainties in the modeling analysis presented here relate to
the ratio of $M_*/R_*$ for the central source, and the inclination of the outbursting disk. 

To make progress on these parameters, 
more in-depth exploration of the high-resolution spectra over a broad wavelength range should be undertaken.
Fitting of a disk model to spectrally resolved individual line profiles and line families should constrain the line broadening
as a function of wavelength, and hence both  $M_*/R_*$ and sin$i$. As illustrated in Figures~\ref{fig:fracflux} and \ref{fig:lambda_t_v},
different wavelengths probe different radii in the disk, and hence different ranges of the expected Keplerian velocity profile.
It should be possible to constrain $v\sin i$, and therefore a combination of $M_*, R_*, \textrm{ and } i$. 
The variation of the velocity width as a function of wavelength could also provide clues regarding 
any non-Keplerian aspects of the rotation profile in the innermost parts of the disk, 
and perhaps even the general nature of viscosity in protostellar disks.
Finally, the high dispersion spectra should help constrain $T_{max}$ by revealing the details of higher excitation, ``hotter" lines.

Also valuable as an inclination constraint are spatially resolved observations of the outer disk, 
for example those using the Atacama Large Millimeter Array (ALMA), as in \cite{perez2020}. 
However, there is evidence that at least some YSO disks are significantly warped, with outer and inner disks
seen at vastly different inclination angles \citep{armitage1997,sakai2019}. 
Constraints on the {\it inner} disk geometry, where we expect the FU Ori phenomenon to manifest,
could come from next generation near-infrared and optical interferometers.

Another need comes from the theoretical perspective. The definition and meaning of $R_*$ could be further considered,
given the effects of boundary region accretion in FU Ori objects, and the expected ``puffing up" of the central star
as it accretes mass, which deposits energy. This has been discussed in \cite{popham96} and \cite{hartmannreview2016}.   

Besides better constraints on uncertain model parameters, there is one observational advance
that would significantly improve the disk modeling effort. The UV spectral region is the most sensitive to viscous heating 
and boundary region emission at the relevant temperatures for FU Ori accretion.  
Our current modeling of the blue optical through near-infrared spectrophotometry is able to provide a constraint on 
the maximum temperature of the disk, mainly through our fitting for the extinction parameter, $A_V$. 
The results from this procedure are highly sensitive to the flux calibration of the bluest parts of our spectrum, 
as well as our assumption regarding $M_*/R_*$.   While our modeling has produced
values of $T_\textrm{max}$, from which we estimate accretion rates,  ultraviolet spectra {\ would more directly} probe 
the hottest and most rapidly rotating, innermost parts of these disks.  {\ Such observations}
would thus better constrain the maximum temperature and the temperature profile of the boundary region,
as well as the maximum velocity, and the velocity profile of this region.

\section{Conclusion}
We have presented new flux-calibrated medium-resolution spectrophotometry and high-resolution spectra of the two outbursting FU Ori objects, HBC 722 and Gaia 17bpi. Using the data, we constructed SEDs from 3900 \AA\ to 2.4 microns to compare against a geometrically thin accretion disk model made up of concentric annuli radiating at temperatures given by a modified Shakura-Sunyaev profile known as the $\a$-disk. To obtain credible intervals for parameters and explore the various degeneracies in the parameter space, we utilized Bayesian Markov Chain Monte Carlo techniques to accomplish this task. Given our sets of parameters obtained from fitting model SEDs to data, we assessed consistency with our spectra, and found good agreement between models and data at that resolution as well, though some over-prediction of line depths. 

We find the accretion rate of HBC 722 to be $\dot{M} = 10^{-4.90} M_\odot \textrm{ yr}^{-1}\; {}^{+0.99}_{-0.40} \textrm{ dex}$ and of Gaia 17bpi to be $\dot{M} = 10^{-6.70} M_\odot \textrm{ yr}^{-1}\; {}^{+0.46}_{-0.36} \textrm{ dex}$, with the latter being an order of magnitude and a half lower than most previously studied FU Ori sources. 
Notably, the radius of Gaia 17bpi was found to be $0.45^{+0.19}_{-0.09}R_\odot$, about 4-5 times lower than that of HBC 722.  {\ Only a young brown dwarf would be expected to have a radius this small.} These model results are ultimately related to the low luminosity of Gaia 17bpi compared to other FU Ori objects. We suggest that in this case, the model may need to include not only the disk but the central {\ source} as well. 


Furthermore, we conclude that constraining the power-law index, $\gamma$, of the temperature profile of the boundary region of the star-disk interface is impossible given the uncertainties of our current data. We therefore can only place constraints on $\gamma$ assuming strong constraints on other parameters. 

We believe that future investigation into FU Ori objects, in order to provide novel points of view that our data could not capture, would require quantitative modeling and fitting to high-resolution spectra and/or ultraviolet spectra. 
SED and spectrophotometric fitting can constrain only some parameters well, such as $A_V$ and $R_*$,
and others very poorly, such as $i$ and $M_*$.  We believe that our credible intervals on $\dot{M}$ are robust.
Overall, our study continues to support the geometrically thin, optically thick $\a$-disk model as reasonable in capturing the basic nature of the FU Ori phenomenon by matching the overall spectral energy distribution from the optical to the near-infrared. However, the continued difficulty in closely matching high-resolution spectra across various wavelengths raises important questions to drive future studies that will add to our understanding of FU Ori objects and their broader role in star formation.

More outburst examples, and detailed studies of the early stages of FU Ori outbursts,
are needed in order to understand the processes involved in outburst origin and evolution.

\section{Acknowledgements}
We are grateful for the allocation of observing time on the Palomar 200" telescope through the Caltech SURF Observing Program. 
We appreciate the observatory staff at both Palomar and WMKO for  excellent operations that enabled the new optical spectroscopic data presented here. 
We thank {\'A}gnes {K{\'o}sp{\'a}l} for providing her published infrared spectrum of HBC 722. 
We also thank Mike Connelley for use of his infrared spectrum of Gaia 17bpi,
which is of higher quality than the one published in our discovery paper.  We acknowledge
useful discussions with Roger W. Romani and his research group, as well as early discussions with Michael Kuhn. 
We thank Lee Hartmann for constructive feedback regarding accretion disk fundamentals. We additionally thank the anonymous referee for insightful comments and feedback that helped improve the final manuscript.

\bibliography{main}{}
\bibliographystyle{aasjournal}
\appendix
\section{Disk Model Parameters}
\label{sec:appendix_tutorial}
The parameters in our model disk are the following:
\begin{itemize}
    \item $M_*$, the mass of the central star.
    \item $R_*$, the radius of the central star, which we effectively take as $r_i$, the radius of the innermost disk annulus. 
    \item  $\dot{M}$, the rate of mass accretion through the disk onto the star.
    \item $i$, the inclination of the disk with respect to our line of sight, where $i = 0^\circ$ is face-on.
    \item $A_V$, the extinction within the $V$-band.
    \item $\gamma$, the power law exponent governing the temperature distribution in the boundary region of the disk at $r < 1.36 R_*$.
    \item $R_\textrm{outer}$, the distance of the outer edge of the disk from the center.
\end{itemize}
In this section, we discuss the effects that each of these parameters has on the resulting temperature profile, the SED, and the spectrum. In Figures \ref{fig:m_m_dot_tradeoff} and \ref{fig:r_and_a_v}, we show model SEDs with fiducial parameters: $M = 0.8M_{\odot}$, $R = 1.6R_{\odot}$, $\dot{M} = 1 \times 10^5 M_{\odot} \:\text{yr}^{-1}$, $i = 45^{\circ}$, $A_V = 3$, $d = 1$ kpc, $R_{\text{outer}} = 100R_{*}$, and $\gamma = 0$. We vary some of these parameters individually to show what effects they have on the overall SEDs, and discuss how these parameters couple and trade off with one another before explaining how we are able to place reasonable constraints on them. In creating these tutorial SEDs, we use blackbodies for simplicity, and note that the reddened SEDs are not perfectly smooth due to the effects of extinction. In the subsections below, we describe each of the parameters in greater detail, and their individual effects on the disk SED. While we plot a wider range of wavelengths in the figures below, we remind the reader that the data we have available only spans from 0.39-2.4 $\mu$m.

\subsection{Stellar Mass, $M_*$}\label{sec:m_star}

\begin{figure}
    \centering
    \includegraphics[scale=0.45]{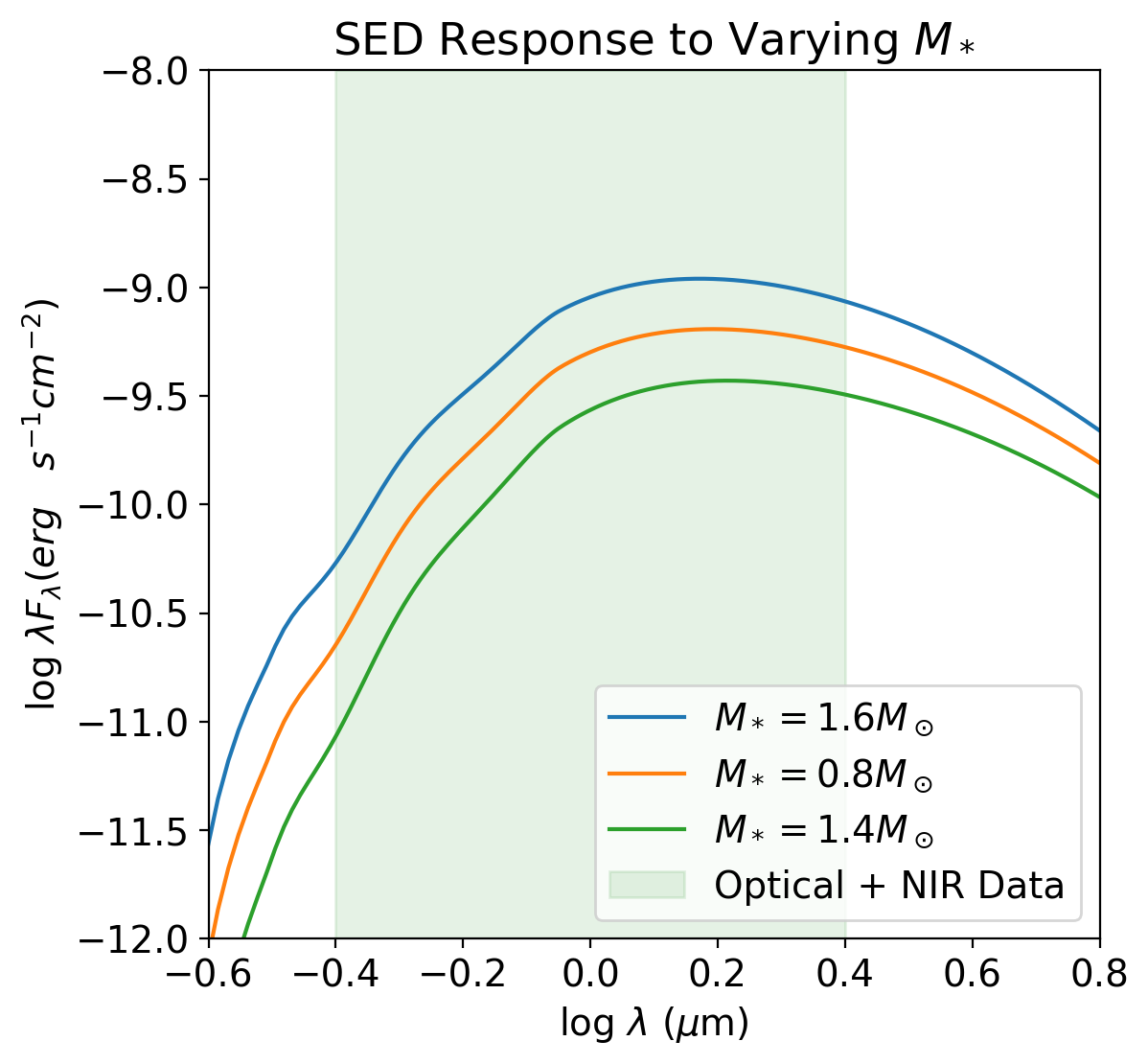}\includegraphics[scale=0.45]{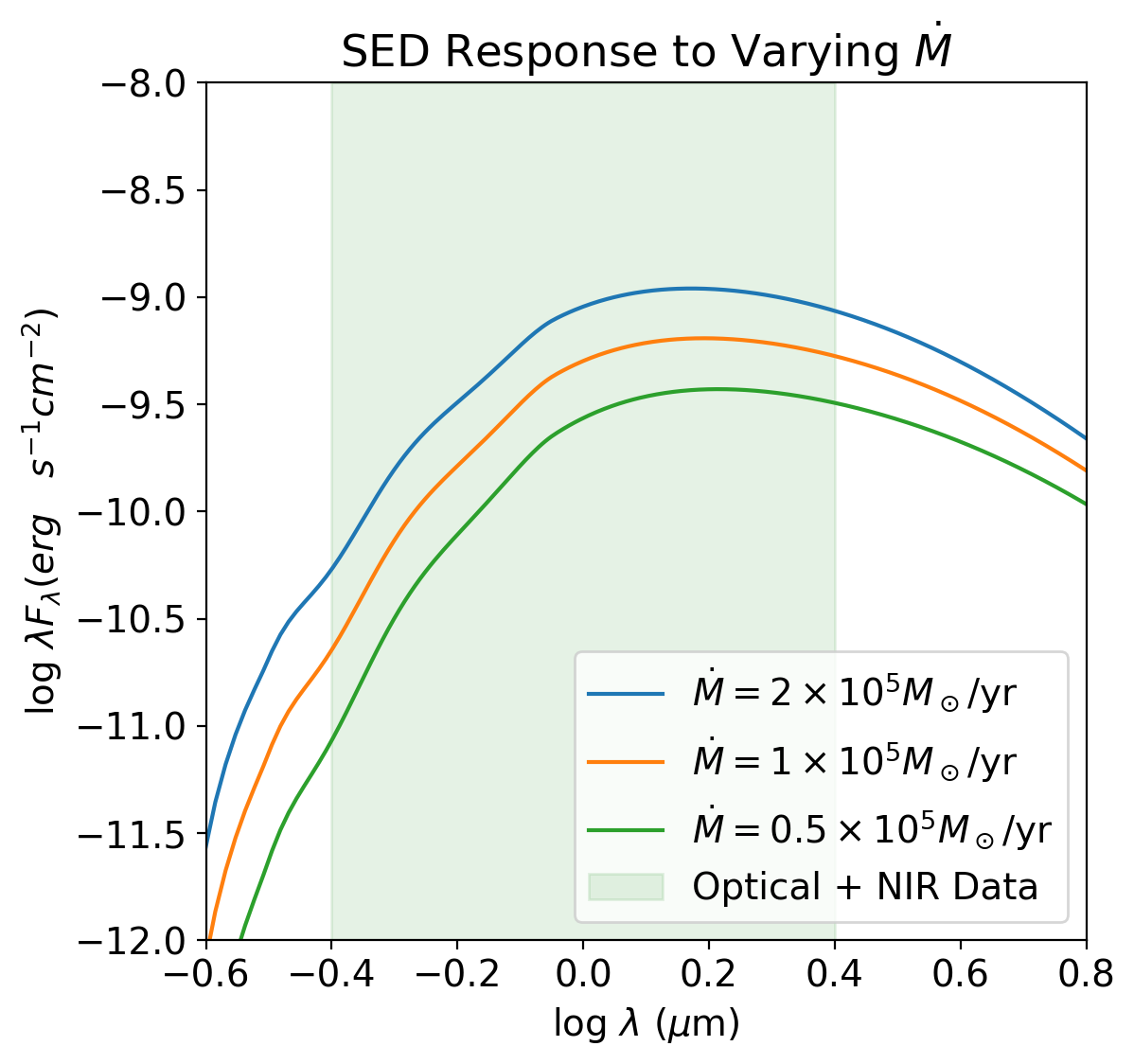}\\
    \caption{SED appearance as a function of stellar mass (left panel) and disk accretion rate (right panel). By increasing either $M$ or $\dot{M}$, the overall flux of an FU Ori disk is increased due to the increasing maximum temperature, 
    and the SED also becomes slightly more blue.
    The two plots visually illustrate what appears mathematically in Equation \ref{eq:new_temp_profiles}: $M$ and $\dot{M}$ directly trade off with one another, with the same SED reproduced by increasing one while decreasing the other. }
    \label{fig:m_m_dot_tradeoff}

\end{figure}

From Equation \ref{eq:new_temp_profiles}, it can be seen that $T^4\propto M_*$, so raising the mass of the central star raises $T_\textrm{max}$, the maximum temperature of the disk. As a result, the SED shifts slightly blueward and the overall flux level increases, as does the source luminosity. In the SED, the $M_*$ parameter directly trades off with the accretion rate $\dot{M}$, as seen in Figure \ref{fig:m_m_dot_tradeoff}, since increasing $\dot{M}$ has the same effect as increasing $M_*$. One way that we can constrain $M_*$ is by fixing the ratio of $M_*/R_*$ to reasonable values as predicted by pre-main sequence stellar evolution theory. This then also allows us to better constrain $\dot{M}$.  

In regard to the spectrum, increasing $M_*$, leads to an increase in $T_{\text{max}}$ and hence spectral signatures of hotter temperatures would appear.
A more subtle change occurs in the spectral  line broadening. Equation \ref{eq:disk_broaden_2}
demonstrates that $\l_\textrm{max} \propto \sqrt{M_*}$. Thus, increasing the mass of the central star makes the broadening effect more pronounced. 
The expected $M_*$ range can be large, but for the typical young stellar object we expect variations in this parameter 
among sources of only order unity (say, $M_*$ = 0.4$M_\odot$ or 0.8$M_\odot$). 
As articulated below, the parameters $M_*$, $R_*$, and $i$
contribute almost equally to line broadening effects, and one does not dominate over the others.

\subsection{Accretion Rate, $\dot{M}$}

From equation \ref{eq:new_temp_profiles}, it can be seen that $T^4\propto \dot{M}$, so the effects on the SED are the same as for $M_*$. Specifically, increasing $\dot{M}$ raises $T_\textrm{max}$, the maximum temperature of the disk. As a result, the SED shifts blueward and the overall flux level increases, along with the total luminosity. 

There are no differences in the broadening effects of the spectrum from increasing $\dot{M}$ alone, since $\dot{M}$ does not appear in Equation \ref{eq:disk_broaden}. The only effects on the spectrum are those associated with the increase in $T_\textrm{max}$, as already discussed in section \ref{sec:m_star}. 

\subsection{Stellar Radius, $R_*$}

\begin{figure}
    \centering
    \includegraphics[scale=0.45]{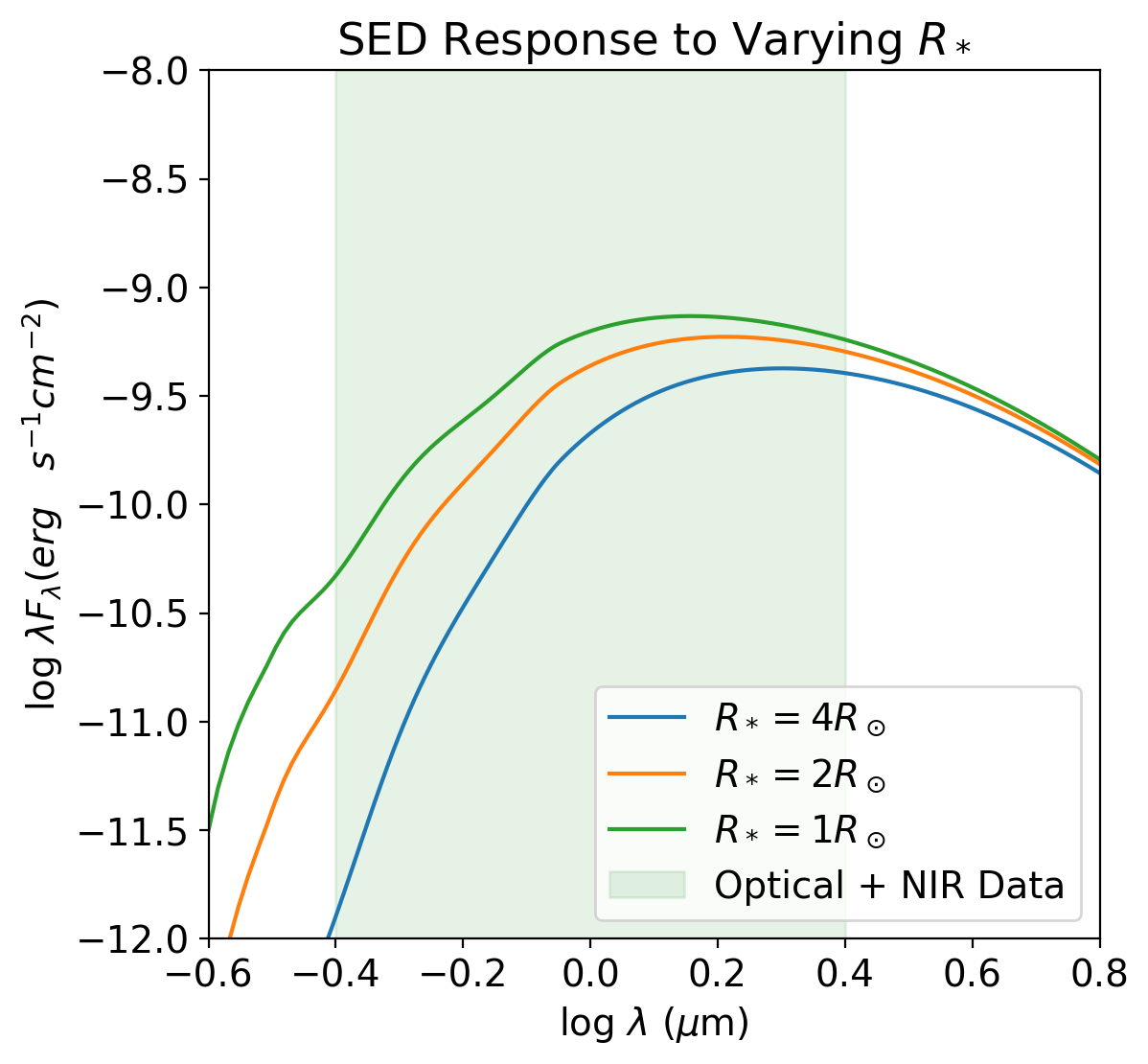}\includegraphics[scale=0.45]{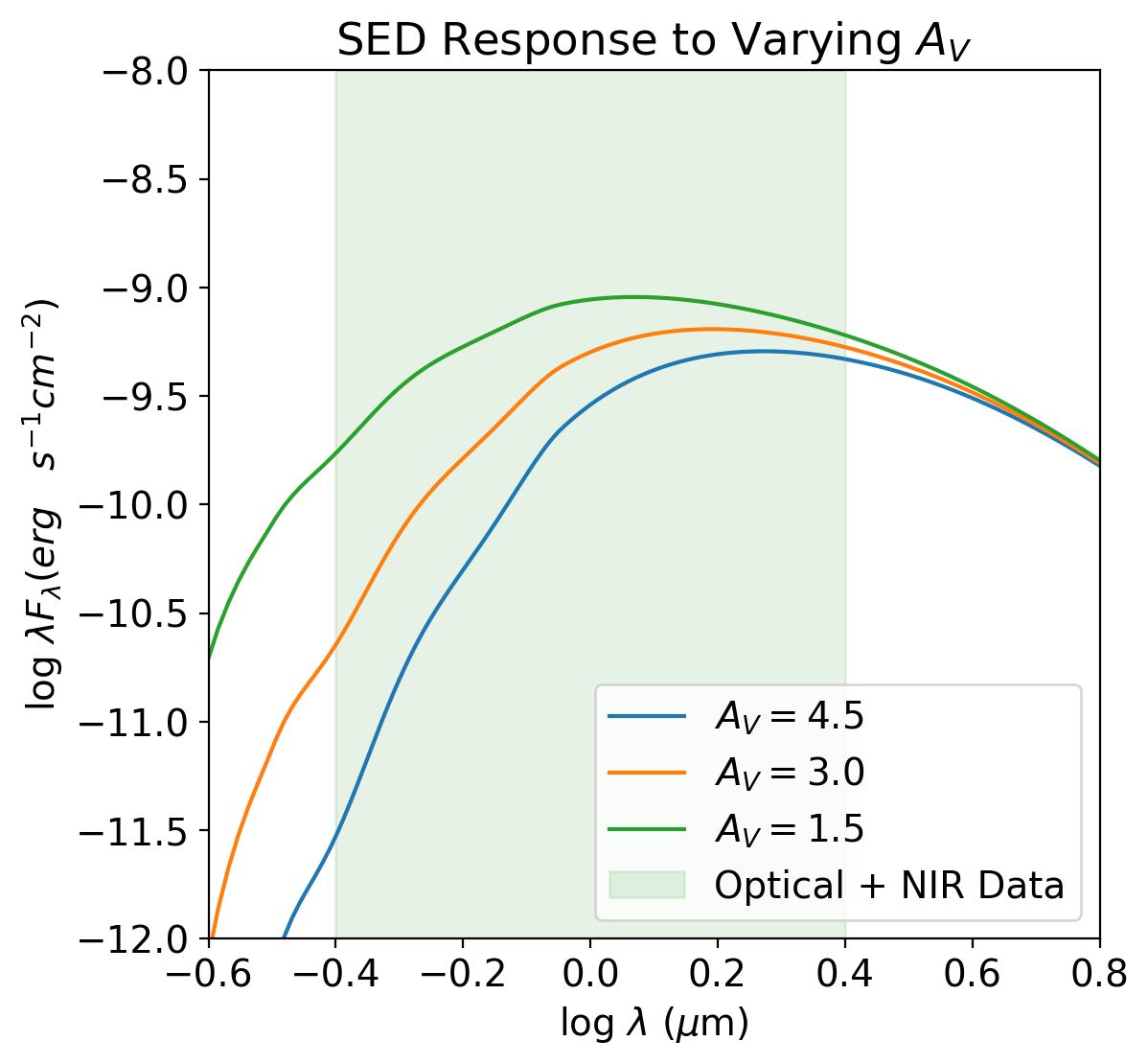}\\
    \caption{SED appearance as a function of stellar radius (left) and extinction (right). Increasing $R$ decreases the overall flux of an FU Ori accretion disk, by decreasing the maximum temperature, leading also to a slightly bluer SED.
    Increasing $A_V$ has the effect of reddening the SED. Because $A_V$ is a property of the \text{line-of-sight} medium and not the FU Ori \text{accretion}, this parameter has no effect on the temperature distribution of the object. Moreover, while these two parameters appear to trade off with one another similar to the two variables in Figure \ref{fig:m_m_dot_tradeoff}, the fact that one parameter changes the temperature distribution of the disk, while the other does not, allows us to decouple their effects (especially when using stellar atmospheres instead of the blackbodies shown here).}
    \label{fig:r_and_a_v}
\end{figure}

From equation \ref{eq:new_temp_profiles}, since the accretion disk starts at values of $r > R_*$, it is clear that $(R_*/r) < 1$. As $T^4 \propto (1-(R_*/r)^{1/2})$, increasing $R_*$ at a given $r$ lowers the temperature at that point, and therefore also the overall $T_\textrm{max}$ of the disk. The overall SED shifts redward when $R_*$ is increased, as demonstrated visually in \ref{fig:r_and_a_v}. In a sense, $R_*$ trades off with $A_V$, but not in a closed-form way as is true with $M_*$ and $\dot{M}$. This trade makes it easier to constrain the values of $R_*$ in \S \ref{sec:results}.

The effects on the spectrum are inverse to those outlined in \S \ref{sec:m_star} regarding $M_*$.
The value of $R_*$ sets the minimum value of the $\sqrt{r}$ term in the denominator of Equation \ref{eq:disk_broaden_2}.
Larger values of $R_*$ will lead to the innermost, hottest annulus being located farther out from the central star, and thus spinning at a lower Keplerian speed. This, in turn, leads to less pronounced spectral line broadening. Just as in the case of $M_*$, a pronounced effect on the overall spectrum comes into play through the temperature decrease that results from increasing $R_*$.

\subsubsection{Mass-to-Radius Ratio, $M_*/R_*$}

\begin{figure}
    \centering
    \includegraphics[scale=0.45]{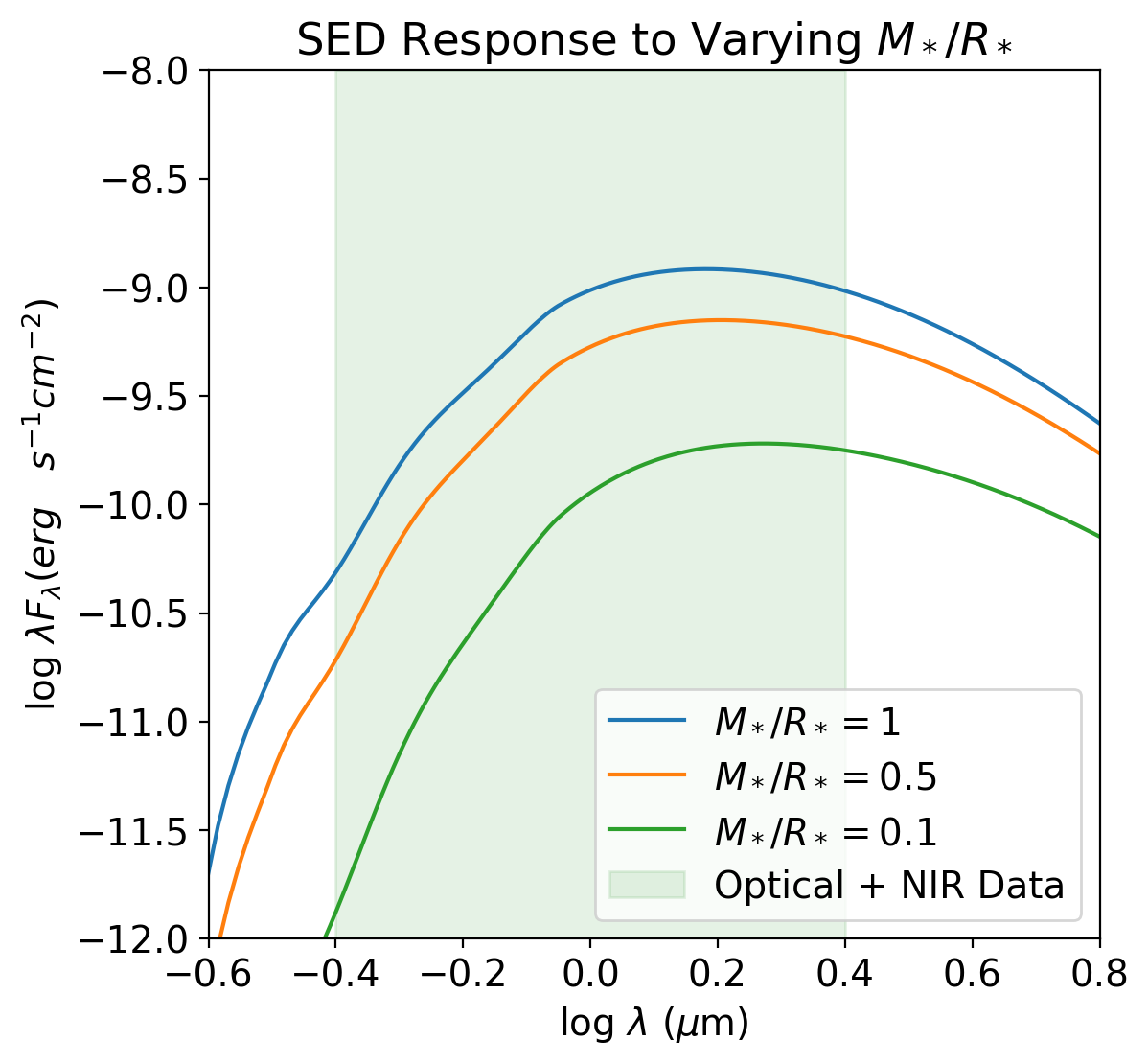}
    \caption{SED appearance as a function of stellar mass-to-radius ratio, $M_*/R_*$ in solar units ($M_\odot/R_\odot$). 
    Values as low as 0.1 are appropriate only for very low mass stars and brown dwarfs, at the youngest ages,
    whereas values as high as 1 are appropriate only for $>1 M_\odot$ stars. We fix this parameter at 0.5 while generating Model 3.}
    \label{fig:m_r_ratio_demo}
\end{figure}

We maintain the fiducial radius of 2.0 $R_\odot$ and vary $M_\odot$ to obtain values of $M_*/R_* = 0.1, 0.5, 1$ in solar units ($M_\odot/R_\odot$). We produce SEDs at each value in Figure \ref{fig:m_r_ratio_demo}, noting that higher values of $M_*/R_*$ lead to an overall bluer SED and lower values lead to a redder SED. In terms of the spectrum, the individual line broadening will be strongly affected due to the dependence of the broadening kernel on this ratio as demonstrated in Equation \ref{eq:disk_broaden_2}. In addition, the overall spectrum is affected due to the strong temperature dependence on this ratio.

\subsection{Visual Extinction, $A_V$}

$A_V$ has no effect on the temperature profile or any other physical properties of the FU Ori system, since the extinction is governed by \text{line-of-sight} effects between the accretion disk system and Earth. However, $A_V$ does affect the observed SED and can thus trade off with other parameters in our model. Values of $A_V$ for most FU Ori objects, as reported by \cite{gramajo2014} and \cite{connelley2018}, are generally higher than 1.5 mag, as the sources are embedded in dusty regions and sometimes circumstellar envelopes. 
Increasing the value of $A_V$ shifts the SED redward, as illustrated in Figure \ref{fig:r_and_a_v}. 
The adopted reddening law also introduces some bumps and wiggles in the SED. 

Finally, it follows that there are no effects on the spectral lines from the $A_V$ parameter, other than an overall relative change in flux level with wavelength due to reddening. 

\subsection{Inclination, $i$}

\begin{figure}
    \centering
    \includegraphics[scale=0.45]{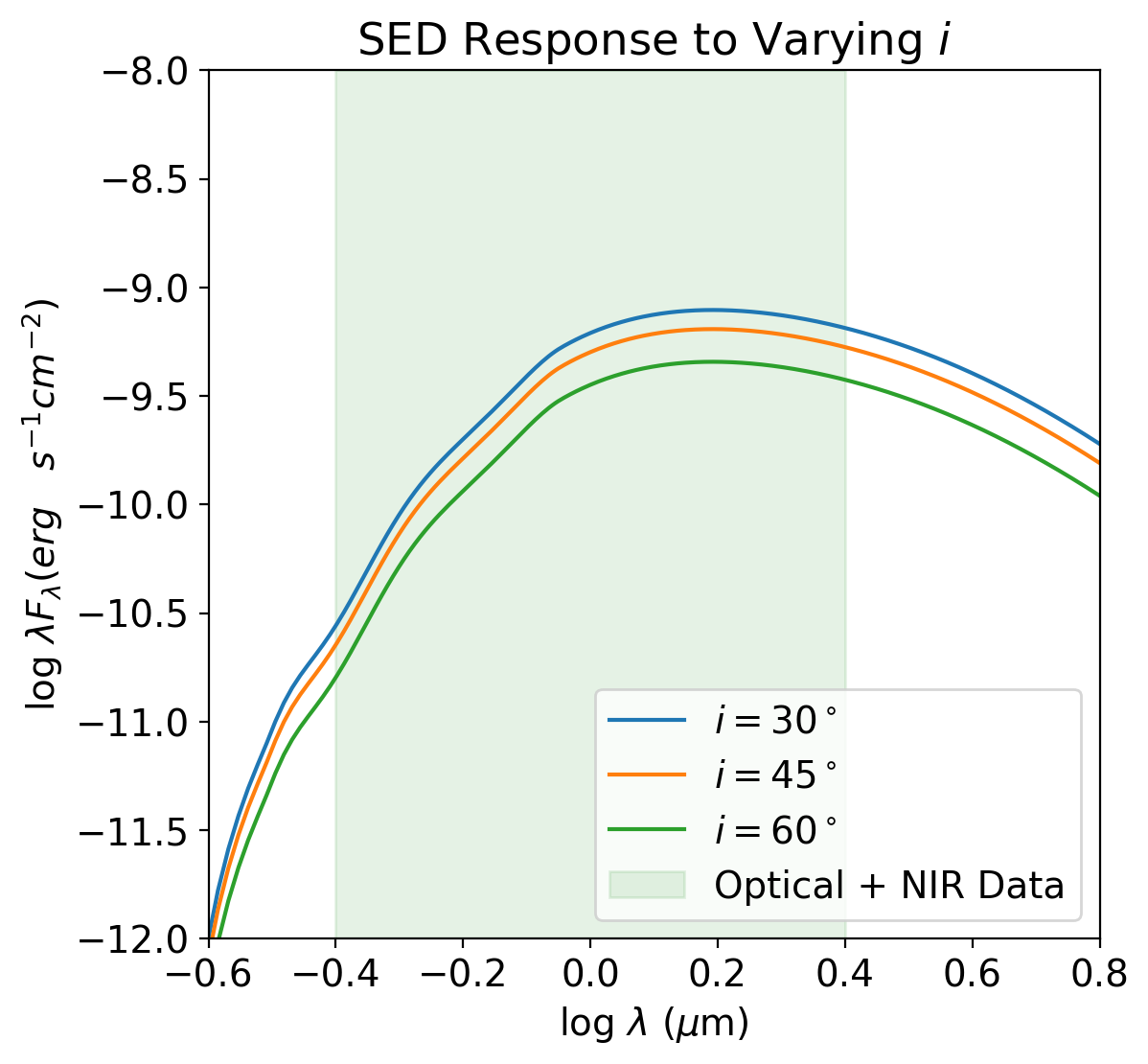}\includegraphics[scale=0.45]{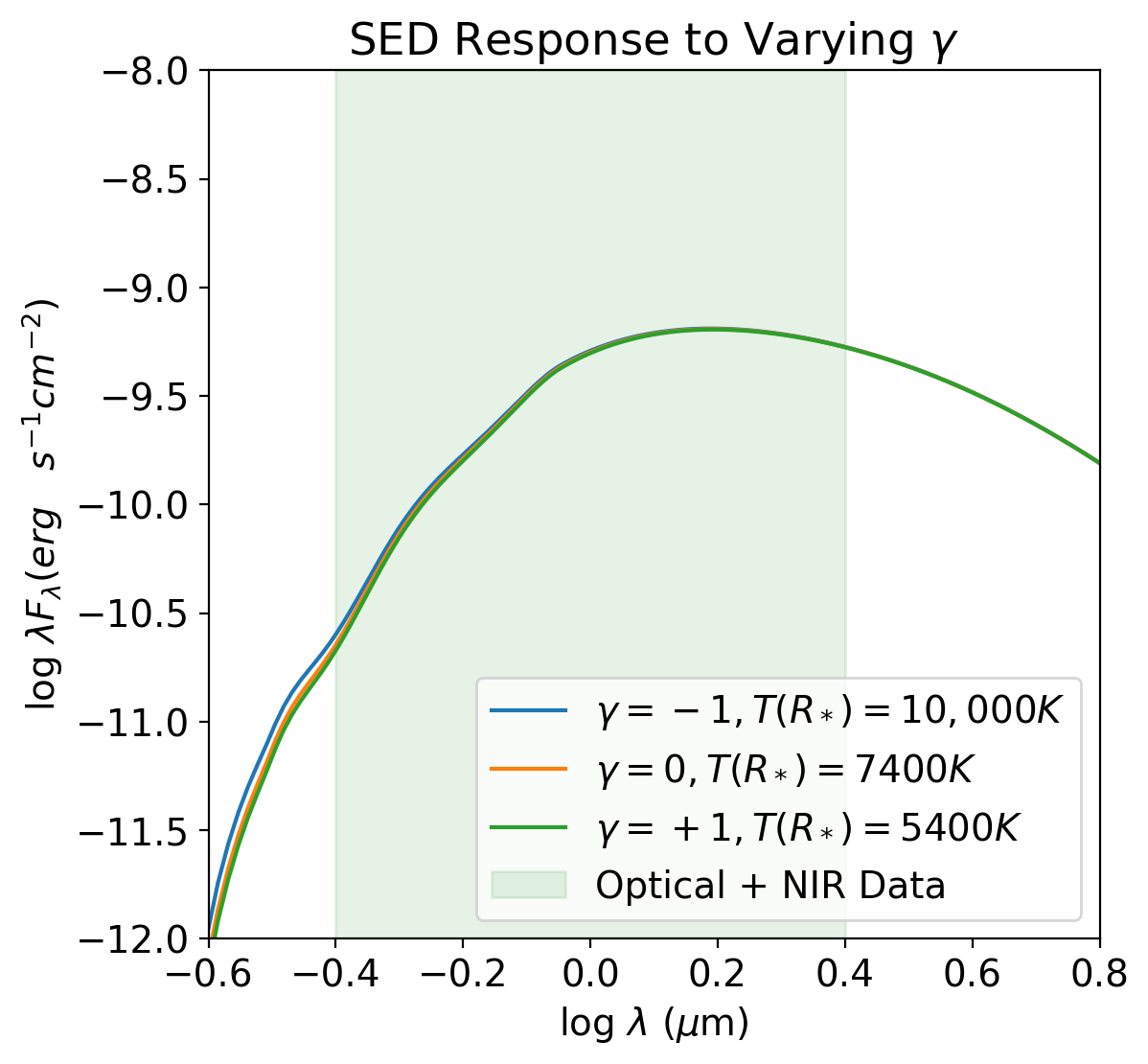}\\
    \caption{SED appearance as a function of inclination (left) and temperature profile in inner disk (right). These parameters have minimal but noticeable effects on the overall SED. The inclination, $i$, translates the SED up and down. The power-law index, $\gamma$, despite significantly changing the temperature at the star-disk interface, has only minute effects on the bluest parts of the SED.} 
    \label{fig:gamma_and_inc}
\end{figure}

The disk inclination parameter has no effect on the temperature profile, but  because $F_\lambda \propto \cos i$ 
(Equation \ref{eq:disk_flux}), 
it does vertically shift the SED  as seen in Figure \ref{fig:gamma_and_inc}. 

Increasing $i$ increases the line broadening in a disk spectrum, 
with $\sin i$ varying by order unity (for example, $\sin 60^\circ  = \sqrt{3}\sin 30^\circ$). 
From Equation \ref{eq:disk_broaden}, it is clear that as $i$ is increased, the broadening is stronger, with more significant double profiles seen towards higher inclinations. 
Varying $i$ affects line broadening on the same order as varying $M_*$ and $R_*$.

\subsection{Boundary Region Power Law Exponent, $\gamma$}

We refer the reader to \S \ref{sec:gamma_parameter} for a full motivation and explanation of the $\gamma$ parameter,
which describes the temperature exponent in the innermost disk regions. 
Figure \ref{eq:new_temp_profiles} shows that $\gamma$ controls the temperature profile in the innermost part of the disk, what we refer to as the ``boundary region". The subsequent effect of $\gamma$ on the SED is shown in Figure \ref{fig:gamma_and_inc}, which is only very minimal, and slightly noticeable in the bluest parts of the SED.  

In terms of spectral signatures, $\gamma$ directly influences the temperature profile and thus the spectral signatures
as a function of wavelength, particularly the shortest wavelength regions of the spectrum.  It does not affect line broadening.

\subsection{Disk Outer Radius, $R_{\textrm{outer}}$}

The outer disk radius describes where we truncate the pure accretion disk. While young star accretion disks typically have outer dusty envelopes as described in \citep{hartmannreview2016}, we only model the thin inner disk since the outer envelopes do not contribute substantially to the observed flux in the SED at optical and infrared wavelengths. 
In Figure \ref{fig:fracflux} we show that annuli farther out than $100R_*$ contribute to the total flux in the optical and near infrared wavelengths at four orders of magnitude less than annuli within $10R_*$ (where $R_*$ is on the same order of magnitude as $R_\odot$). This justifies our choice of $R_{\text{outer}} = 100R_*$ as the fiducial value for this parameter.
\begin{figure}
    \centering
    \includegraphics[scale=0.45]{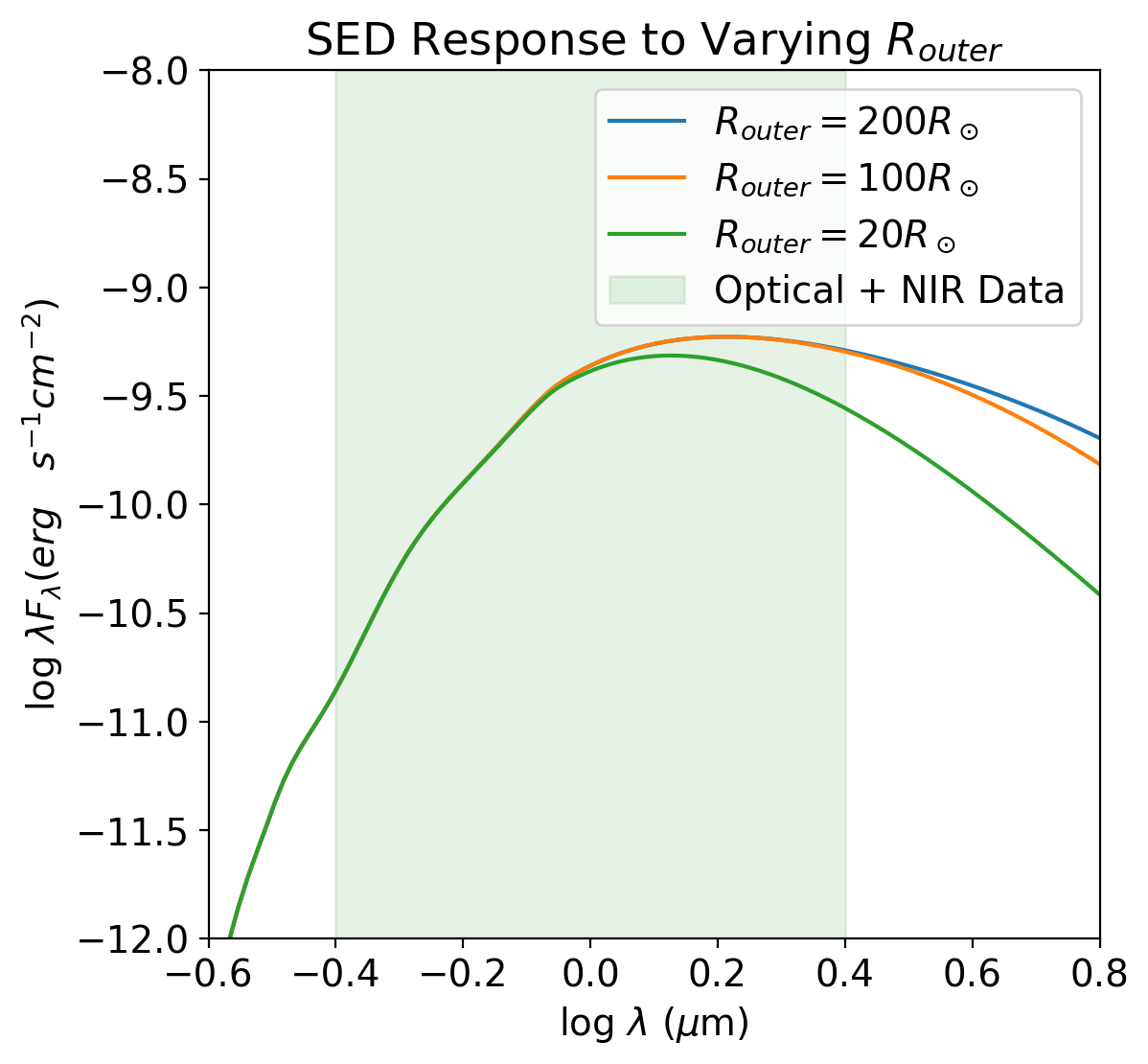}
    \caption{SED appearance as a function of outer disk radius, $R_\textrm{outer}$.  This parameter most strongly affects the SED in the near-infrared. Annuli beyond $100 R_\odot$ have negligible contribution to the overall disk model in the wavelengths of concern in this study.}
    \label{fig:r_outer_demo}
\end{figure}

Our fiducial outer disk radius is large enough to encompass all of the contributions to the optical and
near-infrared flux, so it is essentially infinite for our purposes, even though only about 0.5 AU in practice. 
However, there is reason to believe that it is not the physical outer radius of the disk accretion outburst that matters, as much as a minimum temperature above which the disk is unstable. We thus allow the option of variation in the
outer disk truncation radius (or temperature).


\end{document}